\documentclass[a4paper,11pt]{article}
\pdfoutput=1 

\usepackage{jheppub} 

\usepackage[T1]{fontenc} 
\usepackage{diagbox}
\usepackage{multirow}
\usepackage{slashed}

\usepackage[utf8]{inputenc}

\usepackage{longtable,pdflscape}
\usepackage{stackengine}
\usepackage[export]{adjustbox}

\allowdisplaybreaks

\def\be{\begin{equation}}
\def\ee{\end{equation}}
\def\beq{\begin{eqnarray}}
\def\eeq{\end{eqnarray}}

\newcommand{\gam}{\gamma}

\newcommand{\msbar}{\overline{\text{MS}}}
\def\nn{\nonumber}

\def\mH{\mathcal{H}}
\def\mC{\mathcal{C}}
\def\mQ{\mathcal{Q}}

\def\gev{{\rm GeV}}
\def\mev{{\rm MeV}}

\usepackage{tikz}
\usetikzlibrary{calc}
\usetikzlibrary{trees}
\usetikzlibrary{decorations.pathmorphing}
\usetikzlibrary{decorations.markings}
\usetikzlibrary{arrows.meta}

\tikzset{
	vector/.style={decorate, decoration={snake}, draw},
	provector/.style={decorate, decoration={snake,amplitude=2.5pt}, draw},
	antivector/.style={decorate, decoration={snake,amplitude=-2.5pt}, draw},
	fermion/.style={draw=black, postaction={decorate},
		decoration={markings,mark=at position .5 with {\arrow[draw=black]{>}}}},
	fermion1/.style={draw=black, postaction={decorate},
		decoration={markings,mark=at position .6 with {\arrow[draw=black]{>}}}},
	fermionbar/.style={draw=black, postaction={decorate},
		decoration={markings,mark=at position .5 with {\arrow[draw=black]{<}}}},
	fermionbar1/.style={draw=black, postaction={decorate},
		decoration={markings,mark=at position .6 with {\arrow[draw=black]{<}}}},
	fermionnoarrow/.style={draw=black},
	gluon/.style={decorate, draw=black,
		decoration={coil,amplitude=4pt, segment length=5pt}},
	scalar/.style={dashed,draw=black, postaction={decorate},
		decoration={markings,mark=at position .5 with {\arrow[draw=black]{>}}}},
	scalarbar/.style={dashed,draw=black, postaction={decorate},
		decoration={markings,mark=at position .5 with {\arrow[draw=black]{<}}}},
	scalarnoarrow/.style={dashed,draw=black},
	electron/.style={draw=black, postaction={decorate},
		decoration={markings,mark=at position .5 with {\arrow[draw=black]{>}}}},
	bigvector/.style={decorate, decoration={snake,amplitude=4pt}, draw},
}

\title{\Large \bf \boldmath Probing new physics in class-I $B$-meson decays into heavy-light final states}

\author[a]{Fang-Min Cai,}
\author[a]{Wei-Jun Deng,}
\author[a,1]{Xin-Qiang Li\note{Corresponding author.},}
\author[a]{and Ya-Dong Yang}

\affiliation[a]{Institute of Particle Physics and Key Laboratory of Quark and Lepton Physics~(MOE), Central China Normal University, Wuhan, Hubei 430079, P.~R. China}

\emailAdd{caifangmin@mails.ccnu.edu.cn}
\emailAdd{dengweijun@mails.ccnu.edu.cn}
\emailAdd{xqli@mail.ccnu.edu.cn}
\emailAdd{yangyd@mail.ccnu.edu.cn}

\abstract{With updated experimental data and improved theoretical calculations, several significant deviations are being observed between the Standard Model predictions and the experimental measurements of the branching ratios of $\bar{B}_{(s)}^0\to D_{(s)}^{(*)+} L^-$ decays, where $L$ is a light meson from the set $\{\pi,\rho,K^{(\ast)}\}$. Especially for the two channels $\bar{B}^0\to D^{+}K^-$ and $\bar{B}_{s}^0\to D_{s}^{+}\pi^-$, both of which are free of the weak annihilation contribution, the deviations observed can even reach 4-5$\sigma$. Here we exploit possible new-physics effects in these class-I non-leptonic $B$-meson decays within the framework of QCD factorization. Firstly, we perform a model-independent analysis of the effects from twenty linearly independent four-quark operators that can contribute, either directly or through operator mixing, to the quark-level $b\to c\bar{u} d(s)$ transitions. It is found that, under the combined constraints from the current experimental data, the deviations observed could be well explained at the $1\sigma$ level by the new-physics four-quark operators with $\gam^{\mu}(1-\gam_5)\otimes\gam_{\mu} (1-\gam_5)$ structure, and also at the $2\sigma$ level by the operators with $(1+\gam_5)\otimes(1-\gam_5)$ and $(1+\gam_5)\otimes(1+\gam_5)$ structures. However, the new-physics four-quark operators with other Dirac structures fail to provide a consistent interpretation, even at the $2\sigma$ level. Then, as two specific examples of model-dependent considerations, we discuss the case where the new-physics four-quark operators are generated by either a colorless charged gauge boson or a colorless charged scalar, with their masses fixed both at the $1$~TeV. Constraints on the effective coefficients describing the couplings of these mediators to the relevant quarks are obtained by fitting to the current experimental data.}

\begin{document} 
\maketitle
\flushbottom

\section{Introduction}
\label{sec:intro}

Flavor physics plays always an important role in testing the Standard Model (SM) of particle physics and probing new physics (NP) beyond it~\cite{Buchalla:2008jp,Antonelli:2009ws}. Here, the non-leptonic weak decays of $B$ mesons are of particular interest, since they provide direct access to the fundamental parameters of the Cabibbo-Kobayashi-Maskawa (CKM) matrix~\cite{Cabibbo:1963yz,Kobayashi:1973fv} and further insight into the strong-interaction dynamics involved in these decays. Aiming at such a goal, the BaBar and Belle collaborations~\cite{Bevan:2014iga}, as well as the LHCb experiment~\cite{Bediaga:2012py} have already performed many high-precision measurements of these kinds of decays~\cite{ParticleDataGroup:2020ssz,HFLAV:2019otj}. In addition, new frontiers of precision are expected in the era of Belle II~\cite{Kou:2018nap} and upgraded LHCb~\cite{Bediaga:2018lhg}. 

Confronted with the plethora of high precision measurements made by these dedicated experiments, we are forced to improve as much as possible the accuracy of theoretical predictions about these non-leptonic weak decays. Here the main challenge we are now facing is how to calculate reliably the hadronic matrix elements of four-quark operators in the effective weak Hamiltonian (see subsection~\ref{subsec:effectiveD}). For a long time, the naive factorization (NF) assumption~\cite{Bauer:1986bm} and modifications thereof (see, \textit{e.g.}, refs.~\cite{Neubert:1997uc,Ali:1997nh,Ali:1998eb,Chen:1999nxa} and references therein) were used to estimate the non-leptonic $B$-decay amplitudes. Several more promising strategies built upon either the $SU(3)$ flavor symmetry of strong interactions~\cite{Zeppenfeld:1980ex,Savage:1989ub,Gronau:1994rj} or the factorization theorem, such as the QCD factorization (QCDF)~\cite{Beneke:1999br,Beneke:2000ry,Beneke:2001ev} and its field theoretical formulation, the soft-collinear effective theory~\cite{Bauer:2000ew,Bauer:2000yr,Bauer:2001yt,Beneke:2002ph,Beneke:2002ni}, as well as the perturbative QCD~\cite{Keum:2000ph,Keum:2000wi,Lu:2000em} approach, have been developed to study the same problem. Certain combinations of these approaches have also been adopted in, \textit{e.g.}, refs.~\cite{DescotesGenon:2006wc,Zhou:2015jba,Zhou:2016jkv}.

In this paper, we shall consider the exclusive two-body $\bar{B}_{(s)}^0\to D_{(s)}^{(*)+} L^-$ decays, where $L\in\{\pi,\rho,K^{(\ast)}\}$, within the QCDF framework. For these class-I non-leptonic decays, the spectator antiquark and other light degrees of freedom of the initial $\bar{B}_{(s)}^0$ mesons need to rearrange themselves only slightly to form the heavy $D_{(s)}^{(*)+}$ mesons together with the charm quark created in the weak $b\to c\bar{u}d(s)$ transitions, while the light quark-antiquark pair $\bar{u}d(s)$ must be highly energetic and collinear to form the light meson $L^-$ but with energy of $\mathcal{O}(m_b)$, where $m_b$ denotes the bottom-quark mass. For such a configuration, a factorization formula, valid in the heavy-quark limit up to power corrections of $\mathcal{O}(\Lambda_\mathrm{QCD}/m_b)$,\footnote{Here we treat the bottom and charm quarks as massive while the light quarks as massless. The heavy-quark limit is defined as $m_b,m_c\gg \Lambda_\mathrm{QCD}$ but with $m_c/m_b$ fixed, where $\Lambda_\mathrm{QCD}$ is the QCD intrinsic scale.} can be established for the hadronic matrix elements governing the decays; explicitly, we have~\cite{Beneke:2000ry,Bauer:2001cu}
\begin{equation}\label{eq:QCDF}
\langle D_{(s)}^{(*)+} L^-|\mQ_{i}|\bar{B}_{(s)}^0\rangle=\sum_{j} F_{j}^{B_{(s)}\to D_{(s)}^{(*)}}(m_{L}^{2})\,\int_{0}^{1} du\,T_{ij}(u)\,\Phi_{L}(u)+\mathcal{O}(\Lambda_\mathrm{QCD}/m_b)\,,
\vspace{-0.35cm}
\end{equation}
where the $B_{(s)}\to D_{(s)}^{(*)}$ transition form factors, $F_{j}^{B_{(s)}\to D_{(s)}^{(*)}}$, and the light-cone distribution amplitude (LCDA), $\Phi_{L}(u)$, of the light meson encode all the long-distance strong-interaction effects, both of which can be extracted from experimental data or calculated by using non-perturbative methods like QCD sum rules and/or lattice QCD. The hard kernels $T_{ij}(u)$ receive, on the other hand, contributions only from scales of $\mathcal{O}(m_b)$ in the heavy-quark limit and are therefore calculable perturbatively. At leading order (LO) in the strong coupling $\alpha_s$, eq.~\eqref{eq:QCDF} reproduces the NF result, and both the next-to-leading-order (NLO)~\cite{Beneke:2000ry,Politzer:1991au} and next-to-next-to-leading-order (NNLO)~\cite{Huber:2015bva,Huber:2016xod} corrections to $T_{ij}(u)$ are now known.

As all the four quark flavors in $b\to c\bar{u}d(s)$ transitions are different from each other, these tree-level decays receive contributions from neither the penguin operators nor the penguin topology. There is also no color-suppressed tree topology in these class-I decays. At leading power in $\Lambda_\mathrm{QCD}/m_b$, these decays are dominated by the color-allowed tree topology that receives only vertex corrections, while interactions with the spectator antiquark and the weak annihilation topology are both power-suppressed~\cite{Beneke:2000ry}. In fact, noting that the weak annihilation topology contributes only to $\bar{B}^0\to D^{(*)+} \pi^-$ and $\bar{B}_{s}\to D_{s}^{(*)+} K^{-}$, but not to $\bar{B}^0\to D^{(*)+} K^{-}$ and $\bar{B}_{s}\to D_{s}^{(*)+} \pi^-$, one can use the ratios between the branching fractions of these two kinds of decays to probe the topology. Remarkably, the current experimental data shows already that the impact due to such a topology is negligible~\cite{Fleischer:2010ca}. Other sources of power corrections, such as the higher-twist corrections to the light-meson LCDAs and the exchange of a single soft gluon between the $B_{(s)}\to D_{(s)}^{(*)}$ transitions and the light meson, are also estimated to be quite small~\cite{Beneke:2000ry,Bordone:2020gao}. Therefore, these class-I non-leptonic decays are theoretically clean and the QCDF approach is expected to work well for them. However, with the updated input parameters, the SM predictions~\cite{Huber:2016xod,Bordone:2020gao,Chang:2016eto} are found to be generically higher than the current experimental measurements~\cite{ParticleDataGroup:2020ssz,HFLAV:2019otj} of the branching ratios of $\bar{B}_{(s)}^0\to D_{(s)}^{(*)+} L^-$ decays. Especially for the two channels $\bar{B}^0\to D^{+}K^-$ and $\bar{B}_{s}^0\to D_{s}^{+}\pi^-$, which are free of the weak annihilation contribution, the deviations observed can even reach 4-5$\sigma$, once the updated input parameters as well as the higher-order power and perturbative corrections to the decay amplitudes are taken into account~\cite{Bordone:2020gao}. As emphasized already in refs.~\cite{Huber:2016xod,Bordone:2020gao}, it is quite difficult to accommodate such a clear and significant discrepancy in the SM. In this paper, as an alternative, we shall examine possible NP interpretations of the deviations observed; for recent discussions along this line, see refs.~\cite{Bobeth:2014rda,Brod:2014bfa,Bobeth:2014rra,Lenz:2019lvd,Iguro:2020ndk,Bordone:2021cca}.

In the SM, these class-I decays only receive contributions from the four-quark operators with the Dirac structure $\gam^{\mu} (1-\gam_5)\otimes\gam_{\mu} (1-\gam_5)$,\footnote{Throughout this paper, we adopt the convention where the Dirac structures before and after the symbol ``$\otimes$'' should be inserted into the quark-bilinear currents $(\bar{c}\cdots b)$ and $(\bar{d}(\bar{s})\cdots u)$, respectively.} which originate in the tree-level $W^\pm$ exchanges. Beyond the SM, however, new four-quark operators with different Dirac structures can be generated and contribute potentially to the decays considered, either directly or through operator mixing under renormalization~\cite{Buchalla:1995vs,Buras:1998raa}. The full set of linearly independent four-quark operators with four different quark flavors in all possible extensions of the SM, together with their one- and two-loop QCD anomalous dimension matrices (ADMs), can be found in refs.~\cite{Ciuchini:1997bw,Ciuchini:1998ix,Buras:2000if}. The calculation of $\mathcal{O}(\alpha_s)$ corrections to the matching conditions for the short-distance Wilson coefficients of these operators have also been completed~\cite{Buras:2012gm}. Currently, the only missing ingredient aimed at a full NLO analysis of the class-I decays in any extension of the SM is the evaluation of the hadronic matrix elements of these four-quark operators, also at the NLO in $\alpha_s$. Thus, in this paper, we shall firstly calculate the NLO vertex corrections to the hadronic matrix elements of these four-quark operators within the QCDF framework, and then discuss in a model-independent way possible effects of these NP operators on the class-I decays. As emphasized already in ref.~\cite{Buras:2012gm}, such an NLO analysis in the NP sector is crucial for reducing certain unphysical scale and renormalization scheme dependences present in the absence of these $\mathcal{O}(\alpha_s)$ corrections~\cite{Bobeth:2014rda,Brod:2014bfa,Bobeth:2014rra,Lenz:2019lvd,Iguro:2020ndk,Bordone:2021cca}. It is numerically found that, under the combined constraints from the current experimental data, the deviations observed could be well explained at the $1\sigma$ level by the NP four-quark operators with $\gam^{\mu}(1-\gam_5)\otimes\gam_{\mu} (1-\gam_5)$ structure, and also at the $2\sigma$ level by the operators with $(1+\gam_5)\otimes(1-\gam_5)$ and $(1+\gam_5)\otimes(1+\gam_5)$ structures. However, the NP operators with other Dirac structures fail to provide a consistent interpretation, even at the $2\sigma$ level. As two specific examples of model-dependent considerations, we shall also discuss the case where the NP four-quark operators are generated by either a colorless charged gauge boson or a colorless charged scalar. Constraints on the effective coefficients describing the couplings of these mediators to the relevant quarks are then obtained by fitting to the data.

Our paper is organized as follows. In section~\ref{sec:theory} the theoretical framework is presented. This includes the effective weak Hamiltonian describing the quark-level $b\to c\bar{u}d(s)$ transitions, the calculation of $\mathcal{O}(\alpha_s)$ vertex corrections to the hadronic matrix elements of the twenty linearly independent four-quark operators, and the estimate of the weak annihilation contribution, within the QCDF framework. In section~\ref{sec:Numerical analysis}, we firstly present the updated SM predictions for the branching ratios of these class-I decays and their ratios with respect to the semi-leptonic $\bar{B}_{(s)}^0\to D_{(s)}^{(\ast)+}\ell^-\bar{\nu}_{\ell}$ decay rates evaluated at $q^2=m_L^2$, $R_{(s)L}^{(\ast)}$, and then discuss the NP effects both in a model-independent setup and in the case where the NP operators are generated by either a colorless charged gauge boson or a colorless charged scalar. Our conclusions are finally made in section~\ref{sec:conclusions}. For convenience, the ranges for the NP Wilson coefficients $C_i(m_b)$ allowed by the ratios $R_{(s)L}^{(\ast)}$ are given in the appendix.

\section{Theoretical framework}
\label{sec:theory}

\subsection{Effective weak Hamiltonian}
\label{subsec:effectiveD}

The class-I $\bar{B}_{(s)}^0\to D_{(s)}^{(*)+} L^-$ decays are mediated by the quark-level $b\to c\bar{u} d(s)$ transitions. Once the top quark, the gauge bosons $W^\pm$ and $Z^0$, the Higgs boson, as well as other heavy degrees of freedom present in any extension of the SM are integrated out, the corresponding QCD amplitudes of the decays are computed most conveniently in the framework of effective weak Hamiltonian~\cite{Buchalla:1995vs,Buras:1998raa}, which for the problem at hand reads\footnote{Here we assume that the NP scale $\mu_0$ satisfies the condition $\mu_0\gg m_b$, ensuring therefore that all the NP effects can be accounted for by such a local effective weak Hamiltonian.}
\begin{align}\label{eq:Hamiltonian}
\mH_\text{eff} &= \frac{G_F}{\sqrt{2}}\,V_{cb}V^*_{uq}\,\bigg\{\sum_{i}\mC_i(\mu)\mQ_i(\mu) +\sum_{i,j}\Big[C_{i}^{VLL}(\mu)\mQ_{i}^{VLL}(\mu) + C_{i}^{VLR}(\mu)\mQ_{i}^{VLR}(\mu) \nn \\[0.1cm]
& \hspace{2.6cm} + C_{j}^{SLL}(\mu)\mQ_{j}^{SLL}(\mu) + C_{i}^{SLR}(\mu) \mQ_{i}^{SLR}(\mu) + (L\leftrightarrow R)\Big]\bigg\} + \text{h.c.} \,, 
\end{align}
where $G_F$ is the Fermi constant, and $V_{cb}V^*_{uq}$ ($q=d,s$) the product of the CKM matrix elements. $\mQ_{i}$ ($i=1,2$) are the two SM four-quark current-current operators given in the Buchalla-Buras-Lautenbacher (BBL) basis~\cite{Buchalla:1995vs}, while the remaining ones in eq.~\eqref{eq:Hamiltonian} denote the full set of twenty linearly independent four-quark operators that can contribute, either directly or through operator mixing, to the quark-level $b\to c \bar{u} d(s)$ transitions~\cite{Ciuchini:1997bw,Ciuchini:1998ix,Buras:2000if}. 

The NP four-quark operators can be further split into eight separate sectors, among which there is no mixing~\cite{Buras:2000if,Buras:2012gm}. Firstly, the operators belonging to the two sectors $VLL$ and $VLR$, which are relevant for tree-level contributions mediated by heavy charged gauge bosons in any extension of the SM, can be written, respectively, as~\cite{Buras:2000if,Buras:2012gm}
\begin{align}
\mQ_{1}^{VLL}&=\overline{c}_{\alpha}\gamma^{\mu}(1-\gam_5)b_{\beta}\,\overline{q}_{\beta}\gamma_{\mu}(1-\gam_5)u_{\alpha}\,,\nn \\[0.1cm]
\mQ_{2}^{VLL}&=\overline{c}_{\alpha}\gamma^{\mu}(1-\gam_5)b_{\alpha}\,\overline{q}_{\beta}\gamma_{\mu}(1-\gam_5)u_{\beta}\,,\label{VLL} \\[0.2cm]
\mQ_{1}^{VLR}&=\overline{c}_{\alpha}\gamma^{\mu}(1-\gam_5)b_{\beta}\,\overline{q}_{\beta}\gamma_{\mu}(1+\gam_5)u_{\alpha}\,,\nn \\[0.1cm]
\mQ_{2}^{VLR}&=\overline{c}_{\alpha}\gamma^{\mu}(1-\gam_5)b_{\alpha}\,\overline{q}_{\beta}\gamma_{\mu}(1+\gam_5)u_{\beta}\,,\label{VLR}
\end{align}
where $\alpha$, $\beta$ are the color indices, and $\mQ_{i}^{VLL}$ are identical to the SM operators $\mQ_{i}$ given in the BBL basis~\cite{Buchalla:1995vs}. Secondly, the operators belonging to the two sectors $SLL$ and $SLR$, which are relevant for tree-level contributions generated by new heavy charged scalars, are given, respectively, by~\cite{Buras:2000if,Buras:2012gm}
\begin{align}
\mQ_{1}^{SLL}&=\overline{c}_{\alpha}(1-\gam_5)b_{\beta}\,\overline{q}_{\beta}(1-\gam_5)u_{\alpha}\,,\nn \\[0.1cm]                                                                 
\mQ_{2}^{SLL}&=\overline{c}_{\alpha}(1-\gam_5)b_{\alpha}\,\overline{q}_{\beta}(1-\gam_5)u_{\beta}\,,\nn \\[0.1cm]
\mQ_{3}^{SLL}&=\overline{c}_{\alpha}\sigma^{\mu\nu}(1-\gam_5)b_{\beta}\,\overline{q}_{\beta}\sigma_{\mu\nu}(1-\gam_5)u_{\alpha}\,,\nn \\[0.1cm]
\mQ_{4}^{SLL}&=\overline{c}_{\alpha}\sigma^{\mu\nu}(1-\gam_5)b_{\alpha}\,\overline{q}_{\beta}\sigma_{\mu\nu}(1-\gam_5)u_{\beta}\,,\label{SLL} \\[0.2cm]
\mQ_{1}^{SLR}&=\overline{c}_{\alpha}(1-\gam_5)b_{\beta}\,\overline{q}_{\beta}(1+\gam_5)u_{\alpha}\,,\nn \\[0.1cm]
\mQ_{2}^{SLR}&=\overline{c}_{\alpha}(1-\gam_5)b_{\alpha}\,\overline{q}_{\beta}(1+\gam_5)u_{\beta}\,,\label{SLR} 
\end{align}
where $\sigma^{\mu\nu}=\frac{1}{2}[\gamma^\mu,\gamma^\nu]$. Finally, the operators belonging to the four remaining chirality-flipped sectors $VRR$, $VRL$, $SRR$ and $SRL$ are obtained, respectively, from eqs.~\eqref{VLL}--\eqref{SLR} by making the interchanges $(1\mp\gam_5)\leftrightarrow (1\pm\gam_5)$. It should be noted that, due to parity invariance of strong interactions, the QCD ADMs of the chirality-flipped sectors are identical to that of the original sectors, simplifying therefore the renormalization group (RG) analysis of these operators~\cite{Buras:2000if}. 

\begin{figure}[t]
\begin{center}
\begin{tikzpicture}[line width=1.5pt, scale=1.80, >=Stealth]
\begin{scope}
\draw[vector](0:-1.0)--(0,0);
\node at (-0.4,0.2) {$A^{+}$};
\end{scope}
\begin{scope}[shift={(0.0,0.0)}]
\draw[fermion](-30:1)--(0,0);
\draw[fermionbar](30:1)--(0,0);
\node at (0.8,0.3) {$i_{\alpha}$};
\node at (0.8,-0.3) {$j_{\beta}$};
\node at (3.8,0) {$i\frac{g_2}{\sqrt{2}}V_{ij}\gamma^{\mu}\delta_{\alpha\beta}\,\Big[\Delta^{L}_{ij}(A) P_{L}+\Delta^{R}_{ij}(A) P_{R}\Big]$};
\end{scope}
\begin{scope}[shift={(0.0,-1.5)}]
\draw[dashed](0:-1.0)--(0,0);
\node at (-0.4,0.2) {$H^{+}$};
\end{scope}
\begin{scope}[shift={(0.0,-1.5)}]
\draw[fermion](-30:1)--(0,0);
\draw[fermionbar](30:1)--(0,0);
\node at (0.8,0.3) {$i_{\alpha}$};
\node at (0.8,-0.3) {$j_{\beta}$};
\node at (3.8,0) {$i\frac{g_2}{\sqrt{2}}V_{ij}\delta_{\alpha\beta}\,\Big[\Delta^{L}_{ij}(H) P_{L}+\Delta^{R}_{ij}(H) P_{R}\Big]$};
\end{scope}
\end{tikzpicture}
\caption{\label{Feynrule} Feynman rules for the couplings of a colorless charged gauge boson $A^{+}$ (upper) and a colorless charged scalar $H^{+}$ (lower) to an up- ($i_\alpha$) and a down-type ($j_\beta$) quark, with the strengths normalized to that of the SM tree-level $W^+$ exchange, where $g_2$ is the $SU(2)_L$ gauge coupling and $P_{L(R)}=\frac{1}{2}(1\mp\gamma_5)$ denote the left- and right-handed chirality projectors.}
\end{center}
\end{figure}
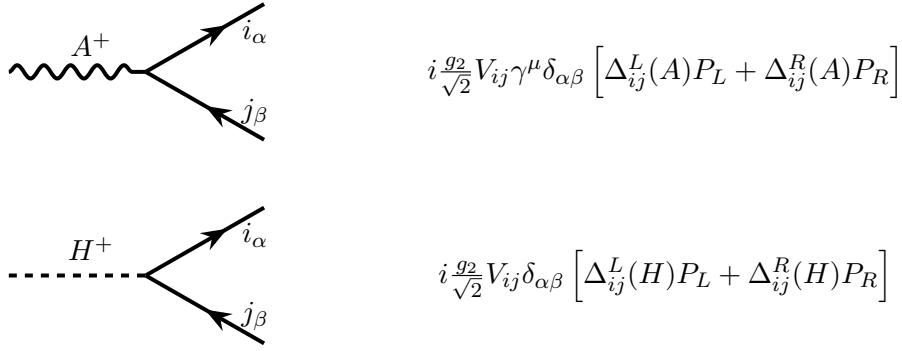

The short-distance Wilson coefficients $\mC_{i}(\mu)$ and $C_{i}(\mu)$ in eq.~\eqref{eq:Hamiltonian} can be reliably calculated by using the RG-improved perturbation theory~\cite{Buchalla:1995vs,Buras:1998raa}. Explicit expressions up to the NNLO in $\alpha_s$ for the SM part can be found, \textit{e.g.}, in ref.~\cite{Gorbahn:2004my}, and will be used throughout this paper. For the NP part, on the other hand, one can easily obtain the NLO results of $C_{i}(\mu_b)$ evaluated at the typical scale $\mu_b\simeq m_b$ that is appropriate for the non-leptonic $B$-meson decays, by solving the RG equations satisfied by these short-distance Wilson coefficients, based on the one- and two-loop QCD ADMs of the NP four-quark operators~\cite{Ciuchini:1997bw,Ciuchini:1998ix,Buras:2000if}, as well as the $\mathcal{O}(\alpha_s)$ corrections to the matching conditions for $C_{i}(\mu_0)$ evaluated at the NP scale $\mu_0$~\cite{Buras:2012gm}. Here, for later convenience, we show in Fig.~\ref{Feynrule} the Feynman rules describing the couplings of both a colorless charged gauge boson $A^+$ and a colorless charged scalar $H^+$ to an up- ($i_\alpha$) and a down-type ($j_\beta$) quark, the strengths of which have been normalized to that of the tree-level $W^+$ exchange in the SM. For further details about the matching and evolution procedures in the case of these tree-level mediators, the readers are referred to ref.~\cite{Buras:2012gm} and references therein. Throughout this paper, we shall assume that the NP Wilson coefficients $C_{i}(\mu)$ as well as the effective couplings $\Delta^{L,R}_{ij}(A)$ and $\Delta^{L,R}_{ij}(H)$ are all real, and take the same values for both the $b\to c \bar{u} d$ and $b\to c \bar{u} s$ transitions.

\subsection{Calculation of one-loop vertex corrections}
\label{subsec:HSK}

To obtain the non-leptonic $B$-decay amplitudes, we must also calculate the hadronic matrix elements of the local four-quark operators present in eq.~\eqref{eq:Hamiltonian}. To this end, we shall adopt the QCDF approach~\cite{Beneke:1999br,Beneke:2000ry,Beneke:2001ev}, within the framework of which the hadronic matrix element of a four-quark operator satisfies the factorization formula given by eq.~\eqref{eq:QCDF}. For the SM contribution, the hard kernels $T_{ij}(u)$ have been calculated up to the NNLO in $\alpha_s$~\cite{Huber:2015bva,Huber:2016xod}, and will be used throughout this paper. For the NP contribution, on the other hand, we shall calculate the one-loop vertex corrections to the hard kernels $T_{ij}(u)$, completing therefore a full NLO analysis of the class-I non-leptonic $\bar{B}_{(s)}^0\to D_{(s)}^{(*)+} L^-$ decays in the case where the short-distance Wilson coefficients of the four-quark operators are also known at the same order~\cite{Buras:2012gm}. Such an NLO analysis in the NP sector is helpful for reducing the dependence of the final decay amplitudes on certain unphysical scale and renormalization scheme~\cite{Buras:2012gm}, as will be detailed in subsection~\ref{subsec:model-dependent}.

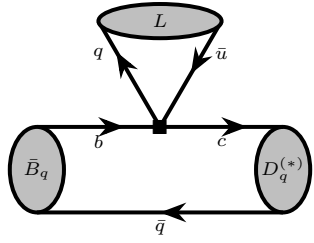
\begin{figure}[t]\fontsize{7.0}{10}
	\begin{center}
	\begin{tikzpicture}[line width=1.5pt, scale=1.62, >=Stealth]
		\begin{scope}
			\draw[fermionbar](0:1)--(0,0);
			\node at (0.5,-0.12) {$b$};
			\node at (1.0,0.0) {$\blacksquare$};
			\draw[fill=lightgray] (0.0,-0.35) circle [x radius=0.22, y radius=0.35, rotate=0];
			\node at (0.0,-0.35) {$\bar{B}_{q}$};
			\draw[fill=lightgray] (2.0,-0.35) circle [x radius=0.22, y radius=0.35, rotate=0];
			\node at (2.0,-0.35) {$D^{(*)}_{q}$};
		\end{scope}ij
		\begin{scope}[shift={(1,0)}]
			\draw[fermionbar](0:1)--(0,0);
			\node at (0.5,-0.12) {$c$};
			\draw[fermion](60:1)--(0,0);
			\draw[fermionbar](120:1)--(0,0);
			\draw[fill=lightgray] (0.0,0.866) circle [x radius=0.5, y radius=0.14, rotate=0];
			\node at (0.0,0.866) {$L$};
			\node at (-0.5,0.6) {$q$};
			\node at (0.5,0.6) {$\bar{u}$};
		\end{scope}ij
		\begin{scope}[shift={(0.0,-0.7)}]
			\draw[fermion](0:2)--(0,0);
			\node at (1,-0.12) {$\bar{q}$};
		\end{scope}ij
	\end{tikzpicture}
	\caption{\label{fig:LO} Leading-order Feynman diagram contributing to the hard kernels $T_{ij}(u)$, where the local four-quark operators are represented by the black square.}
	\end{center}
\end{figure}

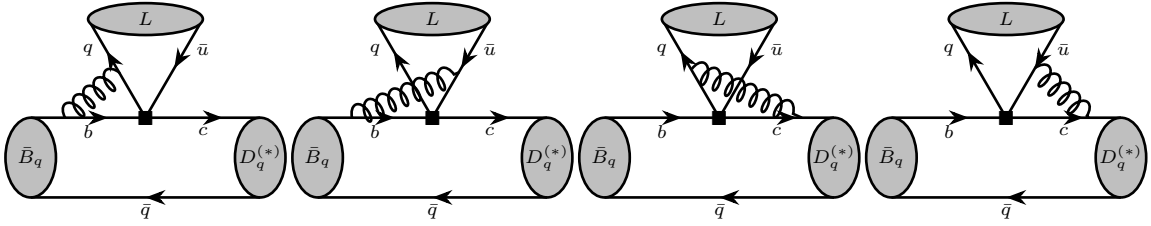
\begin{figure}[t]\fontsize{7.0}{10}
	\begin{center}
	\begin{tikzpicture}[line width=1.0pt, scale=1.51, >=Stealth]
		\begin{scope}
			\draw[fermionbar](0:1)--(0,0);
			\draw[gluon](0.3,0)--(0.78,0.40);
			\node at (0.5,-0.12) {$b$};
			\node at (1.0,0.0) {$\blacksquare $};
			\draw[fill=lightgray] (0.0,-0.35) circle [x radius=0.22, y radius=0.35, rotate=0];
			\node at (0.0,-0.35) {$\bar{B}_{q}$};
			\draw[fill=lightgray] (2.0,-0.35) circle [x radius=0.22, y radius=0.35, rotate=0];
			\node at (2.0,-0.35) {$D^{(*)}_{q}$};
		\end{scope}ij
		\begin{scope}[shift={(1,0)}]
			\draw[fermionbar](0:1)--(0,0);
			\node at (0.5,-0.12) {$c$};
			\draw[fermion](60:1)--(0,0);
			\draw[fermionbar](120:1)--(0,0);
			\draw[fill=lightgray] (0.0,0.866) circle [x radius=0.5, y radius=0.14, rotate=0];
			\node at (0.0,0.866) {$L$};
			\node at (-0.5,0.6) {$q$};
			\node at (0.5,0.6) {$\bar{u}$};
		\end{scope}ij
		\begin{scope}[shift={(0.0,-0.7)}]
			\draw[fermion](0:2)--(0,0);
			\node at (1,-0.12) {$\bar{q}$};
		\end{scope}ij
		
		\begin{scope}[shift={(2.5,0)}]
			\draw[fermionbar](0:1)--(0,0);
			\draw[gluon](0.3,0)--(1.235,0.40);
			\node at (0.5,-0.12) {$b$};
			\node at (1.0,0.0) {$\blacksquare$};
			\draw[fill=lightgray] (0.0,-0.35) circle [x radius=0.22, y radius=0.35, rotate=0];
			\node at (0.0,-0.35) {$\bar{B}_{q}$};
			\draw[fill=lightgray] (2.0,-0.35) circle [x radius=0.22, y radius=0.35, rotate=0];
			\node at (2.0,-0.35) {$D^{(*)}_{q}$};
		\end{scope}ij
		\begin{scope}[shift={(3.5,0)}]
			\draw[fermionbar](0:1)--(0,0);
			\node at (0.5,-0.12) {$c$};
			\draw[fermion](60:1)--(0,0);
			\draw[fermionbar](120:1)--(0,0);
			\draw[fill=lightgray] (0.0,0.866) circle [x radius=0.5, y radius=0.14, rotate=0];
			\node at (0.0,0.866) {$L$};
			\node at (-0.5,0.6) {$q$};
			\node at (0.5,0.6) {$\bar{u}$};
		\end{scope}ij
		\begin{scope}[shift={(2.5,-0.7)}]
			\draw[fermion](0:2)--(0,0);
			\node at (1,-0.12) {$\bar{q}$};
		\end{scope}ij
		
		\begin{scope}[shift={(5,0)}]
			\draw[fermionbar](0:1)--(0,0);
			\node at (0.5,-0.12) {$b$};
			\node at (1.0,0.0) {$\blacksquare$};
			\draw[fill=lightgray] (0.0,-0.35) circle [x radius=0.22, y radius=0.35, rotate=0];
			\node at (0.0,-0.35) {$\bar{B}_{q}$};
			\draw[fill=lightgray] (2.0,-0.35) circle [x radius=0.22, y radius=0.35, rotate=0];
			\node at (2.0,-0.35) {$D^{(*)}_{q}$};
		\end{scope}ij
		\begin{scope}[shift={(6.72,0)}]
			\draw[gluon](156:1.05)--(0.0,0);
		\end{scope}ij
		\begin{scope}[shift={(6,0)}]
			\draw[fermionbar](0:1)--(0,0);
			\node at (0.5,-0.12) {$c$};
			\draw[fermion](60:1)--(0,0);
			\draw[fermionbar](120:1)--(0,0);
			\draw[fill=lightgray] (0.0,0.866) circle [x radius=0.5, y radius=0.14, rotate=0];
			\node at (0.0,0.866) {$L$};
			\node at (-0.5,0.6) {$q$};
			\node at (0.5,0.6) {$\bar{u}$};
		\end{scope}ij
		\begin{scope}[shift={(5,-0.7)}]
			\draw[fermion](0:2)--(0,0);
			\node at (1,-0.12) {$\bar{q}$};
		\end{scope}ij
		
		\begin{scope}[shift={(7.5,0)}]
			\draw[fermionbar](0:1)--(0,0);
			\node at (0.5,-0.12) {$b$};
			\node at (1.0,0.0) {$\blacksquare$};
			\draw[fill=lightgray] (0.0,-0.35) circle [x radius=0.22, y radius=0.35, rotate=0];
			\node at (0.0,-0.35) {$\bar{B}_{q}$};
			\draw[fill=lightgray] (2.0,-0.35) circle [x radius=0.22, y radius=0.35, rotate=0];
			\node at (2.0,-0.35) {$D_{q}^{(*)}$};
		\end{scope}ij
		\begin{scope}[shift={(9.22,0)}]
			\draw[gluon](137:0.64)--(0,0);
		\end{scope}ij
		\begin{scope}[shift={(8.5,0)}]
			\draw[fermionbar](0:1)--(0,0);
			\node at (0.5,-0.12) {$c$};
			\draw[fermion](60:1)--(0,0);
			\draw[fermionbar](120:1)--(0,0);
			\draw[fill=lightgray] (0.0,0.866) circle [x radius=0.5, y radius=0.14, rotate=0];
			\node at (0.0,0.866) {$L$};
			\node at (-0.5,0.6) {$q$};
			\node at (0.5,0.6) {$\bar{u}$};
		\end{scope}ij
		\begin{scope}[shift={(7.5,-0.7)}]
			\draw[fermion](0:2)--(0,0);
			\node at (1,-0.12) {$\bar{q}$};
		\end{scope}ij
	\end{tikzpicture}
	\caption{\label{fig:vertex} ``Non-factorizable'' vertex corrections to the hard kernels $T_{ij}(u)$ at the NLO in $\alpha_s$, where the other captions are the same as in Fig.~\ref{fig:LO}.}
	\end{center}
\end{figure}

As mentioned already in the last section, at leading power in $\Lambda_\mathrm{QCD}/m_b$, these class-I non-leptonic decays are dominated by the color-allowed tree topology with the lowest-order Feynman diagram shown in Fig.~\ref{fig:LO}, and the hard kernels $T_{ij}(u)$ receive only the ``non-factorizable'' vertex corrections~\cite{Beneke:2000ry}, with the corresponding one-loop Feynman diagrams shown in Fig.~\ref{fig:vertex}. It should be noted that, as the light quark-antiquark pair $(\bar{u}q)$ has to be in a color-singlet configuration to produce an energetic light meson $L$ in the leading Fock-state approximation, the hard kernels $T_{ij}(u)$ only receive non-vanishing contributions from the color-singlet operators starting at the zeroth order in $\alpha_s$ and from the color-octet operators starting at the first order in $\alpha_s$, respectively. This implies that $T_{ij}(u)\propto 1+\mathcal{O}(\alpha_s^2) + \cdots$ for the color-singlet and $T_{ij}(u)\propto \mathcal{O}(\alpha_s) + \cdots$ for the color-octet operators, respectively. It is also observed that, although there exist both collinear and infrared divergences in each of the four vertex diagrams shown in Fig.~\ref{fig:vertex}, these divergences cancel when one sums over all the four diagrams, yielding therefore a finite and perturbatively calculable $\mathcal{O}(\alpha_{s})$ correction to the hard kernels $T_{ij}(u)$~\cite{Beneke:2000ry,Beneke:2001ev}. Explicit evaluations of these one-loop vertex diagrams with insertions of the SM current-current operators in the Chetyrkin-Misiak-M\"unz (CMM) basis~\cite{Chetyrkin:1996vx,Chetyrkin:1997gb} can be found, \textit{e.g.}, in refs.~\cite{Beneke:2000ry,Huber:2016xod,Bell:2009nk}. Our results for the one-loop vertex corrections to the hard kernels $T_{ij}(u)$, which arise from insertions of the NP four-quark operators present in eq.~\eqref{eq:Hamiltonian} into the Feynman diagrams shown in Fig.~\ref{fig:vertex}, will be presented below. This, together with the NLO results of the NP Wilson coefficients $C_{i}(\mu_b)$~\cite{Ciuchini:1997bw,Ciuchini:1998ix,Buras:2000if,Buras:2012gm}, completes our analysis at the NLO in $\alpha_{s}$.

\begin{enumerate}
    \item[$\bullet$] For operators with $\gam^{\mu}(1-\gam_5)\otimes\gam_{\mu} (1-\gam_5)$ structure, we have
    \begin{align}
    	&\langle D_{(s)}^{(*)+}(p^\prime)L^-(q)|\overline{c}_{\alpha}\gamma^{\mu}(1-\gam_5)b_{\beta}\,\overline{q}_{\beta}\gamma_{\mu}(1-\gam_5)u_{\alpha}|\bar{B}_{(s)}^0(p)\rangle =\pm \, i f_L \int^1_0 du\,\Phi_L(u) \nn \\[0.1cm]
    	& \hspace{1.6cm} \times \Big[\langle D_{(s)}^+|\bar{c}\slashed{q}b|\bar{B}_{(s)}^0\rangle \cdot T^{VLL}(u,z)-\langle D_{(s)}^{*+}|\bar{c}\slashed{q}\gamma_5b|\bar{B}_{(s)}^0\rangle
    	\cdot T^{VLL}(u,-z)\Big]\, ,
    \end{align}
    where $q=p-p^\prime$,\footnote{Although taking the same symbol, the light-meson momentum $q$ can be clearly distinguished in the context from the quark field $q$ present in the four-quark operator.} and the upper (lower) sign applies when $L$ is a pseudoscalar (vector) meson. $f_L$ and $\Phi_L$ denote respectively the decay constant and the leading-twist LCDA of the light meson $L$, while the reduced matrix elements $\langle D_{(s)}^{(*)+}|\bar{c}\cdots b|\bar{B}_{(s)}^0\rangle$ can be further parameterized in terms of the $B_{(s)}\to D_{(s)}^{(*)}$ transition form factors. The one-loop hard kernel $T^{VLL}(u,z)$ is given by
    \begin{align}\label{t8VLL}
    	T^{VLL}(u,z)=\frac{\alpha_s}{4\pi}\frac{C_F}{N_c} \left[- 6\ln\frac{\mu^2}{m_b^2} - 18 + F^{VLL}(u,z) \right]\,,
    \end{align}
    where $C_F=(N_c^2-1)/(2N_c)$, with $N_c=3$ being the number of colors, and 
    \begin{align}
    	F^{VLL}(u,z) = \left(3 + 2\ln\frac{u}{\bar{u}} \right) \ln z^2 + f^{VLL}(u,z) + f^{VLL}(\bar{u},1/z)\,,
    \end{align}
    with
    \begin{align}
    f^{VLL}(u,z) &= -\frac{u(1-z^2)\left[3(1-u (1-z^2))+z\right]}{\left[1-u(1-z^2)\right]^2}
    \ln[u(1-z^2)] - \frac{z}{1-u(1-z^2)} \nn \\[0.1cm]
    & \hspace{-0.7cm} + 2 \left\{\frac{\ln[u(1-z^2)]}{1-u(1-z^2)} - \ln^2[u(1-z^2)] 
    - \mbox{Li}_2[1-u(1-z^2)] - \left\{u\to\bar{u}\right\} \right \}\,.
    \end{align}
    Here $z=m_c/m_b$, $\bar{u}=1-u$, and the dilogarithm is defined by
    \begin{align}
    \mbox{Li}_2(x)=-\int_0^x \frac{\ln(1-t)}{t}dt\,.
    \end{align}
    
    In the limit $z\to 0$, our results coincide with that for the charmless $B$-meson decays presented in refs.~\cite{Beneke:2001ev,Beneke:2003zv}, where the four-quark operators are also defined in the BBL basis. In addition, we have also checked explicitly that, by using the relations among the short-distance Wilson coefficients~\cite{Beneke:2001at,Gorbahn:2004my,Chetyrkin:1997gb}
    \begin{align}\label{eq:BBL2CMM}
   	 C_{1}^\mathrm{BBL} &= \frac{1}{2}C_{1}^\mathrm{CMM} + \frac{\alpha_s}{4\pi}\left[-\frac{5}{6}C_{1}^\mathrm{CMM} - 2 C_{2}^\mathrm{CMM}\right] + \mathcal{O}(\alpha_s^2)\,,\nn \\[0.2cm] 
   	 C_{2}^\mathrm{BBL} &= -\frac{1}{6}C_{1}^\mathrm{CMM} + C_{2}^\mathrm{CMM} + \frac{\alpha_s}{4\pi}\left[-\frac{11}{18}C_{1}^\mathrm{CMM} + \frac{2}{3}C_{2}^\mathrm{CMM}\right] + \mathcal{O}(\alpha_s^2)\,,
    \end{align}
    corresponding to the four-quark operators defined in the BBL and CMM bases respectively, our results for the hadronic matrix elements $\langle D_{(s)}^{(*)+} L^-|\sum_{i=1,2}C_{i}^{VLL}\mQ_{i}^{VLL}|\bar{B}_{(s)}^0\rangle$, with the operators $\mQ_{i}^{VLL}$ given in the BBL basis, agree up to the NLO in $\alpha_s$ with that presented in refs.~\cite{Huber:2016xod,Bell:2009nk,Beneke:2009ek}, where the calculations are however performed with the four-quark operators defined in the CMM basis. To reproduce the results presented in ref.~\cite{Beneke:2000ry}, on the other hand, one should keep in mind that the LO relations among the short-distance Wilson coefficients are used when transforming from one operator basis to another.
    	
	\item[$\bullet$] For operators with $\gam^{\mu}(1-\gam_5)\otimes\gam_{\mu} (1+\gam_5)$ structure, we obtain
	\begin{align}
		&\langle D_{(s)}^{(*)+}(p^\prime)L^-(q)|\overline{c}_{\alpha}\gamma^{\mu}(1-\gam_5)b_{\beta}\,\overline{q}_{\beta}\gamma_{\mu}(1+\gam_5)u_{\alpha}|\bar{B}_{(s)}^0(p)\rangle = -\, i f_L \int^1_0 du\,\Phi_L(u) \nn \\[0.1cm]
		& \hspace{1.5cm} \times \Big[\langle D_{(s)}^+|\bar{c}\slashed{q}b|\bar{B}_{(s)}^0\rangle \cdot T^{VLR}(u,z)-\langle D_{(s)}^{*+}|\bar{c}\slashed{q}\gamma_5b|\bar{B}_{(s)}^0\rangle
		\cdot T^{VLR}(u,-z)\Big]\, ,
	\end{align}
	where the one-loop hard kernel $T^{VLR}(u,z)$ is now given by
	\begin{align}\label{t8VLR}
		T^{VLR}(u,z)=\frac{\alpha_s}{4\pi}\frac{C_F}{N_c} \left[
		6\ln\frac{\mu^2}{m_b^2} + 6 + F^{VLR}(u,z) \right]\,,
	\end{align}
	with 
	\begin{align}
		F^{VLR}(u,z) = -\left(3 + 2\ln\frac{\bar{u}}{u} \right) \ln z^2 - f^{VLL}(\bar{u},z) - f^{VLL}(u,1/z)\,.
	\end{align}
	It has been checked that, in the limit $z\to 0$, the above results are also reduced to that for the charmless $B$-meson decays given in refs.~\cite{Beneke:2001ev,Beneke:2003zv}.
	
	\item[$\bullet$] For operators with $(1-\gam_5)\otimes (1-\gam_5)$ structure, we have
	\begin{align}
		&\langle D_{(s)}^{(*)+}(p^\prime) L^-(q)|\overline{c}_{\alpha}(1-\gam_5)b_{\beta}\,\overline{q}_{\beta}(1-\gam_5)u_{\alpha}|\bar{B}_{(s)}^0(p)\rangle = i f_L\,\mu_{m} \int^1_0 du\,\Phi_m(u) \nn \\[0.1cm]
		& \hspace{1.6cm} \times \Big[\langle D_{(s)}^+|\bar{c}b|\bar{B}_{(s)}^0\rangle
		\cdot T^{SLL}(u,z) - \langle D_{(s)}^{*+}|\bar{c}\gamma_5 b|\bar{B}_{(s)}^0\rangle
		\cdot T^{SLL}(u,-z)\Big]\, ,
	\end{align}
	where the parameters $\mu_{m}$ are defined, respectively, as $\mu_{p}(\mu)=m_{L}^{2}/[\overline{m}_{u}(\mu)+\overline{m}_{q}(\mu)]$ for a pseudoscalar and $\mu_{v}(\mu)=m_{L}f_L^{\perp}(\mu)/f_L$ for a vector meson, with $\overline{m}_{u,q}(\mu)$ being the running quark masses in the $\mathrm{\overline{MS}}$ scheme and $f_L^{\perp}(\mu)$ the scale-dependent transverse decay constant of a vector meson. When all three-particle contributions are neglected, the twist-3 two-particle LCDA $\Phi_{p}(u)$ is determined completely by the equations of motion, with its asymptotic form given exactly by $\Phi_p(u)=1$~\cite{Beneke:2003zv,Beneke:2000wa}, while $\Phi_{v}(u)$ is related to the twist-2 LCDA $\Phi_{\perp}(u)$ of a transversely polarized vector meson by~\cite{Beneke:2003zv,Ball:1998sk}  
	\begin{align}
	  \Phi_{v}(u) &\equiv \int_0^u dv \frac{\Phi_{\perp}(v)}{1-v} - \int_u^1 dv \frac{\Phi_{\perp}(v)}{v} = 3\sum_{n=0}^{\infty}\alpha_{n,\perp}^L(\mu)P_{n+1}(2u-1)\,,
	\end{align}
	where the second equation is obtained by inserting the Gegenbauer expansion of $\Phi_{\perp}(u)$, with $\alpha_{n,\perp}^L(\mu)$ being the Gegenbauer moments with $\alpha_{0,\perp}^L=1$, and $P_n(x)$ the Legendre polynomials. For further details about these hadronic parameters, the readers are referred to ref.~\cite{Beneke:2003zv} and references therein. 
	
	The reduced matrix elements of the scalar and pseudoscalar currents are related, respectively, to that of the vector and axial-vector currents by 
	\begin{align}
		\langle D_{(s)}^+|\bar{c}b|\bar{B}_{(s)}^0\rangle &= \frac{1}{\overline{m}_{b}(\mu)-\overline{m}_{c}(\mu)}\, \langle  D_{(s)}^+|\bar{c}\slashed{q}b|\bar{B}_{(s)}^0\rangle\,, \\[0.2cm]
		\langle D_{(s)}^{*+}|\bar{c}\gamma_5 b|\bar{B}_{(s)}^0\rangle &= -\frac{1}{\overline{m}_{b}(\mu)+\overline{m}_{c}(\mu)}\,
		\langle D_{(s)}^{*+}|\bar{c}\slashed{q}\gamma_5b|\bar{B}_{(s)}^0\rangle\,.
	\end{align}
	The one-loop hard kernel $T^{SLL}(u,z)$ reads
	\begin{align} \label{t8SLL}
		T^{SLL}(u,z)=\frac{\alpha_s}{4\pi}\frac{C_F}{N_c} \left[-\frac{4(u-\bar{u})(1-z)}{1+z}\ln\frac{\mu^2}{m_b^2} + F^{SLL}(u,z) \right]\, ,
	\end{align}
	where
	\begin{align}
		F^{SLL}(u,z) = 2\left[\frac{(u-\bar{u})(1-z)}{1+z} + \ln\frac{u}{\bar{u}} \right] \ln z^2
		+ f^{SLL}(u,z) + f^{SLL}(\bar{u},1/z)\, ,
	\end{align}
	with 
	\begin{align}
		f^{SLL}(u,z) &= -2\Bigg\lbrace\frac{u(1-z)\left[u(1-z)+2z\right]-1}{1-u(1-z^2)}
		\ln[u(1-z^2)] + \frac{5u}{1+z} + \ln^2[u(1-z^2)]\nn \\[0.1cm] 
		& + \mbox{Li}_2[1-u(1-z^2)] \Bigg\rbrace - \left\{u\to\bar{u}\right\}\, .	
	\end{align}
	
	\item[$\bullet$] For operators with $\sigma^{\mu\nu}(1-\gamma_5)\otimes\sigma_{\mu\nu}(1-\gamma_5)$ structure, we get
	\begin{align}
		&\langle D_{(s)}^{(*)+}(p^\prime) L^{-}(q)|\overline{c}_{\alpha}\sigma^{\mu\nu}(1-\gam_5)b_{\beta}\,\overline{q}_{\beta}\sigma_{\mu\nu}(1-\gam_5)u_{\alpha}|\bar{B}_{(s)}^0(p)\rangle = i f_L\,\mu_{m} \int^1_0 du\,\Phi_m(u) \nn \\[0.1cm]
		& \hspace{1.6cm} \times \Big[\langle D_{(s)}^+|\bar{c}b|\bar{B}_{(s)}^0\rangle
		\cdot T^{TLL}(u,z) - \langle D_{(s)}^{*+}|\bar{c}\gamma_5 b|\bar{B}_{(s)}^0\rangle
		\cdot T^{TLL}(u,-z)\Big]\, ,
	\end{align}
	where the one-loop hard kernel $T^{TLL}(u,z)$ is given by
	\begin{align} \label{t8TLL}
		T^{TLL}(u,z)=\frac{\alpha_s}{4\pi}\frac{C_F}{N_c} \left[
		-48\ln\frac{\mu^2}{m_b^2} + F^{TLL}(u,z) \right]\, ,
	\end{align}
	with
	\begin{align}
		F^{TLL}(u,z) = 8\left[3+\frac{(u-\bar{u})(1-z)}{z+1}\ln\frac{u}{\bar{u}} \right] \ln z^2 + f^{TLL}(u,z) + f^{TLL}(\bar{u},1/z)\, ,
	\end{align}
	and 
    \begin{align}
          f^{TLL}(u,z) &= -\frac{8(4u+3)}{1+z} + \frac{8(1-z)}{1+z}\,\Bigg\lbrace \frac{u\left[(u-2)z^{2}-2z+2-u\right]-1}{1-u(1-z^2)}\ln[u(1-z^2)]\nn \\[0.1cm]
          & + (1-2u)\Big[\ln^2[u(1-z^2)] 
          + \mbox{Li}_2[1-u(1-z^2)]\Big] + \left\{u\to\bar{u}\right\}\Bigg\rbrace\, .
   \end{align}

	$\bullet$ For operators with $(1-\gam_5)\otimes (1+\gam_5)$ structure, we have
	\begin{align}
		&\langle D_{(s)}^{(*)+}(p^\prime) L^-(q)|\overline{c}_{\alpha}(1-\gam_5)b_{\beta}\,\overline{q}_{\beta}(1+\gam_5)u_{\alpha}|\bar{B}_{(s)}^0(p)\rangle =\mp\, i f_L\,\mu_{m} \int^1_0 du\,\Phi_m(u) \nn \\[0.1cm]
		& \hspace{1.6cm} \times \Big[\langle D_{(s)}^{+}|\bar{c}b|\bar{B}_{(s)}^0\rangle
		\cdot T^{SLR}(u,z) - \langle D_{(s)}^{*+}|\bar{c}\gamma_5 b|\bar{B}_{(s)}^0\rangle
		\cdot T^{SLR}(u,-z)\Big]\, ,
	\end{align}
	where the upper (lower) sign applies when $L$ is a pseudoscalar (vector) meson, and the one-loop hard kernel $T^{SLR}(u,z)$ reads
	\begin{align}\label{t8SLR}
		T^{SLR}(u,z)=\frac{\alpha_s}{4\pi}\frac{C_F}{N_c} F^{SLR}(u,z)\, ,
	\end{align}
	with
	\begin{align}
		F^{SLR}(u,z) = 2\ln\frac{u}{\bar{u}}\ln z^2 - 6 + f^{SLR}(u,z) + f^{SLR}(\bar{u},1/z)\,,
	\end{align}
	and 
	\begin{align}
		f^{SLR}(u,z) &= \Bigg\lbrace\frac{u^{2}(z-1)^{2}(3z^{2}+4z+2)-2}{\left[1-u(1-z^2)\right]^{2}}
		\ln[u(1-z^2)] + \frac{z^{2}}{(1+z)^{2}\left[1-u(1-z^2)\right]} \nn\\[0.1cm]
		& \hspace{-1.2cm} + 2\left[\frac{2\ln[u(1-z^2)]}{1-u(1-z^2)} - \ln^2[u(1-z^2)] 
		- \mbox{Li}_2[1-u(1-z^2)]\right] \Bigg\rbrace- \left\{u\to\bar{u}\right\}\, .
	\end{align}
   It is noted that, in the limit $z\to 0$, our results are consistent with that for the charmless $B$-meson decays presented in refs.~\cite{Beneke:2001ev,Beneke:2003zv,Beneke:2009eb}.
\end{enumerate}

The one-loop vertex corrections to the hard kernels $T_{ij}(u)$ with insertions of the chirality-flipped four-quark operators can be easily obtained from the results given above by changing, if necessary, the overall signs of the reduced matrix elements $\langle D_{(s)}^{+}|\bar{c}\cdots b|\bar{B}_{(s)}^0\rangle$ and $\langle D_{(s)}^{*+}|\bar{c}\cdots b|\bar{B}_{(s)}^0\rangle$. It should be noted that our calculations of the hadronic matrix elements of these four-quark operators are performed in the naive dimensional regularization scheme with anti-commuting $\gamma_5$ in $D=4-2\epsilon$ dimensions, which matches exactly the one used for evaluations of the short-distance Wilson coefficients $C_{i}(\mu)$~\cite{Buras:2012gm,Buras:2000if}. This ensures, therefore, the renormalization scheme and scale independence of the non-leptonic decay amplitudes up to the NLO in $\alpha_s$.

\subsection{Estimate of weak annihilation contribution}
\label{subsec:Annihilation}	

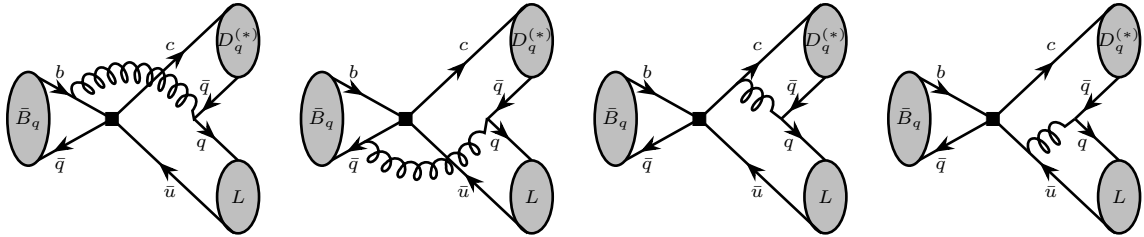
\begin{figure}[t]\fontsize{7.0}{10}
	\begin{center}
	\begin{tikzpicture}[line width=1.0pt, scale=1.38, >=Stealth]
		\begin{scope}
			\draw[fermion](0,0)--(0.8,-0.45);
			\node at (0.3,-0.0) {$b$};
			\draw[fermionbar](0,-0.9)--(0.8,-0.45);
			\node at (0.3,-0.9) {$\bar{q}$};
			\node at (0.8,-0.45) {$\blacksquare $};
			\draw[fill=lightgray] (0.0,-0.45) circle [x radius=0.2, y radius=0.45, rotate=0];
			\node at (0.0,-0.45) {$\bar{B}_{q}$};
			\draw[fill=lightgray] (2.0,-1.2) circle [x radius=0.2, y radius=0.35, rotate=0];
			\node at (2.0,-1.2) {$L$};
			\draw[fermionbar](0.8,-0.45)--(1.962,-1.55);
			\node at (1.35,-1.15) {$\bar{u}$};
			\draw[fill=lightgray] (2.0,0.3) circle [x radius=0.2, y radius=0.35, rotate=0];
			\node at (2.0,0.3) {$D^{(*)}_{q}$};
			\draw[fermion1](0.8,-0.45)--(1.962,0.65);
			\node at (1.35,0.25) {$c$};
			\draw[fermion](1.58,-0.45)--(2.0,-0.85);
			\node at (1.65,-0.7) {$q$};
			\draw[fermionbar](1.58,-0.45)--(2.0,-0.05);
			\node at (1.68,-0.15) {$\bar{q}$};
			\draw[gluon](0.45,-0.25) arc (140:18:0.66cm);
		\end{scope}ij
		
		\begin{scope}[shift={(2.8,0)}]
			\draw[fermion](0,0)--(0.8,-0.45);
			\node at (0.3,-0.0) {$b$};
			\draw[fermionbar](0,-0.9)--(0.8,-0.45);
			\node at (0.3,-0.9) {$\bar{q}$};
			\node at (0.8,-0.45) {$\blacksquare $};
			\draw[fill=lightgray] (0.0,-0.45) circle [x radius=0.2, y radius=0.45, rotate=0];
			\node at (0.0,-0.45) {$\bar{B}_{q}$};
			\draw[fill=lightgray] (2.0,-1.2) circle [x radius=0.2, y radius=0.35, rotate=0];
			\node at (2.0,-1.2) {$L$};
			\draw[fermionbar1](0.8,-0.45)--(1.962,-1.55);
			\node at (1.35,-1.15) {$\bar{u}$};
			\draw[fill=lightgray] (2.0,0.3) circle [x radius=0.2, y radius=0.35, rotate=0];
			\node at (2.0,0.3) {$D^{(*)}_{q}$};
			\draw[fermion](0.8,-0.45)--(1.962,0.65);
			\node at (1.35,0.25) {$c$};
			\draw[fermion](1.58,-0.45)--(2.0,-0.85);
			\node at (1.65,-0.7) {$q$};
			\draw[fermionbar](1.58,-0.45)--(2.0,-0.05);
			\node at (1.68,-0.15) {$\bar{q}$};
			\draw[style={decorate, draw=black,
				decoration={coil,aspect=0.8,amplitude=3.0pt, segment length=5.7pt}}] (0.4,-0.675) arc [start angle=217, end angle=345.465, radius=0.665];
		\end{scope}ij
		
		\begin{scope}[shift={(5.6,0)}]
			\draw[fermion](0,0)--(0.8,-0.45);
			\node at (0.3,-0.0) {$b$};
			\draw[fermionbar](0,-0.9)--(0.8,-0.45);
			\node at (0.3,-0.9) {$\bar{q}$};
			\node at (0.8,-0.45) {$\blacksquare $};
			\draw[fill=lightgray] (0.0,-0.45) circle [x radius=0.2, y radius=0.45, rotate=0];
			\node at (0.0,-0.45) {$\bar{B}_{q}$};
			\draw[fill=lightgray] (2.0,-1.2) circle [x radius=0.2, y radius=0.35, rotate=0];
			\node at (2.0,-1.2) {$L$};
			\draw[fermionbar](0.8,-0.45)--(1.962,-1.55);
			\node at (1.35,-1.15) {$\bar{u}$};
			\draw[fill=lightgray] (2.0,0.3) circle [x radius=0.2, y radius=0.35, rotate=0];
			\node at (2.0,0.3) {$D^{(*)}_{q}$};
			\draw[fermion](0.8,-0.45)--(1.962,0.65);
			\node at (1.35,0.25) {$c$};
			\draw[fermion](1.58,-0.45)--(2.0,-0.85);
			\node at (1.65,-0.7) {$q$};
			\draw[fermionbar](1.58,-0.45)--(2.0,-0.05);
			\node at (1.68,-0.15) {$\bar{q}$};
			\draw[gluon](1.15,-0.118)--(1.58,-0.45);
		\end{scope}ij
		
		\begin{scope}[shift={(8.4,0)}]
			\draw[fermion](0,0)--(0.8,-0.45);
			\node at (0.3,-0.0) {$b$};
			\draw[fermionbar](0,-0.9)--(0.8,-0.45);
			\node at (0.3,-0.9) {$\bar{q}$};
			\node at (0.8,-0.45) {$\blacksquare $};
			\draw[fill=lightgray] (0.0,-0.45) circle [x radius=0.2, y radius=0.45, rotate=0];
			\node at (0.0,-0.45) {$\bar{B}_{q}$};
			\draw[fill=lightgray] (2.0,-1.2) circle [x radius=0.2, y radius=0.35, rotate=0];
			\node at (2.0,-1.2) {$L$};
			\draw[fermionbar](0.8,-0.45)--(1.962,-1.55);
			\node at (1.35,-1.15) {$\bar{u}$};
			\draw[fill=lightgray] (2.0,0.3) circle [x radius=0.2, y radius=0.35, rotate=0];
			\node at (2.0,0.3) {$D^{(*)}_{q}$};
			\draw[fermion](0.8,-0.45)--(1.962,0.65);
			\node at (1.35,0.25) {$c$};
			\draw[fermion](1.58,-0.45)--(2.0,-0.85);
			\node at (1.65,-0.7) {$q$};
			\draw[fermionbar](1.58,-0.45)--(2.0,-0.05);
			\node at (1.68,-0.15) {$\bar{q}$};
			\draw[gluon](1.15,-0.78)--(1.58,-0.45);
		\end{scope}ij
	\end{tikzpicture}
    \caption{\label{fig:anni} Annihilation diagrams contributing to the class-I $\bar{B}_{(s)}^0\to D_{(s)}^{(*)+} L^-$ decays at $\mathcal{O}(\alpha_s)$. The other captions are the same as in Fig.~\ref{fig:LO}.}
    \end{center}
\end{figure}

We now proceed to discuss the power-suppressed weak annihilation contribution to the class-I $\bar{B}_{(s)}^0\to D_{(s)}^{(*)+} L^-$ decays, with the corresponding Feynman diagrams shown in Fig.~\ref{fig:anni}. It must be emphasized that, due to the presence of endpoint singularities, the weak annihilation topology cannot be computed self-consistently within the QCDF framework~\cite{Beneke:2000ry,Beneke:2001ev}. Nevertheless, we shall still follow the conventions used in refs.~\cite{Beneke:2001ev,Beneke:2003zv,Beneke:2006hg} to make an estimate of the weak annihilation effect in these class-I decays. Instead of considering all the four-quark operators present in eq.~\eqref{eq:Hamiltonian}, we shall focus only on the SM current-current operators. Our purpose is to demonstrate that, even with the weak annihilation contribution taken into account, the deviations observed in the branching ratios of these class-I decays could not be explained in the SM, as will be shown numerically in subsection~\ref{subsec:SMupdate}. 

Following the same conventions as used in refs.~\cite{Beneke:2001ev,Beneke:2003zv,Beneke:2006hg}, one can write the weak annihilation contribution to the decay amplitude of a class-I non-leptonic decay as
\begin{align}\label{eq:anniusual}
\mathcal{A}_\mathrm{ann}(\bar{B}_{(s)}^0\to D_{(s)}^{(*)+} L^-) = \frac{C_F}{N_c^2}\,C_2(\mu_h)\,A_1^i(\mu_h)\,B_{D_{(s)}^{(*)+} L^-}\,, 
\end{align}
where 
\begin{align}\label{eq:Bfactor}
	B_{D_{(s)}^{(*)+} L^-} = \pm\, i\,\frac{G_F}{\sqrt{2}}\,V_{cb}V^*_{uq}\,f_{B_{(s)}^0}\,f_{D_{(s)}^{(*)+}}\,f_{L}\,, 
\end{align}
with the upper sign applied when both final-state mesons are pseudoscalar or longitudinally polarized vector mesons, and the lower when one of them is a vector meson. The Wilson coefficient $C_2(\mu_h)$ and the building blocks $A_1^i(\mu_h)$ should be evaluated at an intermediate scale $\mu_h=\sqrt{m_b\,\Lambda_h}$, with $\Lambda_h=0.5~\gev$~\cite{Beneke:2003zv}. Since the treatment of weak annihilation topology within the QCDF framework is model-dependent anyway, we shall assume that the building blocks $A_1^i(\mu_h)$ take the same expressions as for the charmless $B$-meson decays~\cite{Beneke:2001ev,Beneke:2003zv,Beneke:2006hg}, although the asymptotic forms of the $D_{(s)}^{(*)}$ LCDAs are quite different from that of a light charged charmless meson~\cite{Beneke:2000ry}. Explicitly, we have~\cite{Beneke:2001ev,Beneke:2003zv,Beneke:2006hg}
\begin{align}
	A_1^i(\mu_h) &\approx 2\pi\alpha_s(\mu_h) \bigg[9\left(X_A - 4 + \frac{\pi^2}{3}\right) 
	+ r_\chi^{M_1}(\mu_h)\,r_\chi^{M_2}(\mu_h)\,X_A^2 \bigg]\,,
\end{align}
when both final-state mesons are pseudoscalar,
\begin{align}
	A_1^i(\mu_h) &\approx 6\pi\alpha_s(\mu_h) \bigg[3\left(X_A - 4 + \frac{\pi^2}{3} \right)
	+ r_\chi^{M_1}(\mu_h)\,r_\chi^{M_2}(\mu_h)\left(X_A^2-2 X_A\right)\bigg]\,,
\end{align}
when one of them is a pseudoscalar and the other a vector meson, whereas
\begin{align}	
	A_1^i(\mu_h) &\approx 18\pi\alpha_s(\mu_h) \bigg[\left(X_A - 4 + \frac{\pi^2}{3}\right)
	+ r_\chi^{M_1}(\mu_h)\,r_\chi^{M_2}(\mu_h)(X_A-2)^2 \bigg]\,, 
\end{align}
when both of them are longitudinally polarized vector mesons. Here the model parameter $X_A$ is parameterized by the prescription~\cite{Beneke:2001ev,Beneke:2003zv,Beneke:2006hg}
\begin{align}
	\int^1_0 \frac{du}{u} \to X_A = \left( 1 + \varrho_A\,e^{i\varphi_A} \right)\ln\frac{m_{B_{(s)}^0}}{\Lambda_h}\,,
\end{align}
with $\varrho_A \le 2$ and $\varphi_A\in[0,2\pi]$, which means that we have assigned a $200\%$ uncertainty to the default value obtained with $\varrho_A=0$. The ratios $r_\chi^{M}$ are defined as 
\begin{align}
	r_\chi^{P}(\mu)=\frac{2m_P^2}{\overline{m}_b(\mu)\left[\overline{m}_1(\mu)+\overline{m}_2(\mu)\right]}=\frac{2\mu_{p}(\mu)}{\overline{m}_b(\mu)}\,, \quad 
	r_\chi^{V}(\mu)=\frac{2m_V}{\overline{m}_b(\mu)}\,\frac{f_V^\perp (\mu)}{f_V}=\frac{2\mu_{v}(\mu)}{\overline{m}_b(\mu)}\,,
\end{align}
for a pseudoscalar ($P$) and a vector ($V$) meson respectively, where $\overline{m}_{1,2}$ are the current quark masses of the two valence constituents of the meson considered. In view of the large uncertainties brought by the phenomenological parameters $\varrho_A$ and $\varphi_A$, it is generally expected that such a model-dependent treatment should give the correct order of magnitude of the weak annihilation effect in $B$-meson decays into both the charmless~\cite{Beneke:2001ev,Beneke:2003zv,Beneke:2006hg} and the heavy-light final states~\cite{Beneke:2000ry,Lenz:2019lvd}.  

In order to separate the weak annihilation contribution from that of the dominant color-allowed tree topology, we introduce the effective coefficients $b_1(D_{(s)}^{(*)+} L^-)$ defined by
\begin{align}\label{eq:b1}
b_1(D_{(s)}^{(*)+} L^-) = \frac{C_F}{N_c^2}\,C_2(\mu_h)\,A_1^i(\mu_h)\,\frac{B_{D_{(s)}^{(*)+} L^-}}{A_{D_{(s)}^{(*)+} L^-}}\,,
\end{align}
with
\begin{align}
A_{D_{(s)}^+ P^-} &= i\,\frac{G_F}{\sqrt{2}}\,V_{cb}V^\ast_{uq}\,f_{P^-}\,F_0^{B_{(s)}\to D_{(s)}}(m_{P^-}^2)\,\big(m_{B_{(s)}}^2-m_{D_{(s)}^+}^2\big)\,,\label{eq:PP} \\[0.2cm] 
A_{D_{(s)}^{*+} P^-} &= -i\,\frac{G_F}{\sqrt{2}}\,V_{cb}V^\ast_{uq}\,f_{P^-}\,A_0^{B_{(s)}\to D_{(s)}^\ast}(m_{P^-}^2)\,2m_{D_{(s)}^{\ast+}}\,\big(\epsilon^\ast\cdot p\big)\,, \label{eq:VP} \\[0.2cm]
A_{D_{(s)}^+ V^-} &= -i\,\frac{G_F}{\sqrt{2}}\,V_{cb}V^\ast_{uq}\,f_{V^-}\,F_+^{B_{(s)}\to D_{(s)}}(m_{V^-}^2)\,2m_{V^-}\,\big(\eta^\ast\cdot p\big)\,, \label{eq:PV} \\[0.2cm] 
A_{D_{(s)}^{*+} V^-} &= i\,\frac{G_F}{\sqrt{2}}\,V_{cb}V^\ast_{uq}\,f_{V^-}\, \frac{1}{2m_{D_{(s)}^{*+}}}\Bigg[\big(m_{B_{(s)}}^2-m_{D_{(s)}^{*+}}^2-m_{V^-}^2\big)\big(m_{B_{(s)}}+m_{D_{(s)}^{*+}}\big)\,\nn \\[0.1cm]
&\qquad\qquad \times A_1^{B_{(s)}\to D^*_{(s)}}(m_{V^-}^2) - \frac{4 m_{B_{(s)}}^2 |\vec{q}|^2}{m_{B_{(s)}}+m_{D_{(s)}^{*+}}}\,A_2^{B_{(s)}\to D^*_{(s)}}(m_{V^-}^2)\Bigg]\,, \label{eq:VV}
\end{align}
where 
\begin{align}
	|\vec{q}|=\frac{1}{2 m_{B_{(s)}}}\, \sqrt{\left(m_{B_{(s)}}^{2}-m_{D_{(s)}^{*+}}^{2}-m_{V^-}^{2}\right)^{2} - 4 m_{D_{(s)}^{*+}}^{2} m_{V^-}^{2}}
\end{align}
is the momentum of the two final-state mesons in the parent rest frame. Then, including also the LO contributions from the four-quark operators present in eq.~\eqref{eq:Hamiltonian}, we can write the total decay amplitude of a given channel as~\cite{Beneke:2000ry,Huber:2016xod}:\footnote{It should be noted that the effective weak annihilation coefficients $b_1(D_{(s)}^{(*)+} L^-)$ are relevant only for the six decay modes shown in Table~\ref{tab:annicompare}.}
\begin{align}
	{\mathcal{A}}(\bar{B}_{(s)}^0 \to D_{(s)}^{(*)+} L^-) &= A_{D_{(s)}^{(*)+} L^-}\,\left[a_1(D_{(s)}^{(*)+} L^-)+b_1(D_{(s)}^{(*)+} L^-)\right]\,.
\end{align}
Note that, due to angular momentum conservation, the polarization vectors $\epsilon^\mu$ of $D_{(s)}^{*+}$ and $\eta^\mu$ of $V$ in the final states take only the longitudinal part in eqs.~\eqref{eq:VP} and \eqref{eq:PV}. The decay amplitudes of $\bar{B}_{(s)}^0\to D_{(s)}^{*+} V^-$ modes are more complicated and, to leading power in $\Lambda_\mathrm{QCD}/m_b$, dominated also by the longitudinal polarization, with the transverse parts being suppressed by $\mathcal{O}(m_V/m_{B_{(s)}})$; their explicit expressions could be found, \textit{e.g.}, in ref.~\cite{Beneke:2000ry}. The effective coefficients $a_1(D_{(s)}^{*+} L^-)$ can be expressed in terms of the short-distance Wilson coefficients $C_i(\mu)$ as well as the perturbatively calculable hard kernels $T_{ij}(u)$ convoluted with the light-meson LCDAs $\Phi_{L,m}(u)$. For the SM contributions, both the NLO~\cite{Beneke:2000ry,Politzer:1991au} and the NNLO~\cite{Huber:2015bva,Huber:2016xod} corrections to $a_1(D_{(s)}^{*+} L^-)$ are known. Combining our calculations of the one-loop vertex corrections to $T_{ij}(u)$ as well as the $\mathcal{O}(\alpha_s)$ corrections to the matching conditions for the short-distance Wilson coefficients~\cite{Buras:2012gm}, the effective coefficients $a_1(D_{(s)}^{*+} L^-)$ associated with the complete set of NP four-quark operators present in eq.~\eqref{eq:Hamiltonian} are now known up to the NLO in $\alpha_s$. 

\section{Numerical results and discussions}
\label{sec:Numerical analysis}

\subsection{Input parameters}
\label{subsec:input}


\begin{table}[t]
\begin{center}	
\let\oldarraystretch=\arraystretch
\renewcommand*{\arraystretch}{1.26}
{\tabcolsep=0.785cm\begin{tabular}{|cccc|c|}
\hline\hline
\multicolumn{4}{|l|}{\textbf{QCD and electroweak parameters~~~\cite{ParticleDataGroup:2020ssz}}}
\\
\hline
  $G_F [10^{-5}\gev^{-2}]$
& $\alpha_s(m_Z)$
& $m_Z [\gev]$
& $m_W [\gev]$
\\
  $1.1663787$
& $0.1179 \pm 0.0010$
& $91.1876$
& $80.379$
\\
\hline
\end{tabular}}

{\tabcolsep=0.48cm \begin{tabular}{|ccccc|}
\multicolumn{5}{|l|}{\hspace{0.20cm} \textbf{Quark masses [GeV]~~~\cite{ParticleDataGroup:2020ssz,ATLAS:2014wva}}}
\\
\hline
  $m_t^{\rm pole}$
& $\overline{m}_b(\overline{m}_b)$
& $\overline{m}_c(\overline{m}_c)$
& $\overline{m}_s(2\, \rm GeV)$
& $2\overline{m}_s/(\overline{m}_u+\overline{m}_d)$
\\
  $172.76 \pm 0.30$
& $4.18_{-0.02}^{+0.03}$
& $1.27 \pm 0.02$
& $0.093_{-0.005}^{+0.011}$
& $27.3_{-1.3}^{+0.7}$
\\
\hline
\end{tabular}}

{\tabcolsep=1.184cm \begin{tabular}{|ccc|}
\multicolumn{3}{|l|}{\hspace{-0.45cm} \textbf{CKM matrix elements~~~\cite{Charles:2004jd,ckm2018}}}
\\
\hline
  $|V_{ud}|$
& $|V_{us}|$
& $|V_{cb}|[10^{-3}]$
\\
  $0.9744129_{-0.0000513}^{+0.0000096}$
& $0.224791_{-0.000098}^{+0.000170} $
& $42.41_{-1.51}^{+0.40}$
\\
\hline
\end{tabular}}
{\tabcolsep=0.67cm \begin{tabular}{|cccc|}
		\multicolumn{4}{|l|}{\hspace{0.10cm} \textbf{Lifetimes and masses of $B_{(s)}^0$ and $D_{(s)}^{(*)+}$ mesons ~~~\cite{ParticleDataGroup:2020ssz,HFLAV:2019otj}}}
		\\
		\hline
		$\tau_{B^0} [{\rm ps}]$
		& $m_{B^0} [\mev]$
		& $m_{D^{+}} [\mev]$
		& $m_{D^{\ast +}} [\mev]$
		\\
		$1.519 \pm 0.004$
		& $5279.65\pm 0.12$
		& $1869.66\pm 0.05$
		& $2010.26\pm 0.05$
		\\
		$\tau_{B_s^0} [{\rm ps}]$
		& $m_{B_s^0} [\mev]$
		& $m_{D_{s}^{+}} [\mev]$
		& $m_{D_{s}^{\ast +}} [\mev]$
		\\
		$1.516 \pm 0.006$
		& $5366.88\pm 0.14$
		& $1968.35\pm 0.07$
		& $2112.2\pm 0.4$
		\\
		\hline
\end{tabular}}

{\tabcolsep=0.985 cm \begin{tabular}{|cccc|}
		\multicolumn{4}{|l|}{\hspace{-0.18cm} \textbf{Decay constants of $B_{(s)}^0$ and $D_{(s)}^{(*)+}$ mesons [MeV]~~~\cite{Rosner:2015wva,Lubicz:2017asp,Pullin:2021ebn,FlavourLatticeAveragingGroup:2019iem}}}
		\\
		\hline
	      $f_{B^0} $
		& $f_{D^+} $
		& $f_{D^{\ast+}} $
		& $f^\perp_{D^{\ast+}}$
		\\
		$190.9 \pm 4.1$
		& $211.9 \pm 1.1$
		& $223.5\pm 8.4$
		& $202\pm 16$
		\\
	      $f_{B^0_s} $
		& $f_{D_s^+} $
		& $f_{D_s^{\ast+}}$
		& $f^\perp_{D_s^{\ast+}}$
		\\
	      $227.2\pm 3.4$
		& $249.0\pm 1.2$
		& $268.8\pm 6.6$
		& $256^{+16}_{-17}$
		\\
		\hline
\end{tabular}}
{\tabcolsep=0.263cm \begin{tabular}{|ccccc|c|}
\multicolumn{6}{|l|}{\hspace{0.48cm} \textbf{Masses, decay constants, and Gegenbauer moments of light mesons}}
\\
\hline
\multicolumn{1}{|c|}{}
& $\pi^-$
& $K^-$
& $\rho^-$
& $K^{\ast-}$
&
\\
\hline
\multicolumn{1}{|c|}{$m_{L} [\mev]$}
& $139.57$
& $493.68$
& $775.26$
& $891.67$
& \textbf{{\cite{ParticleDataGroup:2020ssz}}}
\\
\hline
\multicolumn{1}{|c|}{$f_{L} [\mev]$}
& $130.2 \pm 1.7$
& $155.6 \pm 0.4$
& $216 \pm 6$
& $211 \pm 7$
& 
\\
\multicolumn{1}{|c|}{$f_{L}^{\perp} [\mev]$}
& --
& --
& $160 \pm 11$
& $163 \pm 8$
& \textbf{\cite{Rosner:2015wva,Straub:2015ica,Dimou:2012un}}
\\
\hline
\multicolumn{1}{|c|}{$\alpha_1^L$}
& --
& $-0.0525^{+0.0033}_{-0.0031}$
& --
& $0.06 \pm 0.04$
&
\\
\multicolumn{1}{|c|}{$\alpha_2^L$}
& $0.116^{+0.019}_{-0.020}$
& $0.106^{+0.015}_{-0.016}$
& $0.17 \pm 0.07$
& $0.16 \pm 0.09$
& \textbf{\cite{Straub:2015ica,Dimou:2012un,Arthur:2010xf,Bali:2019dqc}}
\\
\hline\hline
\end{tabular}}
\caption{\label{tab:inputs} Summary of the theoretical input parameters used throughout this paper. The values of the CKM matrix elements are taken from the CKMfitter tree-only fit results as of Summer 18~\cite{Charles:2004jd,ckm2018}. The transverse decay constants $f^\perp_{D_s^{(\ast)+}}$ and $f_{L}^{\perp}$ are given at the scales $1.27~\gev$~\cite{Pullin:2021ebn} and $1~\gev$~\cite{Dimou:2012un}, respectively. The Gegenbauer moments of light pseudoscalar and vector mesons are evaluated at $\mu=2~\gev$ and $\mu=1~\gev$, respectively.}
\end{center}
\end{table}


To update the SM predictions of $\bar{B}_{(s)}^0\to D_{(s)}^{(*)+} L^-$ decays presented in ref.~\cite{Huber:2016xod}, we should firstly update the theoretical input parameters, which include the strong coupling constant $\alpha_s$, the quark masses, the CKM matrix elements, the lifetimes of $B_{(s)}^0$ mesons, as well as the hadronic parameters like the $B_{(s)}\to D_{(s)}^{(*)}$ transition form factors and the decay constants and Gegenbauer moments of light mesons. We use the two-loop relation between pole and $\msbar$ mass~\cite{Chetyrkin:2000yt}, to convert the top-quark pole mass $m_t^{\text{pole}}$ to the scale-invariant mass $\overline{m}_t(\overline{m}_t)$. The three-loop running for $\alpha_s$ is used throughout this paper. For convenience, we collect in Table~\ref{tab:inputs} all the input parameters used throughout this paper. To obtain the theoretical uncertainty of an observable, we vary each input parameter within its $1\sigma$ range and then add each individual uncertainty in quadrature. In addition, we have included the uncertainty due to the variation of the renormalization scale $\mu_b\in [m_b/2,2m_b]$.

For the $B\to D^{(\ast)}$ transition form factors, we take the ``$\mathrm{L}_{w\geq1}+\mathrm{SR}$'' fit results obtained in ref.~\cite{Bernlochner:2017jka}, in which both $\mathcal{O}(\Lambda_\mathrm{QCD}/m_{b,c})$ and $\mathcal{O}(\alpha_{s})$ contributions as well as the uncertainties in the predictions of the form-factor ratios at $\mathcal{O}(\Lambda_\mathrm{QCD}/m_{b,c})$ are consistently included within the framework of heavy quark effective theory (HQET).\footnote{For other similar analyses including the missing higher-order pieces in the relations among the HQET form factors, the readers are referred to refs.~\cite{Bordone:2019guc,Bordone:2019vic,Gambino:2019sif,Bigi:2017jbd,Jaiswal:2017rve,Jaiswal:2020wer,Iguro:2020cpg} and references therein.} For the $B_{s}\to D_{s}^{(*)}$ transition form factors, on the other hand, we use the improved lattice QCD determinations presented in refs.~\cite{McLean:2019qcx,Harrison:2021tol}, while the experimental values of the differential semi-leptonic $\bar{B}_{s}^0\to D_{s}^{(\ast)+}\ell^{-}\bar\nu_{\ell}$ decay rates are taken from the LHCb collaboration~\cite{Aaij:2020hsi,LHCb:2020hpv}.\footnote{It should be noted that the unitarity bounds applied in the lattice~\cite{McLean:2019qcx,Harrison:2021tol} and experimental~\cite{Aaij:2020hsi,LHCb:2020hpv} determinations of the $B_{s}\to D_{s}^{(*)}$ transition form factors are much looser than imposed in refs.~\cite{Bernlochner:2017jka,Bordone:2019vic,Bordone:2019guc}.} As a comparison, we list in Table~\ref{tab:FFs} our results for some of these transition form factors evaluated at $q^2=m_{K^-}^2$ or $q^2=m_{\pi^-}^2$, together with the ones used in refs.~\cite{Huber:2016xod,Bordone:2020gao}. It can be seen that our results for these transition form factors are all consistent within errors with that presented in ref.~\cite{Bordone:2020gao}, while being much more precise than the ones used in ref.~\cite{Huber:2016xod}. This justifies our choices of the form factors given in refs.~\cite{Bernlochner:2017jka,McLean:2019qcx,Harrison:2021tol}, rather than adopting the more complete analysis performed in the heavy-quark-expansion framework, where the form-factor uncertainties including $\mathcal{O}(\Lambda_\mathrm{QCD}^2/m_{c}^2)$ corrections, the strong unitarity bounds, and a consistent treatment of the flavor symmetry (breaking) are all taken into account in the global fit~\cite{Bordone:2019guc,Bordone:2019vic}.

\begin{table}[t]
	\begin{center}
		\tabcolsep0.73cm
		\let\oldarraystretch=\arraystretch
		\renewcommand*{\arraystretch}{1.6}	
		\begin{tabular}{lccccc}
			\hline \hline
			Form factor  & This work & Ref.~\cite{Bordone:2020gao}& Ref.~\cite{Huber:2016xod} \\
			\hline
			$F_0^{B\to D}(m_{K^-}^2)$
			& $0.671\pm0.011$	
			& $0.672\pm0.011$
			& $0.670\pm0.031$ \\
			$A_0^{B\to D^{*}}(m_{K^-}^2)$
			& $0.664\pm0.018$
			& $0.708\pm0.038$
			& $0.654\pm0.068$\\ 
			$F_0^{B_{s}\to D_{s}}(m_{\pi^-}^2)$
			& $0.666\pm0.012$
			& $0.673\pm0.011$
			& $0.700\pm0.100$\\
			$A_0^{B_{s}\to D_{s}^{*}}(m_{\pi^-}^2)$
			& $0.630\pm0.069$
			& $0.689\pm0.064$
			& $0.520\pm0.060$ \\
			\hline \hline
		\end{tabular}
		\caption{\label{tab:FFs} Numerical results for some of the $B_{(s)}\to D_{(s)}^{(*)}$ transition form factors evaluated at $q^2=m_{K^-}^2$ or $q^2=m_{\pi^-}^2$, together with the ones used in refs.~\cite{Huber:2016xod,Bordone:2020gao}.}
	\end{center}
\end{table}

\subsection{Updated predictions for branching ratios}
\label{subsec:SMupdate}

\begin{table}[t]
\begin{center}
\tabcolsep0.09cm
\let\oldarraystretch=\arraystretch
\renewcommand*{\arraystretch}{1.6}	
\begin{tabular}{lccccccc}
\hline \hline
Decay mode  & $\mathrm{LO}$ & $\mathrm{NLO}$ & $\mathrm{NNLO}$ & $\mathrm{NNLO^{\#}}$ & Ref.~\cite{Huber:2016xod} & Ref.~\cite{Bordone:2020gao} & Exp.~\cite{ParticleDataGroup:2020ssz,HFLAV:2019otj} \\
\hline
$\bar{B}^0\to D^+\pi^-$
& $4.20$
& $4.45_{-0.40}^{+0.25}$
& $4.58_{-0.38}^{+0.22}$
& $4.74_{-0.69}^{+0.61}$
& $3.93_{-0.42}^{+0.43}$
&
& $2.65\pm0.15$ \\
$\bar{B}^0\to D^{\ast+}\pi^-$
& $3.77$
& $4.00_{-0.40}^{+0.29}$
& $4.13_{-0.39}^{+0.27}$
& $4.26_{-0.80}^{+0.75}$
& $3.45_{-0.50}^{+0.53}$
&
& $2.58\pm0.13$ \\ 
$\bar{B}^0\to D^{+}\rho^-$
& $10.98$
& $11.64_{-1.18}^{+0.88}$
& $11.96_{-1.15}^{+0.82}$
& $12.28_{-1.63}^{+1.40}$
& $10.42_{-1.20}^{+1.24}$
&
& $7.6\pm1.2$ \\
$\bar{B}^0\to D^{\ast+}\rho^-$
& $10.32$
& $10.95_{-1.55}^{+1.40}$
& $11.28_{-1.56}^{+1.40}$
& $11.61_{-2.01}^{+1.88}$
& $9.24_{-0.71}^{+0.72}$
&
& $6.0\pm0.8$ \\
\hline
$\bar{B}^0\to D^+K^-$
& $3.18$
& $3.37_{-0.29}^{+0.17}$
& $3.48_{-0.28}^{+0.14}$
& 
& $3.01_{-0.31}^{+0.32}$
& $3.26\pm0.15$
& $2.19\pm0.13$ \\ 
$\bar{B}^0\to D^{\ast+}K^-$
& $2.82$
& $3.00_{-0.29}^{+0.20}$
& $3.10_{-0.28}^{+0.19}$
& 
& $2.59_{-0.37}^{+0.39}$
& $3.27_{-0.34}^{+0.39}$
& $2.04\pm0.47$ \\ 
$\bar{B}^0\to D^{+}K^{\ast-}$
& $5.48$
& $5.80_{-0.62}^{+0.48}$
& $5.94_{-0.61}^{+0.46}$
& 
& $5.25_{-0.63}^{+0.65}$
&
& $4.6\pm0.8$ \\
\hline
$\bar{B}_s^0\to D_s^+\pi^-$
& $4.23$
& $4.49_{-0.41}^{+0.27}$
& $4.61_{-0.39}^{+0.23}$
& 
& $4.39_{-1.19}^{+1.36}$
& $4.42\pm0.21$
& $3.23\pm0.18$\\
$\bar{B}_s^0\to D_s^{*+}\pi^-$
& $3.51$
& $3.73_{-0.84}^{+0.88}$
& $3.84_{-0.85}^{+0.90}$
& 
& $2.24_{-0.50}^{+0.56}$
& $4.30_{-0.80}^{+0.90}$
& $2.4_{-0.6}^{+0.7}$\\
\hline
$\bar{B}_s^0\to D_s^{+}K^-$
& $3.21$
& $3.41_{-0.30}^{+0.18}$
& $3.52_{-0.29}^{+0.15}$
& $3.69_{-0.65}^{+0.60}$
& $3.34_{-0.90}^{+1.04}$
&
& $2.41\pm0.16$\\
$\bar{B}_s^0\to D_s^{*+}K^-$
& $2.62$
& $2.79_{-0.61}^{+0.65}$
& $2.88_{-0.63}^{+0.66}$
& $3.02_{-0.97}^{+0.99}$
& $1.67_{-0.37}^{+0.42}$
&
& $1.63\pm0.50$ \\
\hline \hline
\end{tabular}
\caption{\label{tab:br} Updated SM predictions for the branching ratios (in units of $10^{-3}$ for $b\to c\bar{u}d$ and $10^{-4}$ for $b\to c\bar{u}s$ transitions) of $\bar{B}_{(s)}^0\to D_{(s)}^{(\ast)+}L^-$ decays at different orders in $\alpha_s$, together with the results obtained in refs.~\cite{Huber:2016xod,Bordone:2020gao} as a comparison. The column marked by $\mathrm{NNLO^{\#}}$ represents our results obtained with the weak annihilation contribution included. For the channel $\bar{B}^0\to D^{\ast+}\rho^-$, only the longitudinal polarization amplitude is considered. The experimental data is taken from refs.~\cite{ParticleDataGroup:2020ssz,HFLAV:2019otj}, with the longitudinal polarization fraction of $\bar{B}^0\to D^{\ast+}\rho^-$ decay taken from ref.~\cite{Csorna:2003bw}. For the decay modes $\bar{B}_s^0\to D_s^+\pi^-$ and $\bar{B}_s^0\to D_s^+K^-$, the previous LHCb results have been superseded by the latest updates~\cite{LHCb:2021qbv} when performing the averages given in ref.~\cite{HFLAV:2019otj}. The measured branching ratios of $\bar{B}_{s}^0$ decays should also be multiplied by a factor $1-y_s^2$, with $y_s=0.062\pm0.004$~\cite{HFLAV:2019otj}, when compared with the corresponding theoretical predictions~\cite{DeBruyn:2012wj}.} 
\end{center}
\end{table}

Our updated SM predictions for the branching ratios of $\bar{B}_{(s)}^0\to D_{(s)}^{(*)+} L^-$ decays at different orders in $\alpha_s$ are given in Table~\ref{tab:br}, together with the results obtained in refs.~\cite{Huber:2016xod,Bordone:2020gao} as a comparison. The experimental data is taken from the Particle Data Group~\cite{ParticleDataGroup:2020ssz} and/or the Heavy Flavor Averaging Group~\cite{HFLAV:2019otj}. For the decay modes $\bar{B}_s^0\to D_s^+\pi^-$ and $\bar{B}_s^0\to D_s^+K^-$, we have also replaced the previous LHCb results by the latest updates~\cite{LHCb:2021qbv} when performing the averages given in ref.~\cite{HFLAV:2019otj}. It can be seen that our updated results are generally higher than the current experimental data, even at the LO, and the higher-order perturbative corrections always add constructively to the LO results.\footnote{Due to the sizable decay width difference, $y_s\equiv\frac{\Delta\Gamma_s}{2\Gamma_s}=0.062\pm0.004$~\cite{HFLAV:2019otj}, the measured branching ratios of $\bar{B}_{s}^0$ decays should be multiplied by a factor $1-y_s^2$, when compared with the theoretical predictions~\cite{DeBruyn:2012wj}.} Especially for the decay modes $\bar{B}_{(s)}^0\to D_{(s)}^{(*)+}\pi^-$ and $\bar{B}_{(s)}^0\to D_{(s)}^{(*)+}K^-$, the difference in central values is at $30$-$70\%$ level and, after taking into account the theoretical and experimental uncertainties, the deviation can even reach about 4-5$\sigma$. It must be pointed out that such a large deviation has been observed for the first time in ref.~\cite{Bordone:2020gao}, where the values of $B_{(s)}\to D_{(s)}^{(*)}$ transition form factors were taken from ref.~\cite{Bordone:2019guc}. Compared with the results presented in ref.~\cite{Huber:2016xod}, our updated central values of the branching ratios of $\bar{B}^0$ decays are increased by about $16\%$ for $D^+$ and $20\%$ for $D^{\ast+}$ final states, respectively. This is mainly due to the following two reasons: Firstly, our input of the CKM matrix element $|V_{cb}|$ is about $7.4\%$ larger than the one used in ref.~\cite{Huber:2016xod}, where the value of $|V_{cb}|$ extracted from exclusive semi-leptonic $B$-meson decays as of 2016 was used instead. Secondly, our inputs for the $B\to D$ and $B\to D^{*}$ transition form factors, whose theoretical information available since the analysis of ref.~\cite{Huber:2016xod} has been systematically taken into account~\cite{Bernlochner:2017jka}, are now about $4.7\%$ and $6.5\%$ larger than the corresponding ones used in ref.~\cite{Huber:2016xod}, when evaluated at $q^2=0$. It is also observed that the theoretical uncertainties of the branching ratios of $\bar{B}_{s}^0\to D_{s}^{(*)+}\pi^-$ and $\bar{B}_{s}^0\to D_{s}^{(*)+} K^-$ decays are significantly reduced with respect to that obtained in ref.~\cite{Huber:2016xod}. This is mainly due to the improved lattice determinations of the $B_{s}\to D_{s}^{(*)}$ transition form factors~\cite{McLean:2019qcx,Harrison:2021tol}. Especially for the two channels $\bar{B}_{s}^0\to D_{s}^{*+}\pi^-$ and $\bar{B}_{s}^0\to D_{s}^{*+} K^-$, our updated results are about $70\%$ larger than that given in ref.~\cite{Huber:2016xod}, due mainly to the increase of the transition form factor $A_0^{B_{s}\to D_{s}^{*}}$ by about $\sim21\%$ (see also Table~\ref{tab:FFs}). On the other hand, our central values of the branching ratios are slightly larger for the $\bar{B}_{(s)}^0\to D_{(s)}^{+} L^-$ but smaller for the $\bar{B}_{(s)}^0\to D_{(s)}^{*+} L^-$ decays than the corresponding ones presented in ref.~\cite{Bordone:2020gao}, while being in perfect agreement within errors. This is also attributed mainly to the different inputs of the CKM matrix element $|V_{cb}|$ and the $B_{(s)}\to D_{(s)}^{(*)}$ transition form factors.

\begin{table}[t]
	\begin{center}
		\tabcolsep0.64cm
		\let\oldarraystretch=\arraystretch
		\renewcommand*{\arraystretch}{1.6}	
		\begin{tabular}{lccccccc}
			\hline \hline
			Decay mode  &$|b_1|$ & $|a_1/(a_1+b_1)|$ & $|a_1/(a_1+b_1)|^\mathrm{exp.}$ \\
			\hline
			$\bar{B}^0\to D^+\pi^-$
			& $\phantom{-}0.019_{\,-0.051}^{\,+0.051}$	
			& $\phantom{-}0.982\pm 0.056$
			& $\phantom{-}1.040_{\,-0.022}^{\,+0.022}$\\
			$\bar{B}^0\to D^{\ast+}\pi^-$
			& $\phantom{-}0.017_{\,-0.064}^{\,+0.065}$
			& $\phantom{-}0.984\pm 0.075$
			& $\phantom{-}1.016_{\,-0.032}^{\,+0.031}$ \\ 
			$\bar{B}^0\to D^{+}\rho^-$
			& $\phantom{-}0.015_{\,-0.038}^{\,+0.038}$
			& $\phantom{-}0.987\pm 0.043$
			&     \\
			$\bar{B}^0\to D^{\ast+}\rho^-$
			& $\phantom{-}0.015_{\,-0.044}^{\,+0.045}$
			& $\phantom{-}0.986\pm 0.050$
			&      \\
			$\bar{B}_s^0\to D_s^{+}K^-$
			& $\phantom{-}0.026_{\,-0.068}^{\,+0.068}$
			&$\phantom{-}0.976\pm 0.072$
			& $\phantom{-}1.003_{\,-0.020}^{\,+0.021}$ \\
			$\bar{B}_s^0\to D_s^{*+}K^-$
			& $\phantom{-}0.025_{\,-0.095}^{\,+0.095}$
			& $\phantom{-}0.977\pm 0.106$
			& $\phantom{-}1.048_{\,-0.046}^{\,+0.043}$ \\
			\hline \hline
		\end{tabular}
		\caption{\label{tab:annicompare} Our estimates of the weak annihilation coefficients $|b_1(D_{(s)}^{(*)+} L^-)|$ defined by eq.~\eqref{eq:b1}, with the default values obtained by setting $\rho_{A}=0$ and the errors by varying $\varrho_A$ and $\varphi_A$ within the intervals $\rho_{A}\in [0,2]$ and $\phi_{A}\in[0,2\pi]$ respectively. The ratios $|a_1(D_{(s)}^{(*)+} L^-)/(a_1(D_{(s)}^{(*)+} L^-)+b_1(D_{(s)}^{(*)+} L^-))|$ obtained both from a direct estimate within the QCDF framework and by following the method proposed in ref.~\cite{Fleischer:2010ca} are also shown in the third and the last column, respectively.}
	\end{center}
\end{table}

As can be seen from Table~\ref{tab:br}, our estimate of the weak annihilation contribution, although being plagued by large uncertainties due to the model parameters $\varrho_A$ and $\varphi_A$, always contributes constructively to the dominant color-allowed tree amplitude, with its effect being less than $5\%$ on the final branching ratios. Thus, the weak annihilation effect does not help to reconcile the deviations observed in the class-I non-leptonic decays~\cite{Huber:2016xod,Bordone:2020gao}. To see the relative size of the weak annihilation contribution in these decays, we show in Table~\ref{tab:annicompare} our estimates of the effective coefficients $|b_1(D_{(s)}^{(*)+} L^-)|$ defined by eq.~\eqref{eq:b1} and the ratios $|a_1(D_{(s)}^{(*)+} L^-)/(a_1(D_{(s)}^{(*)+} L^-)+b_1(D_{(s)}^{(*)+} L^-))|$. The latter can also be extracted from the measured ratios of branching fractions~\cite{HFLAV:2019otj}, $\mathrm{Br}(\bar{B}^0\to D^{(*)+} K^-)/\mathrm{Br}(\bar{B}^0\to D^{(*)+} \pi^-)$ for $|a_1(D^{(*)+} \pi^-)/(a_1(D^{(*)+} \pi^-)+b_1(D^{(*)+} \pi^-))|$ and $\mathrm{Br}(\bar{B}_{s}^0\to D_{s}^{(*)+} \pi^-)/\mathrm{Br}(\bar{B}_{s}^0\to D_{s}^{(*)+} K^-)$ for $|a_1(D_{s}^{(*)+} K^-)/(a_1(D_{s}^{(*)+} K^-)+b_1(D_{s}^{(*)+} K^-))|$, after correcting the factorizable $SU(3)$-breaking corrections~\cite{Fleischer:2010ca}. The values obtained in such a way are shown in the last column of Table~\ref{tab:annicompare} as a comparison. It can be seen that both methods give similar magnitudes of the weak annihilation contributions, with our estimated results being positive while the extracted values negative, and no sign of an enhanced weak annihilation topology is shown in these class-I non-leptonic decays, as is generally expected both within the QCDF framework~\cite{Beneke:2000ry,Lenz:2019lvd} and based on the $SU(3)$-flavor symmetry of strong interactions~\cite{Fleischer:2010ca}. As a consequence, from now on, we shall no longer consider the weak annihilation contribution. 

\subsection{Updated predictions for $|a_1(D_{(s)}^{(\ast)+}L^-)|$ and $R_{(s)L}^{(\ast)}$}
\label{subsec:Ratios}

The CKM matrix element $|V_{cb}|$ and the $B_{(s)}\to D_{(s)}^{(\ast)}$ transition form factors are key inputs aimed at precise theoretical predictions for the absolute branching ratios of $\bar{B}_{(s)}^0\to D_{(s)}^{(\ast)+}L^-$ decays, and their uncertainties are still non-negligible. As can be seen from Table~\ref{tab:FFs}, while the $B_{(s)}\to D_{(s)}$ transition form factors have reached a precision of less than $\sim2\%$~\cite{Bernlochner:2017jka,McLean:2019qcx}, the uncertainties of the  $B_{(s)}\to D_{(s)}^*$ counterparts are still large, especially for $A_0^{B_{s}\to D_{s}^{*}}$ with only a $\sim11\%$ precision~\cite{Harrison:2021tol}. Furthermore, different choices of $|V_{cb}|$ also affect the final results of the absolute branching ratios~\cite{Huber:2016xod,Bordone:2020gao}. In order to minimize the impacts of these input parameters, one can consider the ratios of the non-leptonic $\bar{B}_{(s)}^0\to D_{(s)}^{(\ast)+}L^-$ decay rates with respect to the corresponding differential semi-leptonic $\bar{B}_{(s)}^0\to D_{(s)}^{(\ast)+}\ell^-\bar{\nu}_{\ell}$ decay rates evaluated at $q^2=m_L^2$, where $\ell$ refers to either an electron or a muon, and $q^2$ is the four-momentum squared transferred to the lepton pair. In this way, one obtains~\cite{Bjorken:1988kk,Neubert:1997uc,Beneke:2000ry}\footnote{Here we assume that the semi-leptonic $\bar{B}_{(s)}^0\to D_{(s)}^{(\ast)+}\ell^-\bar{\nu}_{\ell}$ decays do not receive any NP contributions beyond the SM, as indicated by the current experimental data~\cite{ParticleDataGroup:2020ssz,HFLAV:2019otj,Jung:2018lfu}.}
\begin{align} \label{eq:nonlep2semilep}
	R_{(s)L}^{(\ast)} &\equiv \frac{\Gamma(\bar{B}_{(s)}^0\to D_{(s)}^{(\ast)+}L^-)}{d\Gamma(\bar{B}_{(s)}^0\to D_{(s)}^{(\ast)+}\ell^-\bar{\nu}_{\ell})/dq^2\mid_{q^2=m_L^2}}\,= \, 6\pi^2\,|V_{uq}|^2\,f_L^2\,|a_1(D_{(s)}^{(\ast)+}L^-)|^2\, X_L^{(\ast)}\,,
\end{align}
which by construction are free of the uncertainty related to $|V_{cb}|$. Neglecting the masses of light leptons, we have exactly $X_L=X_L^{\ast}=1$ for a vector meson $L$, valid both for the sum of and separately for the longitudinal and the transverse polarization of the $D_{(s)}^{\ast+}$ mesons in the final state. This is due to the kinematic equivalence between the production of the lepton pair via the SM weak current with $\gamma^\mu(1-\gamma_5)$ structure in semi-leptonic decays and that of a vector meson with four-momentum $q^\mu$ in non-leptonic decays~\cite{Neubert:1997uc,Beneke:2000ry}. For a light pseudoscalar meson $L$, on the other hand, $X_L^{(\ast)}$ depend on both the form-factor ratios and the kinematic factors (see eq.~(68) in ref.~\cite{Neubert:1997uc} for their explicit expressions), and deviate numerically from 1 at a few percent level or below with our inputs for the $B_{(s)}\to D_{(s)}^{(*)}$ transition form factors. Eq.~\eqref{eq:nonlep2semilep} offers, therefore, a way to compare the values of the effective coefficients $|a_1(D_{(s)}^{(\ast)+}L^-)|$ fitted from the experimental data with their theoretical predictions based on the QCDF approach, which are collected in Table~\ref{tab:Absa1}, together with the available results presented in refs.~\cite{Huber:2016xod,Bordone:2020gao}. In addition, we give in Table~\ref{tab:nonlep2semilep} the values of the ratios $R_{(s)L}^{(*)}$ extracted from the current experimental data as well as our updated theoretical predictions at different orders in $\alpha_s$, which will be used later to analyze the NP effects in these class-I non-leptonic decays. 

\begin{table}[t]
	\tabcolsep0.055cm
	\let\oldarraystretch=\arraystretch
	\renewcommand*{\arraystretch}{1.6}
	\begin{center}
		\begin{tabular}{lccccccc}
			\hline \hline
			$|a_1(D_{(s)}^{(\ast)+}L^-)|$& ${\rm LO}$  &NLO  & NNLO & Ref.~\cite{Huber:2016xod} & Ref.~\cite{Bordone:2020gao} & Exp. \\ \hline
			$|a_1(D^{+}\pi^-)|$
			& $1.028$
			& $\phantom{-}1.059_{\,-0.019}^{\,+0.017}$
			& $\phantom{-}1.073_{\,-0.010}^{\,+0.005}$
			& $\phantom{-}1.073_{\,-0.014}^{\,+0.012}$
			&$\phantom{-}1.0727_{-0.0140}^{+0.0125}$
			&$\phantom{-}0.88\pm0.04$ \\ 
			$|a_1(D^{\ast+}\pi^-)|$
			& $1.028$
			& $\phantom{-}1.059_{\,-0.019}^{\,+0.017}$
			& $\phantom{-}1.075_{\,-0.011}^{\,+0.006}$
			& $\phantom{-}1.071_{\,-0.014}^{\,+0.013}$
			& $\phantom{-}1.0713_{-0.0137}^{+0.0128}$
			& $\phantom{-}0.92\pm0.04$ \\ 
			$|a_1(D^{+}\rho^-)|$
			& $1.028$
			& $\phantom{-}1.059_{\,-0.019}^{\,+0.017}$
			&$\phantom{-}1.073_{\,-0.010}^{\,+0.005}$
			& $\phantom{-}1.072_{\,-0.014}^{\,+0.012}$
			& 
			& $\phantom{-}0.92\pm0.08$ \\ 
			$|a_1(D^{\ast+}\rho^-)|$
			& $1.028$
			& $\phantom{-}1.059_{\,-0.019}^{\,+0.017}$
			& $\phantom{-}1.075_{\,-0.011}^{\,+0.006}$
			& $\phantom{-}1.071_{\,-0.014}^{\,+0.013}$
			& 
			& $\phantom{-}0.80\pm0.06$ \\ 
			\hline
			$|a_1(D^{+}K^-)|$
			& $1.028$
			& $\phantom{-}1.059_{\,-0.019}^{\,+0.018}$
			& $\phantom{-}1.075_{\,-0.011}^{\,+0.007}$
			& $\phantom{-}1.070_{\,-0.013}^{\,+0.010}$
			& $\phantom{-}1.0702_{-0.0128}^{+0.0101}$
			& $\phantom{-}0.92\pm0.04$ \\ 
			$|a_1(D^{\ast+}K^-)|$
			& $1.028$
			& $\phantom{-}1.059_{\,-0.019}^{\,+0.018}$
			& $\phantom{-}1.078_{\,-0.012}^{\,+0.009}$
			& $\phantom{-}1.069_{\,-0.013}^{\,+0.010}$
			& $\phantom{-}1.0687_{-0.0125}^{+0.0103}$
			& $\phantom{-}0.94\pm0.11$ \\ 
			$|a_1(D^{+}K^{\ast-})|$
			& $1.028$
			& $\phantom{-}1.058_{\,-0.019}^{\,+0.017}$
			& $\phantom{-}1.071_{\,-0.009}^{\,+0.004}$
			& $\phantom{-}1.070_{\,-0.013}^{\,+0.010}$
			& 
			& $\phantom{-}1.02\pm0.10$ \\ 
			\hline
			$|a_1(D_{s}^{+}\pi^-)|$
			& $1.028$
			& $\phantom{-}1.059_{-0.019}^{+0.017}$
			& $\phantom{-}1.073_{-0.010}^{+0.005}$
			& $\phantom{-}1.073_{-0.014}^{+0.012}$
			& $\phantom{-}1.0727_{-0.0140}^{+0.0125}$
			& $\phantom{-}0.90\pm0.04$ \\ 
			$|a_1(D_{s}^{*+}\pi^-)|$
			& $1.028$
			& $\phantom{-}1.059_{-0.019}^{+0.017}$
			& $\phantom{-}1.075_{-0.011}^{+0.006}$
			& $\phantom{-}1.071_{-0.014}^{+0.013}$
			& $\phantom{-}1.0713_{-0.0137}^{+0.0128}$
			& $\phantom{-}0.83\pm0.13$ \\ 
			$|a_1(D_{s}^{+}K^-)|$
			& $1.028$
			& $\phantom{-}1.059_{-0.019}^{+0.018}$
			& $\phantom{-}1.075_{-0.011}^{+0.007}$
			& $\phantom{-}1.070_{-0.013}^{+0.010}$
			& $\phantom{-}1.0702_{-0.0128}^{+0.0101}$
			& $\phantom{-}0.89\pm0.05$ \\ 
			$|a_1(D_{s}^{*+}K^-)|$
			& $1.028$
			& $\phantom{-}1.059_{-0.019}^{+0.018}$
			& $\phantom{-}1.078_{-0.012}^{+0.009}$
			&$\phantom{-}1.069_{\,-0.013}^{\,+0.010}$
			& $\phantom{-}1.0687_{-0.0125}^{+0.0103}$
			& $\phantom{-}0.79\pm0.14$ \\ 
			\hline \hline
		\end{tabular}
		\caption{\label{tab:Absa1} Theoretical and experimental values of the effective coefficients $|a_1(D_{(s)}^{(\ast)+}L^-)|$. The experimental errors are estimated by adding the uncertainties of the non-leptonic branching ratios and the semi-leptonic differential decay rates in quadrature. Note that, at leading power in $\Lambda_\mathrm{QCD}/m_b$, $|a_1(D_{(s)}^{(\ast)+}L^-)|$ calculated within the QCDF framework depend only on the light meson $L$.}
	\end{center}
\end{table}

\begin{table}[t]
\tabcolsep0.43cm
\let\oldarraystretch=\arraystretch
\renewcommand*{\arraystretch}{1.6}
\begin{center}
\begin{tabular}{lccccc}
\hline \hline
$R_{(s)L}^{(*)}$ & ${\rm LO}$ & ${\rm NLO}$ & ${\rm NNLO}$ &  Exp. & Deviation~($\sigma$)\\
\hline
  $R_{\pi}$
& $\phantom{-}1.01$
& $\phantom{-}1.07_{-0.05}^{+0.04}$
& $\phantom{-}1.10_{-0.03}^{+0.03}$
& $\phantom{-}0.74\pm0.06$ 
& $5.4$ \\ 
  $R_{\pi}^{\ast}$
& $\phantom{-}1.00$
& $\phantom{-}1.06_{-0.05}^{+0.04}$
& $\phantom{-}1.10_{-0.04}^{+0.03}$
& $\phantom{-}0.80\pm0.06$ 
& $4.3$ \\ 
  $R_{\rho}$
& $\phantom{-}2.77$
& $\phantom{-}2.94_{-0.19}^{+0.19}$
& $\phantom{-}3.02_{-0.18}^{+0.17}$
& $\phantom{-}2.23\pm0.37$ 
& $1.9$ \\ 
\hline
  $R_{K}$
& $\phantom{-}0.78$
& $\phantom{-}0.83_{-0.03}^{+0.03}$
& $\phantom{-}0.85_{-0.02}^{+0.01}$
& $\phantom{-}0.62\pm0.05$
& $4.5$ \\ 
  $R_{K}^{\ast}$
& $\phantom{-}0.72$
& $\phantom{-}0.76_{-0.03}^{+0.03}$
& $\phantom{-}0.79_{-0.02}^{+0.01}$
& $\phantom{-}0.60\pm0.14$ 
& $1.3$ \\ 
  $R_{K^\ast}$
& $\phantom{-}1.41$
& $\phantom{-}1.49_{-0.11}^{+0.11}$
& $\phantom{-}1.53_{-0.10}^{+0.10}$
& $\phantom{-}1.38\pm0.25$ 
& $0.6$ \\ 
\hline
  $R_{s\pi}$
& $\phantom{-}1.01$
& $\phantom{-}1.07_{-0.05}^{+0.04}$
& $\phantom{-}1.10_{-0.03}^{+0.03}$
& $\phantom{-}0.77\pm0.07$ 
& $4.3$ \\ 
$R^*_{s\pi}$
& $\phantom{-}1.00$
& $\phantom{-}1.06_{-0.05}^{+0.05}$
& $\phantom{-}1.10_{-0.04}^{+0.03}$
& $\phantom{-}0.65_{-0.19}^{+0.22}$ 
& $2.2$ \\ 
$R_{sK}$
& $\phantom{-}0.78$
& $\phantom{-}0.82_{-0.03}^{+0.03}$
& $\phantom{-}0.85_{-0.02}^{+0.01}$
& $\phantom{-}0.58\pm0.06$ 
& $4.4$ \\
$R^*_{sK}$
& $\phantom{-}0.71$
& $\phantom{-}0.75_{-0.03}^{+0.03}$
& $\phantom{-}0.78_{-0.02}^{+0.02}$
& $\phantom{-}0.42\pm0.14$ 
& $2.5$ \\
\hline \hline
\end{tabular}
\caption{\label{tab:nonlep2semilep} Theoretical and experimental values of the ratios $R_{(s)L}^{(*)}$, in units of $\rm GeV^{2}$ for $b\to c\bar{u}d$ and $10^{-1}\rm GeV^{2}$ for $b\to c\bar{u}s$ transitions, respectively. The levels of deviations between the NNLO predictions and the current experimental data are shown in the last column.}
\end{center}
\end{table}

From Table~\ref{tab:Absa1}, one can see that our results for the effective coefficients $|a_1(D_{(s)}^{(\ast)+}L^-)|$ at the NNLO in $\alpha_s$ are well consistent with that obtained in ref.~\cite{Huber:2016xod}, up to slight variations induced by the updated input parameters from $\alpha_{s}(m_{Z})$, the Gegenbauer moments, as well as the quark masses.\footnote{Here we adopt the $\msbar$ scheme for the bottom- and charm-quark masses, which means that the mass ratio $z=\overline{m}_c(\mu)/\overline{m}_b(\mu)$ and the logarithmic terms in the hard kernels $T_{ij}(u)$ should be understood as $\ln{[\mu^2/\overline{m}_b(\mu)^2]}$, with the renormalization scale chosen at $\mu_b=\overline{m}_b(\overline{m}_b)$.} As emphasized already in refs.~\cite{Huber:2016xod,Beneke:2000ry}, an essentially universal value of $|a_1(D_{(s)}^{(\ast)+}L^-)|\simeq1.07~(1.06)$ at the NNLO~(NLO) is predicted within the QCDF framework, which is however consistently higher than the central values fitted from the current experimental data. As shown in the last column of Table~\ref{tab:nonlep2semilep}, the deviations observed in $\bar{B}_{(s)}^0\to D_{(s)}^{(*)+}\pi^-$ and $\bar{B}_{(s)}^0\to D_{(s)}^{(*)+}K^-$ decay modes are particularly remarkable, with some of them reaching even up to 4-5$\sigma$. This is attributed to the increased theoretical predictions~\cite{Huber:2016xod} and, at the same time, the decreased experimental measurements~\cite{ParticleDataGroup:2020ssz,HFLAV:2019otj} of the absolute branching ratios, together with their reduced uncertainties, as compared to the previous analysis performed at the NLO in $\alpha_s$ within the same framework~\cite{Beneke:2000ry}. 

As pointed out already in refs.~\cite{Huber:2016xod,Bordone:2020gao}, it is quite difficult to understand the large deviations observed in these class-I non-leptonic $B$-meson decays in the SM, by simply considering the higher-order power and perturbative corrections to the decay amplitudes based on the QCDF approach~\cite{Beneke:2000ry,Chang:2017sdl}. Thus, as an alternative, we shall in the next subsections resort to possible NP explanations of these deviations, firstly in a model-independent setup by considering the NP effects from twenty linearly independent four-quark operators present in eq.~\eqref{eq:Hamiltonian}, and then within two model-dependent scenarios where the NP four-quark operators are mediated by either a colorless charged gauge boson or a colorless charged scalar. See also refs.~\cite{Bobeth:2014rda,Brod:2014bfa,Bobeth:2014rra,Lenz:2019lvd,Iguro:2020ndk,Bordone:2021cca} for recent discussions along this line.

\subsection{Model-independent analysis}
\label{subsec:model-independent}

With our prescription for the effective weak Hamiltonian given by eq.~\eqref{eq:Hamiltonian}, possible NP effects would be signaled by the non-vanishing NP Wilson coefficients $C_{i}$ that accompany the corresponding NP four-quark operators. As a model-independent analysis, we shall use the ten ratios $R_{(s)L}^{(\ast)}$ collected in Table~\ref{tab:nonlep2semilep} to constrain these NP Wilson coefficients $C_{i}$, both at the characteristic scale $\mu_b=m_{b}$ (low-scale scenario) and at the electroweak scale $\mu_W=m_{W}$ (high-scale scenario).

\subsubsection{Low-scale scenario}

\begin{figure}[t]\vspace{-1.2cm}
	\centering
	\centerline{\hspace{0.5cm}
		\includegraphics[width=0.55\textwidth]{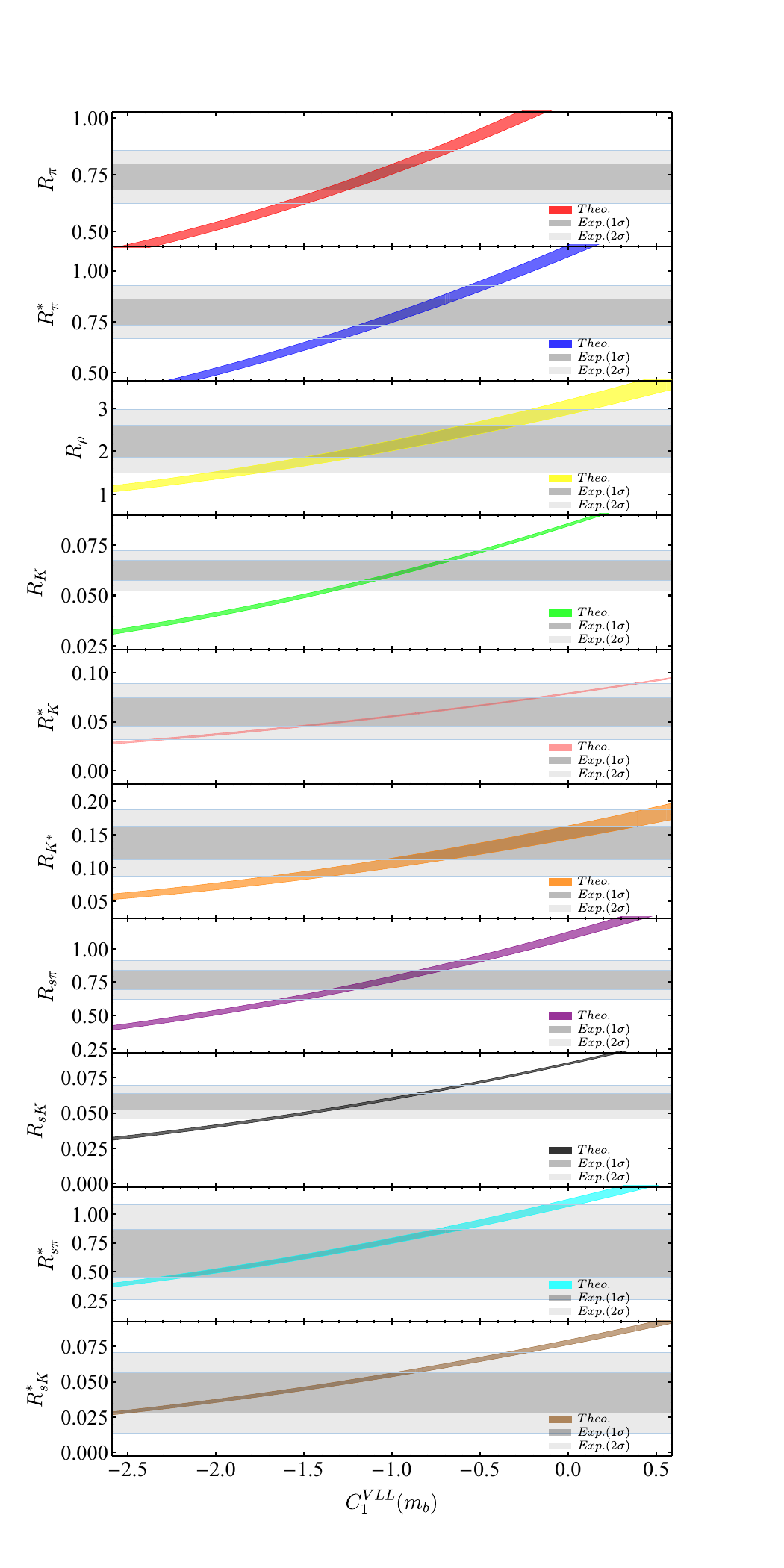}
		\hspace{-1.0cm}
		\includegraphics[width=0.55\textwidth]{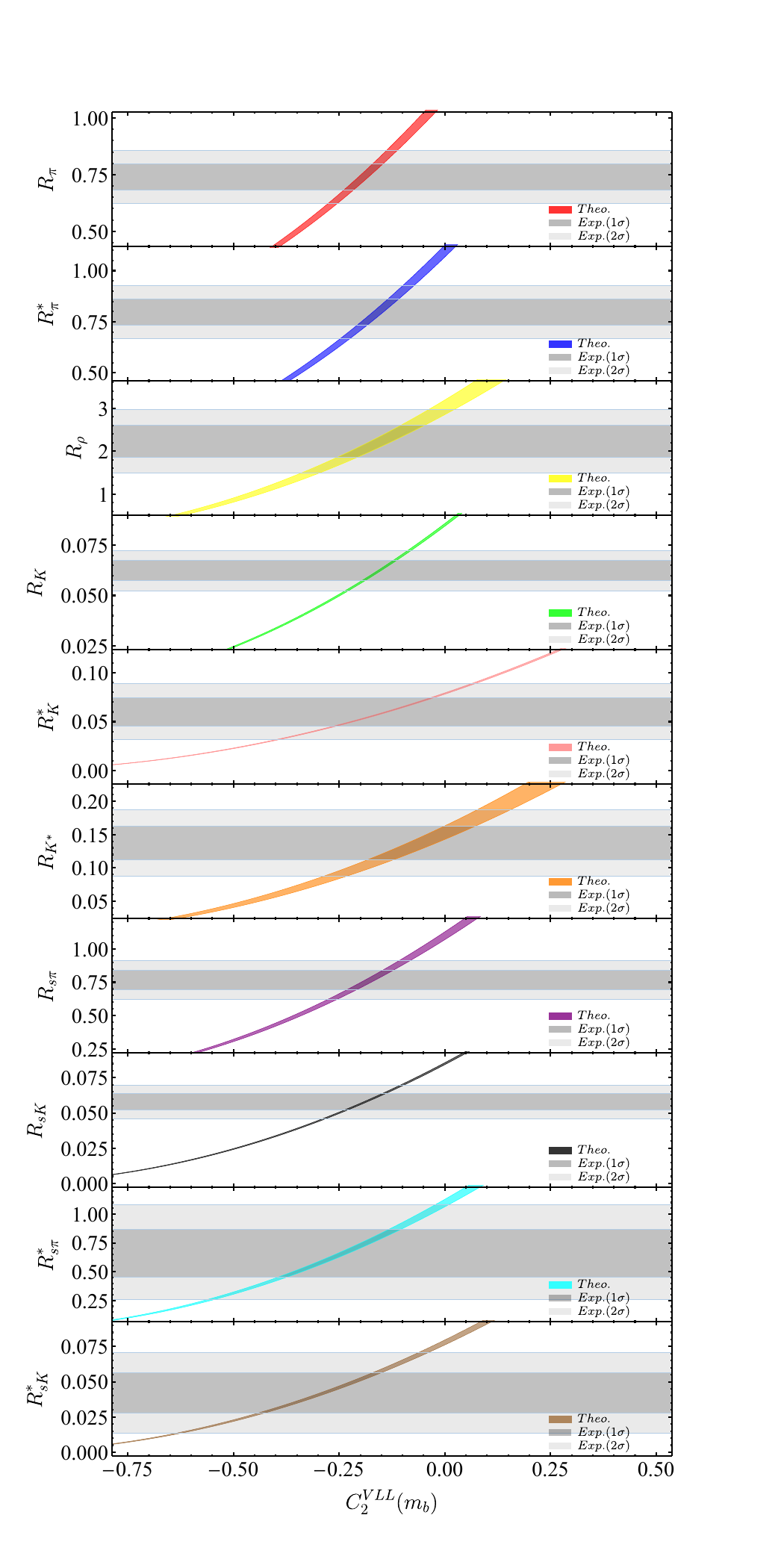}}
	\vspace{-0.5cm}
	\caption{\label{MIVLL} Constraints on the NP Wilson coefficients $C_{1}^{VLL}(m_b)$ (left) and $C_{2}^{VLL}(m_b)$ (right) from the ratios $R_{\pi}$ (red), $R_{\pi}^{*}$ (blue), $R_{\rho}$ (yellow), $R_{K}$ (green), $R_{K}^{*}$ (pink), $R_{K^{*}}$ (orange), $R_{s\pi}$ (purple), $R_{sK}$ (black), $R_{s\pi}^{\ast}$ (cyan), as well as $R_{sK}^{\ast}$ (brown), respectively. The horizontal bounds represent the experimental ranges within $1\sigma$ (dark gray) and $2\sigma$ (light gray) error bars.}
\end{figure}
%
\begin{figure}[t]\vspace{-1.2cm}
	\centering
	\centerline{\hspace{0.5cm}
		\includegraphics[width=0.55\textwidth]{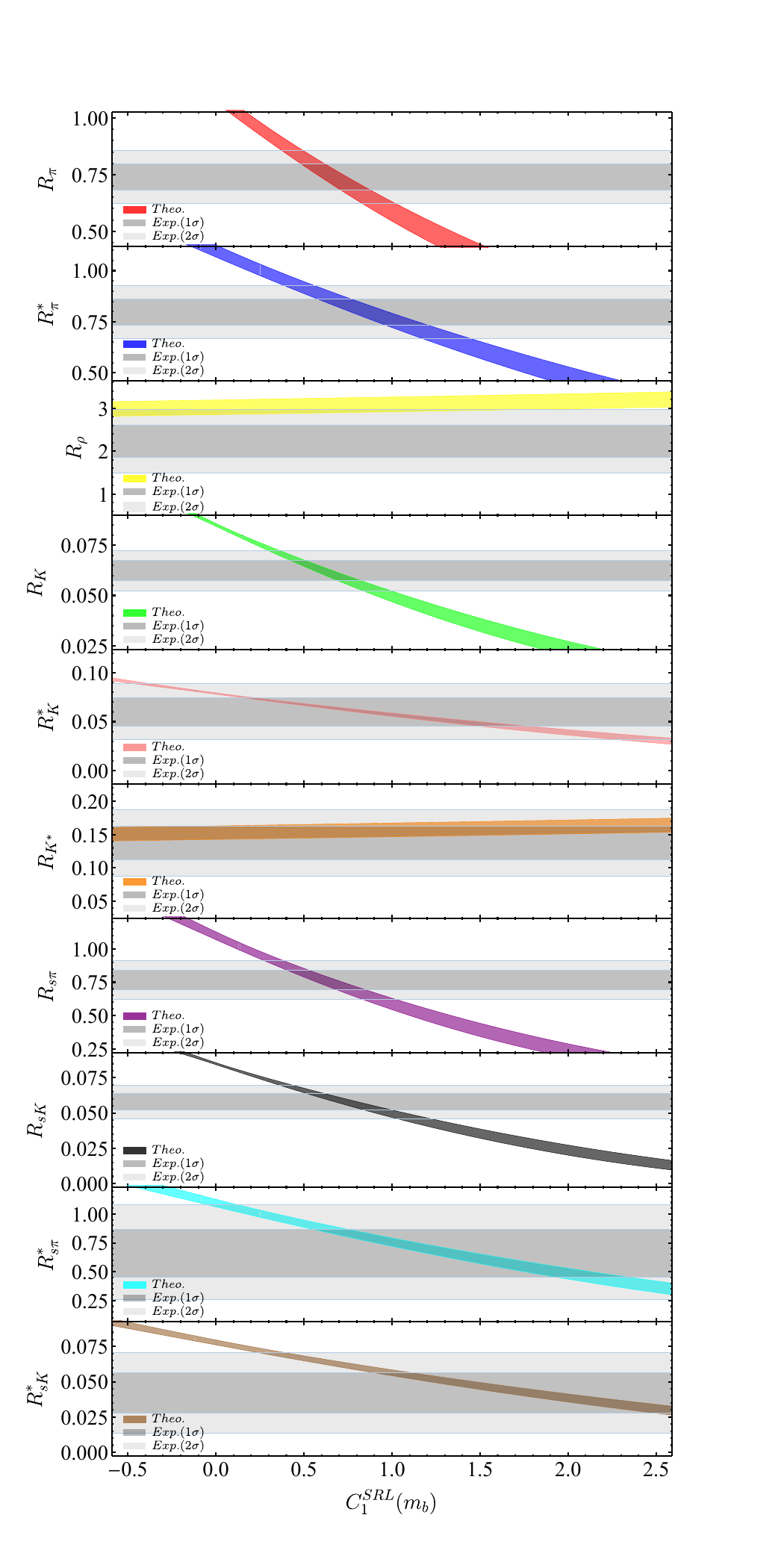}
		\hspace{-1.0cm}
		\includegraphics[width=0.55\textwidth]{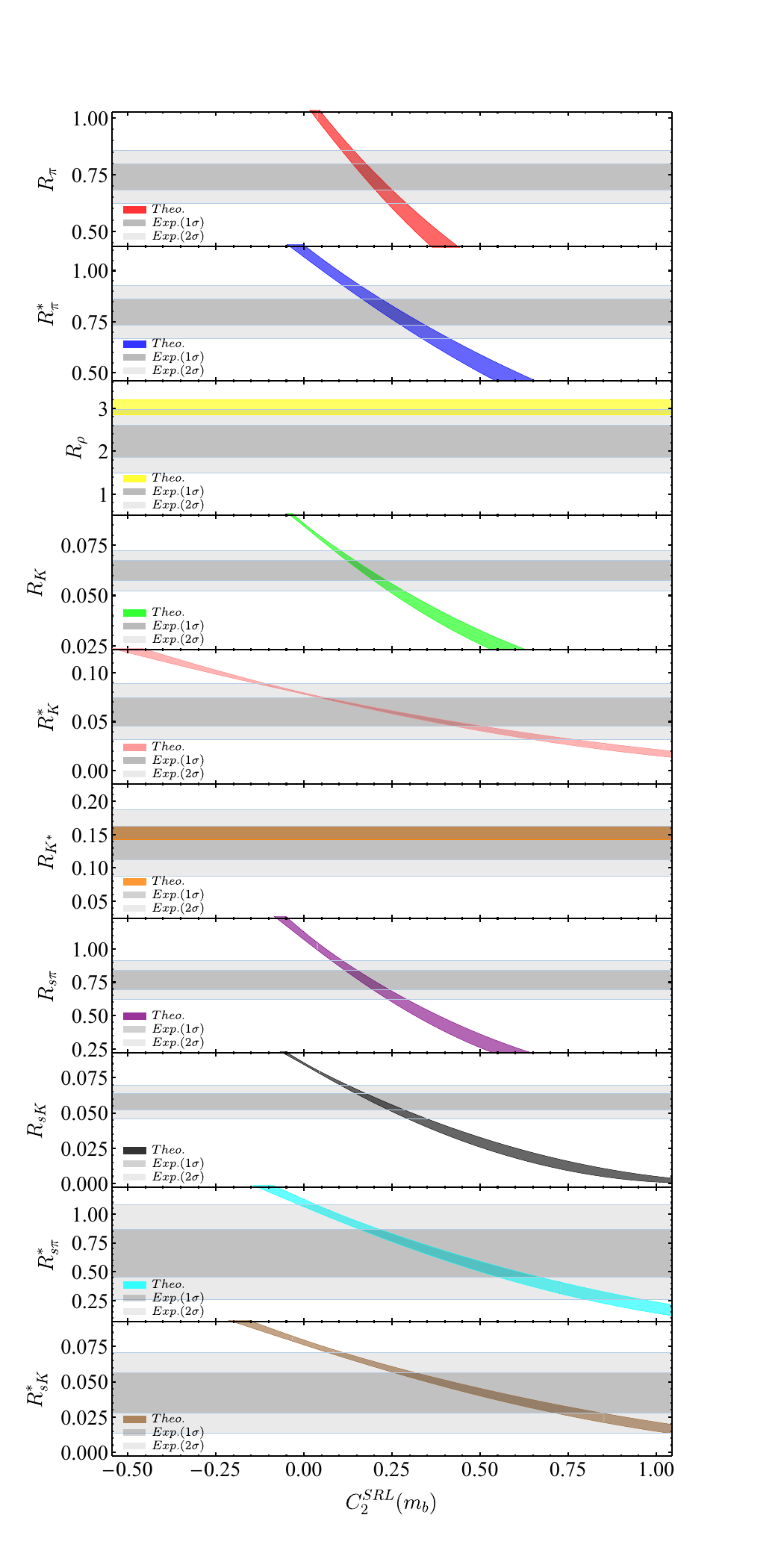}}
	\vspace{-0.5cm}	
	\caption{\label{MISRL} Same as in Fig.~\ref{MIVLL} but for the NP Wilson coefficients $C_{1}^{SRL}(m_b)$ (left) and $C_{2}^{SRL}(m_b)$ (right).}
\end{figure}
%
\begin{figure}[t]\vspace{-1.2cm}
	\centering
	\centerline{\hspace{0.5cm}
		\includegraphics[width=0.55\textwidth]{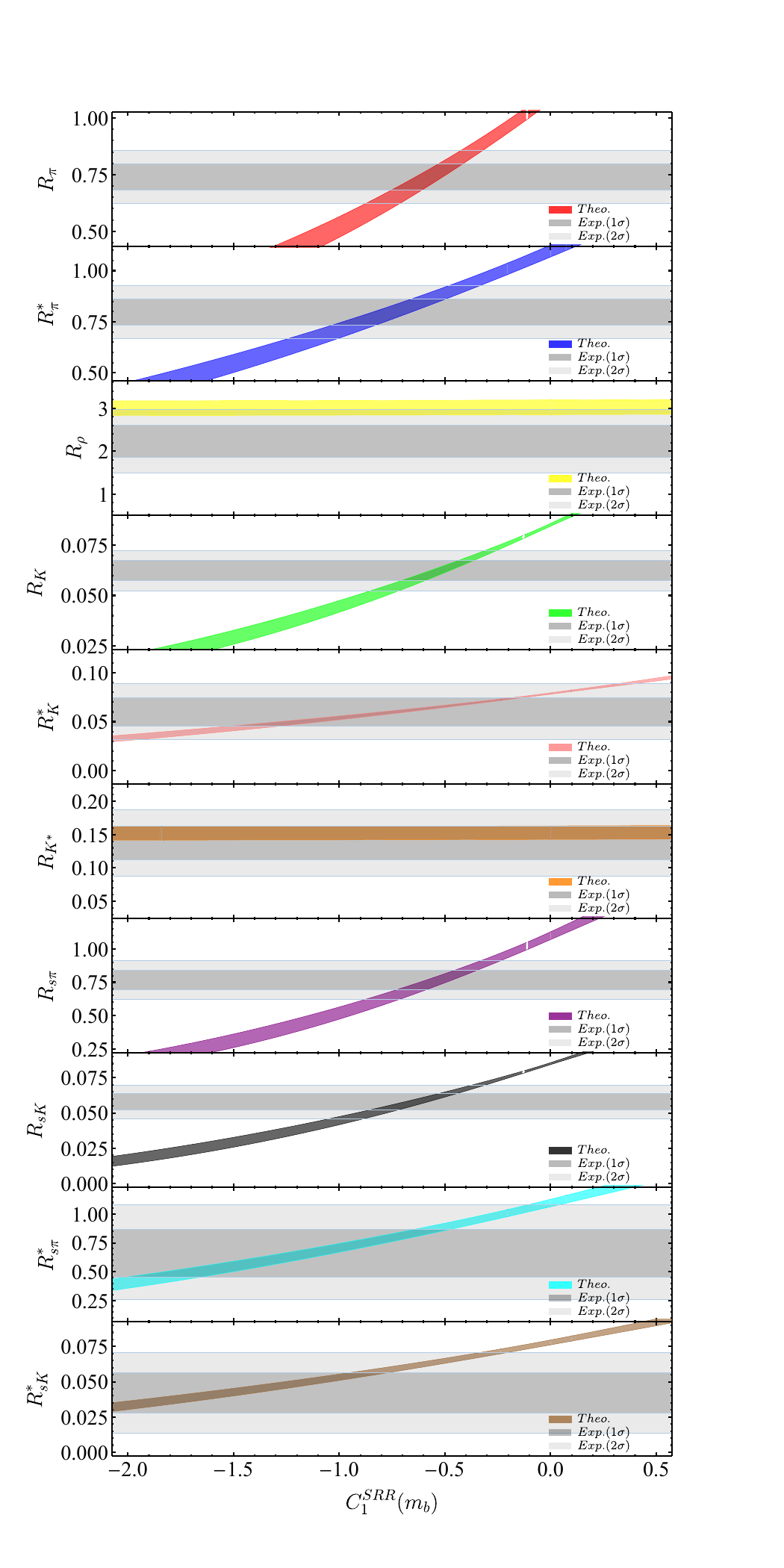}
		\hspace{-1.0cm}
		\includegraphics[width=0.55\textwidth]{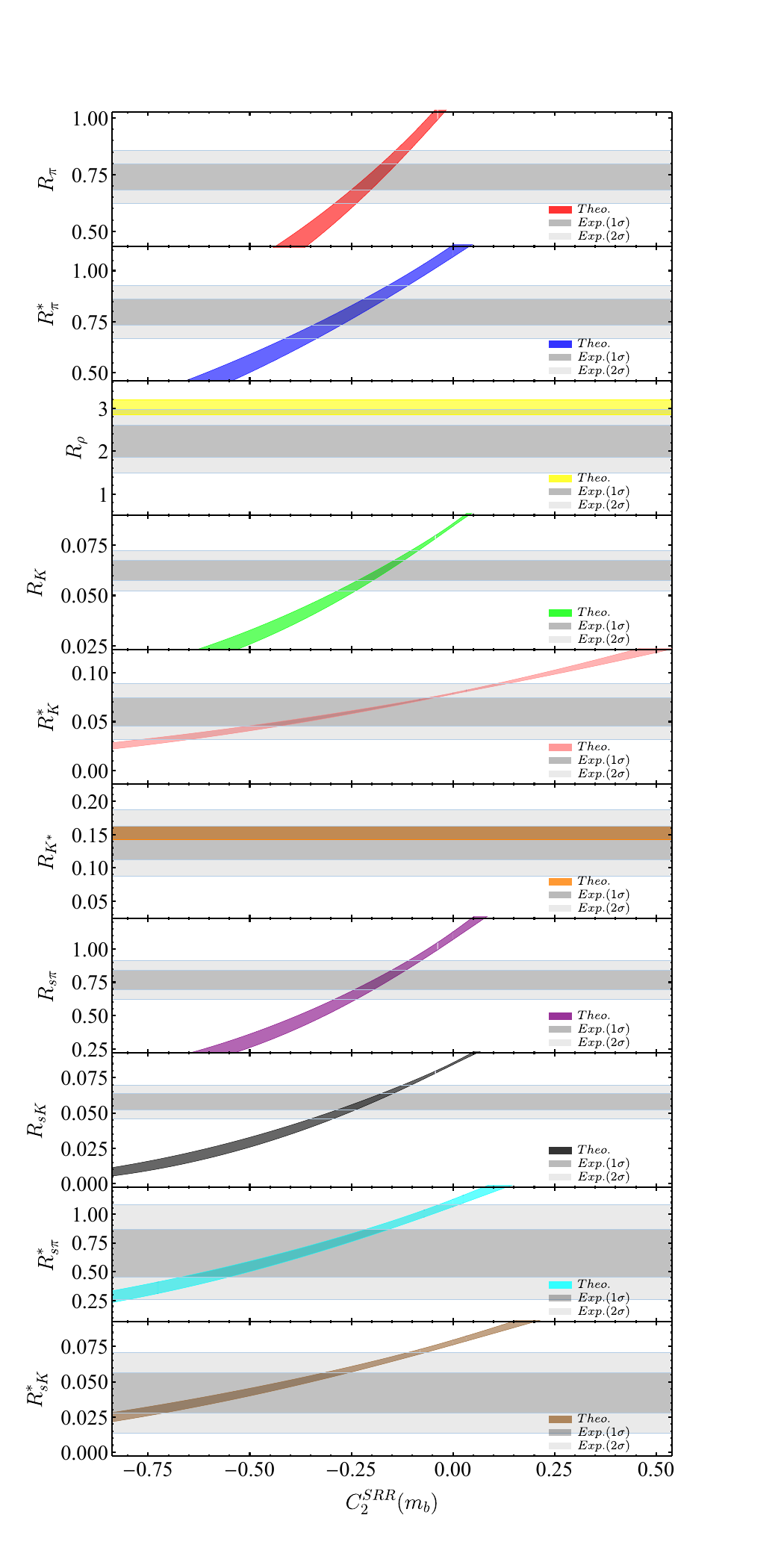}}
	\vspace{-0.5cm}	
	\caption{\label{MISRR} Same as in Fig.~\ref{MIVLL} but for the NP Wilson coefficients $C_{1}^{SRR}(m_b)$ (left) and $C_{2}^{SRR}(m_b)$ (right).}
\end{figure}
%
\begin{figure}[t]\vspace{-1.2cm}	
	\centering
	\centerline{\hspace{0.5cm}
		\includegraphics[width=0.55\textwidth]{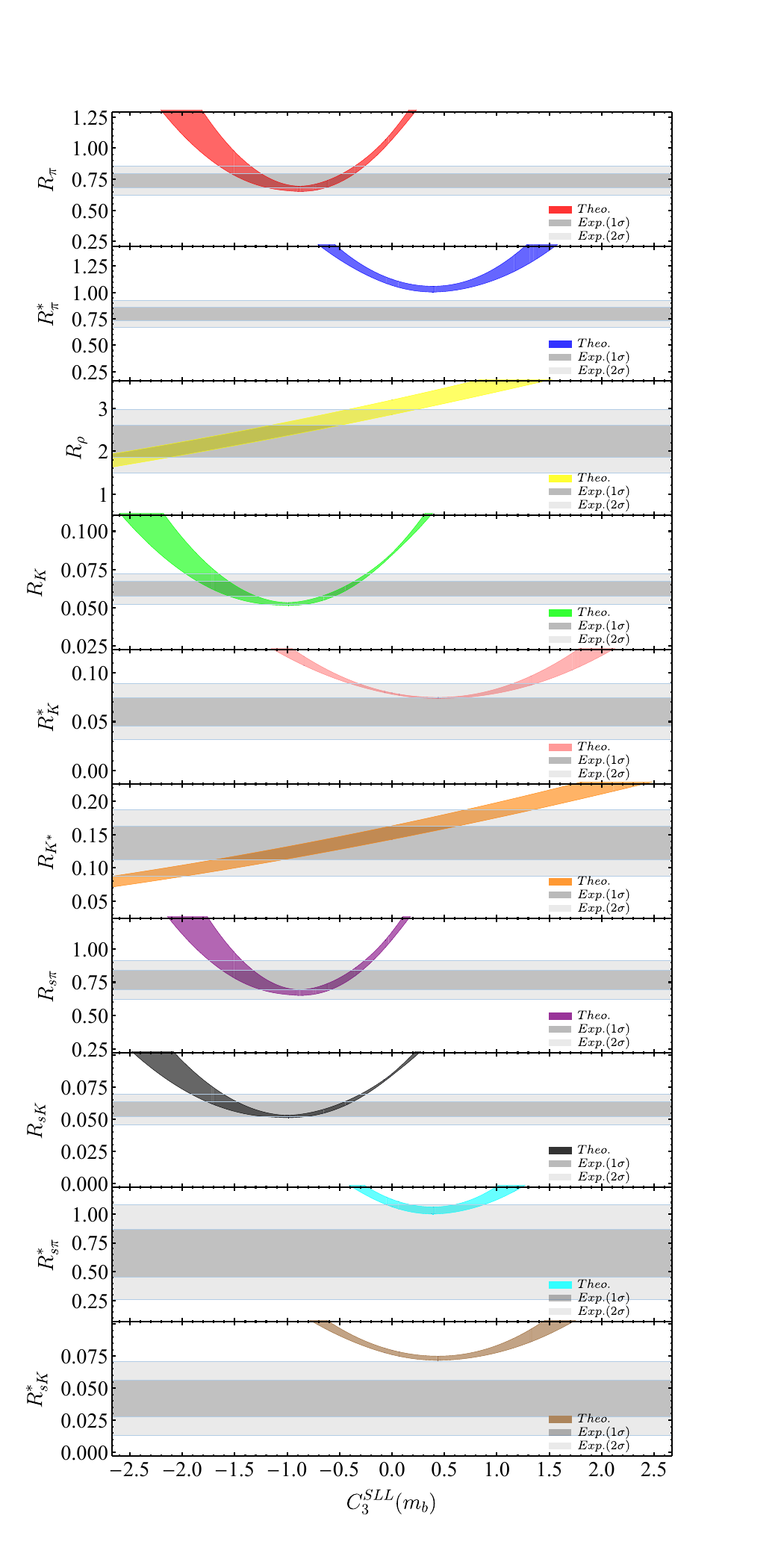}
		\hspace{-1.0cm}
		\includegraphics[width=0.55\textwidth]{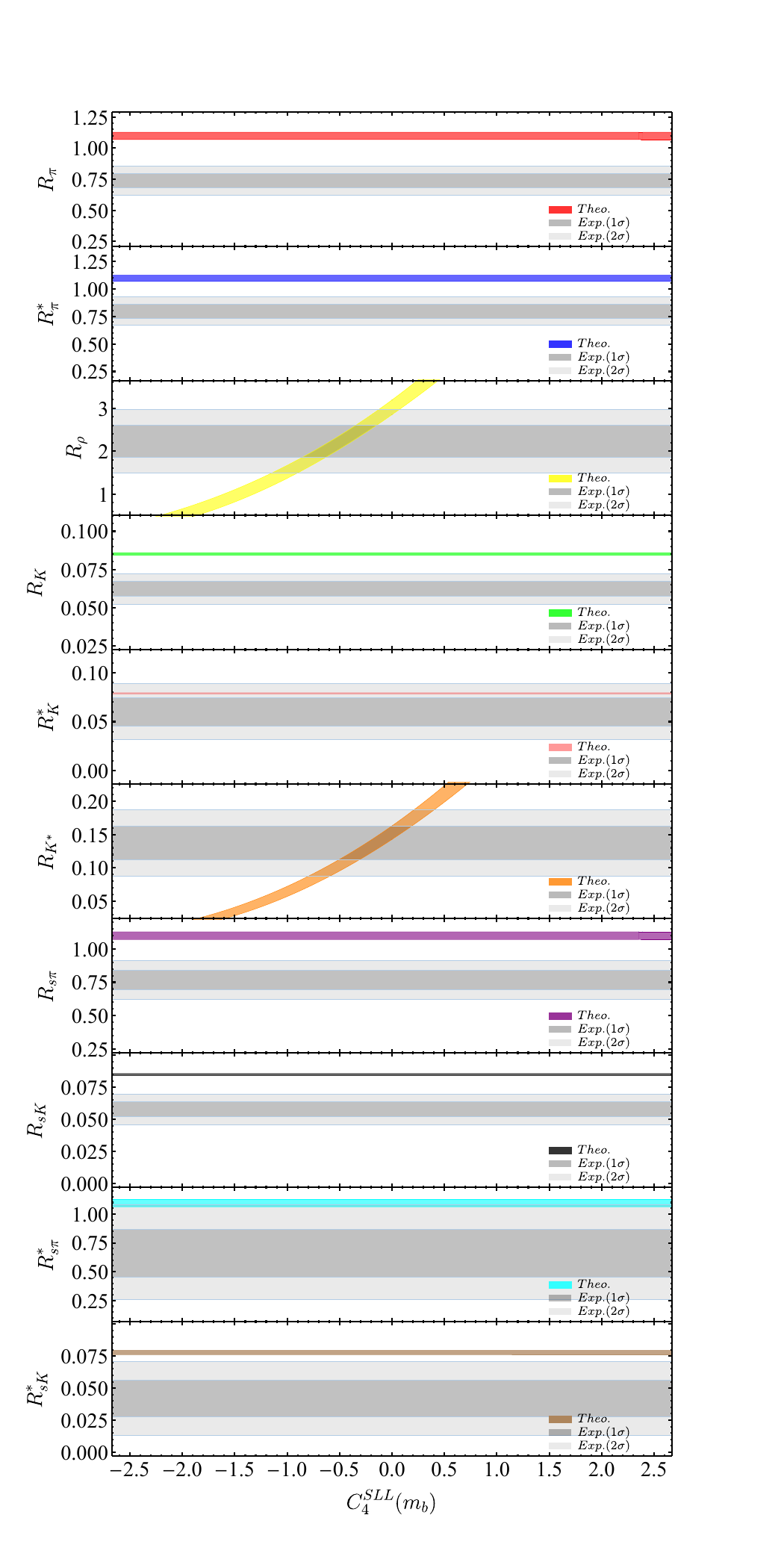}}
	\vspace{-0.5cm}
	\caption{\label{MI_TLL&TRR} Same as in Fig.~\ref{MIVLL} but for the NP Wilson coefficients $C_{3}^{SLL}(m_b)$ (left) and $C_{4}^{SLL}(m_b)$ (right).}
\end{figure}

\begin{figure}[t]\vspace{-1.2cm}	
	\centering
	\centerline{\hspace{0.5cm}
		\includegraphics[width=0.55\textwidth]{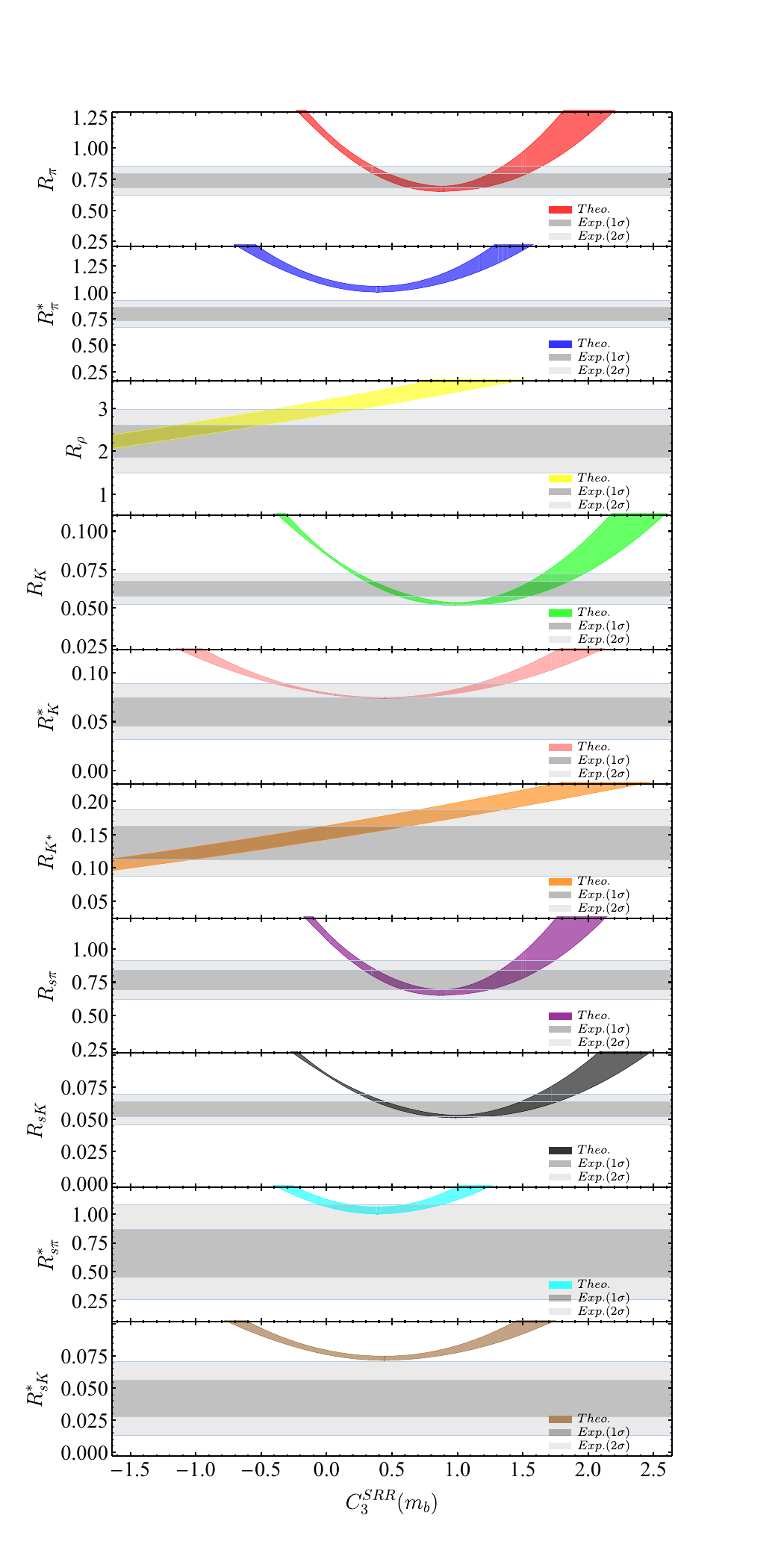}
		\hspace{-1.0cm}
		\includegraphics[width=0.55\textwidth]{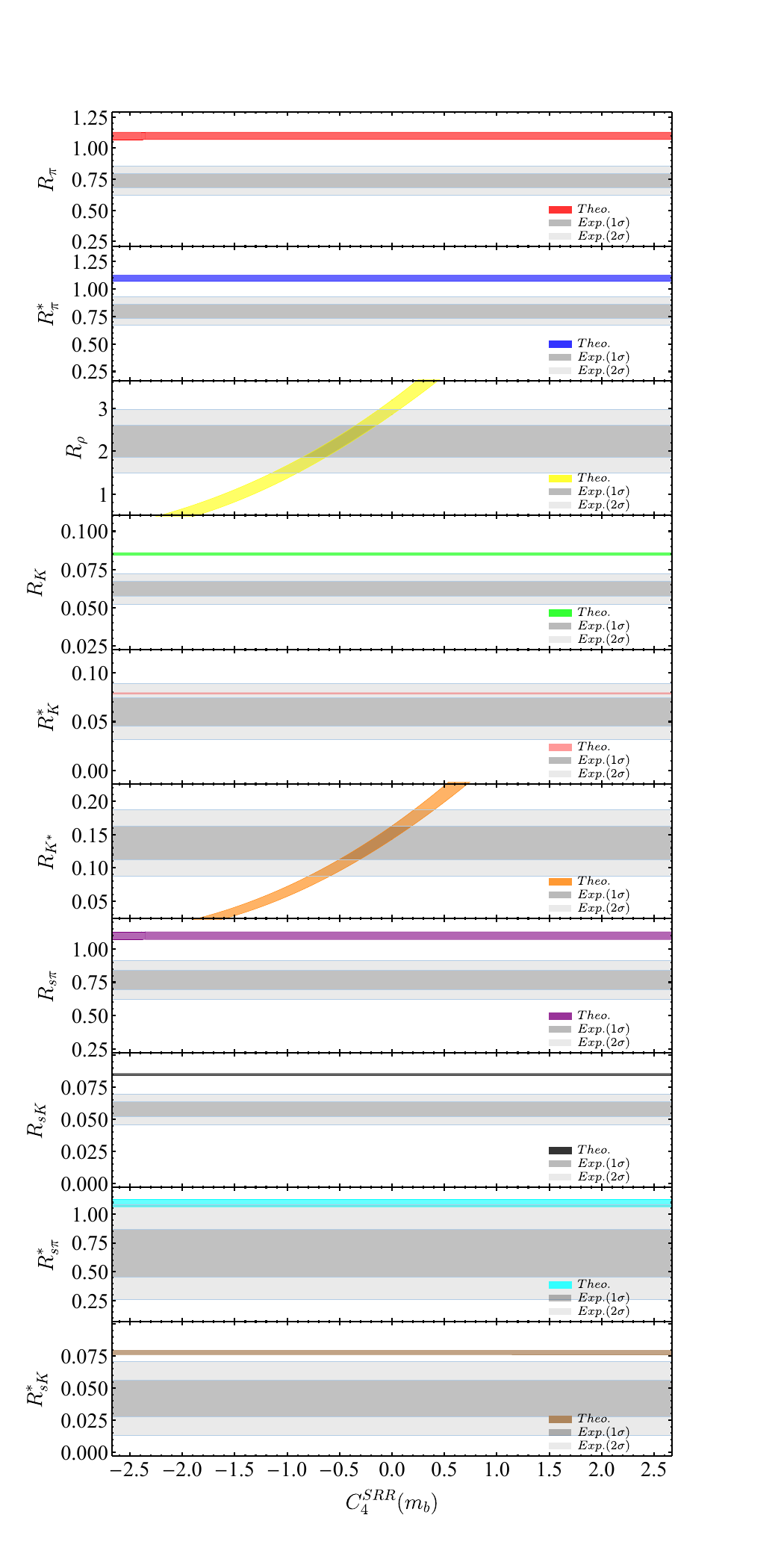}}
	\vspace{-0.5cm}
	\caption{\label{MI_TLL&TRR2} Same as in Fig.~\ref{MIVLL} but for the NP Wilson coefficients $C_{3}^{SRR}(m_b)$ (left) and $C_{4}^{SRR}(m_b)$ (right).}
\end{figure}

Firstly, let us consider the case where only a single NP four-quark operator is present in eq.~\eqref{eq:Hamiltonian}. The resulting constraints on the corresponding NP Wilson coefficient $C_i(m_b)$ from the ratios $R_{(s)L}^{(\ast)}$ are shown in Figs.~\ref{MIVLL}--\ref{MI_TLL&TRR2}. The allowed ranges for $C_i(m_b)$ under the individual and combined constraints from $R_{(s)L}^{(\ast)}$ varied within $1\sigma$ (68.27$\%$ confidence level (C.L.)) and $2\sigma$ (95.45$\%$ C.L.) error bars are collected in Table~\ref{tab:constraint} given in the appendix.\footnote{For the cases where there exist two different solutions for $C_i(m_b)$ allowed by these constraints, we quote only the one closer to the SM point where $\mC_{1}(m_b)=-0.143$ and $\mC_{2}(m_b)=1.058$, while all $C_i(m_b)=0$.} In this case, the following observations can be made:
\begin{enumerate}
	\item[$\bullet$] As can be seen from Fig.~\ref{MIVLL}, all the deviations observed in $\bar{B}_{(s)}^0\to D_{(s)}^{(\ast)+}L^-$ decays could be explained simultaneously by the two NP four-quark operators with $\gam^{\mu}(1-\gam_5)\otimes\gam_{\mu} (1-\gam_5)$ structure. As these two operators appear already in the SM, this means that we can account for the observed deviations collected in Table~\ref{tab:nonlep2semilep} by a shift to the SM Wilson coefficients $\mC_{1}$ and/or $\mC_{2}$. The final allowed ranges for the NP Wilson coefficients $C_{1}^{VLL}(m_b)$ and $C_{2}^{VLL}(m_b)$ under the combined constraints from the ten ratios $R_{(s)L}^{(\ast)}$ varied within $1\sigma$ error bars are found to be
	\begin{equation}\label{eq:C1C2VLLmb}
		C_{1}^{VLL}(m_b)\in[-1.04,-0.849]\,, \qquad C_{2}^{VLL}(m_b)\in[-0.181,-0.148]\,.
	\end{equation} 
    One can see that the constraint on $C_{2}^{VLL}(m_b)$ is much stronger than on $C_{1}^{VLL}(m_b)$. This is due to the fact that $C_{2}^{VLL}(m_b)$ gives the leading contribution to the effective coefficients $a_1(D_{(s)}^{(\ast)+}L^-)$, while $C_{1}^{VLL}(m_b)$ is suppressed by $1/N_c$ at the LO and further by $C_F/4\pi$ at the NLO in $\alpha_s$, within the QCDF framework~\cite{Beneke:2000ry,Huber:2016xod}.
	
	\item[$\bullet$] From Figs.~\ref{MISRL} and \ref{MISRR}, one can see that the NP four-quark operators with either $(1+\gam_5)\otimes(1-\gam_5)$ or $(1+\gam_5)\otimes(1+\gam_5)$ structure could also be used to account for the observed deviations but now at the $2\sigma$ level, with the corresponding allowed ranges for the NP Wilson coefficients given, respectively, by
	\begin{align}\label{eq:C1C2SRLSRRmb}
		& C_{1}^{SRL}(m_b)\in[0.390,0.989]\,, && C_{2}^{SRL}(m_b)\in[0.112,0.283]\,,\nn \\[0.2cm]
		& C_{1}^{SRR}(m_b)\in[-0.848,-0.335]\,, && C_{2}^{SRR}(m_b)\in[-0.283,-0.112]\,.
	\end{align}
    The much weaker constraints on $C_{1}^{SRL(R)}(m_b)$ with respect to on $C_{2}^{SRL(R)}(m_b)$ are also due to the fact that the latter always provide the leading contributions to the hard kernels $T_{ij}(u)$. For the decay modes where $L$ is a light pseudoscalar meson, the hadronic matrix elements of these (pseudo-)scalar four-quark operators, although being formally power-suppressed, would be chirally-enhanced by the factors $2\mu_p(\mu)/(\overline{m}_b(\mu)\mp\overline{m}_c(\mu))$ and hence be not much suppressed numerically for realistic bottom- and charm-quark masses~\cite{Beneke:2001ev,Beneke:2003zv}. This explains the important role played by these (pseudo-)scalar four-quark operators in non-leptonic $B$-meson decays both within the SM~\cite{Beneke:2001ev,Beneke:2003zv}\footnote{Within the SM, the (pseudo-)scalar four-quark operators originate from the Fierz transformation of the penguin operators $Q_{6,8}$; for more details see, \textit{e.g.}, refs.~\cite{Beneke:2001ev,Beneke:2003zv}.} and in various NP models~\cite{Beneke:2009eb,Bobeth:2014rra,Chang:2008tf,Cheng:2003im,Cheng:2004jf,Das:2004hq,Hatanaka:2007mp}.
	
	\item[$\bullet$] As can be seen from Table~\ref{tab:constraint}, the remaining NP four-quark operators with other Dirac structures present in eq.~\eqref{eq:Hamiltonian} are already ruled out by the combined constraints from the ten ratios $R_{(s)L}^{(\ast)}$ collected in Table~\ref{tab:nonlep2semilep}, even at the $2\sigma$ level. As the matrix elements $\langle L^{-}|\bar{q}(1\pm\gamma_5)u|0\rangle\equiv0$ for a light charged vector meson $L^-$, there is no LO contribution to the decay amplitudes of $\bar{B}_{(s)}^0\to D_{(s)}^{+}\rho^-$ and $\bar{B}_{(s)}^0\to D_{(s)}^{+}K^{\ast-}$ decays from the NP four-quark operators with $(1\pm\gam_5)\otimes(1\pm\gam_5)$ and $(1\pm\gam_5)\otimes(1\mp\gam_5)$ structures. This explains why the two ratios $R_{\rho}$ and $R_{K^{*}}$ receive insignificant contributions from these operators (see also the third and the sixth plot in Figs.~\ref{MISRL} and \ref{MISRR}). For the NP four-quark operators with $\sigma^{\mu\nu}(1\pm\gamma_5)\otimes\sigma_{\mu\nu}(1\pm\gamma_5)$ structures, on the other hand, the ratios $R_{\pi}$, $R_{\pi}^{*}$, $R_{K}$ and $R_{K}^{*}$ receive only negligible contributions from the NP Wilson coefficients $C_{4}^{SLL}(m_b)$ and $C_{4}^{SRR}(m_b)$ due to $\langle L^{-}|\bar{q}\sigma^{\mu\nu}(1\pm\gamma_5)u|0\rangle\equiv0$ for a light charged pseudoscalar meson $L^-$, while contributions from $C_{3}^{SLL}(m_b)$ and $C_{3}^{SRR}(m_b)$ depend crucially on whether the final-state heavy mesons are $D_{(s)}^+$ or $D_{(s)}^{\ast+}$, as shown in Figs.~\ref{MI_TLL&TRR} and \ref{MI_TLL&TRR2}. As a consequence, the tensor four-quark operators also fail to provide a simultaneous explanation of the ten ratios $R_{(s)L}^{(\ast)}$ collected in Table~\ref{tab:nonlep2semilep}, even at the $2\sigma$ level.
	
	\item[$\bullet$] Due to the relatively larger experimental uncertainties of the three ratios $R_{\rho}$, $R_{K}^{*}$, and $R_{K^{*}}$, their constraints on the NP Wilson coefficients are much weaker. More precise measurements of these decay modes are, therefore, expected from the LHCb~\cite{Bediaga:2018lhg} and Belle II~\cite{Kou:2018nap} experiments, which will be helpful to further discriminate the NP contributions from $C_{i}^{VLL}(m_b)$, $C_{i}^{SRL}(m_b)$, and $C_{i}^{SRR}(m_b)$.
\end{enumerate}

We now consider the case where two NP four-quark operators with the same Dirac but different color structures are present in eq.~\eqref{eq:Hamiltonian}, and allow the corresponding two NP Wilson coefficients to vary simultaneously. To obtain the allowed regions for the NP Wilson coefficients, we follow the strategies used in refs.~\cite{Jung:2012vu,Li:2013vlx}: each point in the NP parameter space corresponds to a theoretical range constructed for the ratios $R_{(s)L}^{(\ast)}$ in the point, with the corresponding theoretical uncertainty taken also into account. If this range has overlap with the 2$\sigma$ range of the experimental data on $R_{(s)L}^{(\ast)}$, this point is then assumed to be allowed. Here the theoretical uncertainty at each point in the NP parameter space is obtained in the same way as in the SM, \textit{i.e.}, by varying each input parameter within its respective range and then adding the individual uncertainty in quadrature. Such a treatment is motivated by the observation that, while the experimental data yields approximately a Gaussian distribution for the branching ratios of $\bar{B}_{(s)}^0\to D_{(s)}^{(\ast)+}L^-$ decays, a theoretical calculation does not. As the latter depends on a set of hadronic input parameters like the heavy-to-heavy transition form factors as well as the decay constants and Gegenbauer moments of the light mesons, for which no probability distribution is known, it is more suitable to assume that these theory parameters have no particular distribution but are only constrained in certain allowed ranges with an equal weighting, irrespective of how close they are from the edges of the allowed ranges~\cite{Hocker:2001xe,Charles:2004jd}.

\begin{figure}[t]
	\centering
	\includegraphics[width=1.0\textwidth]{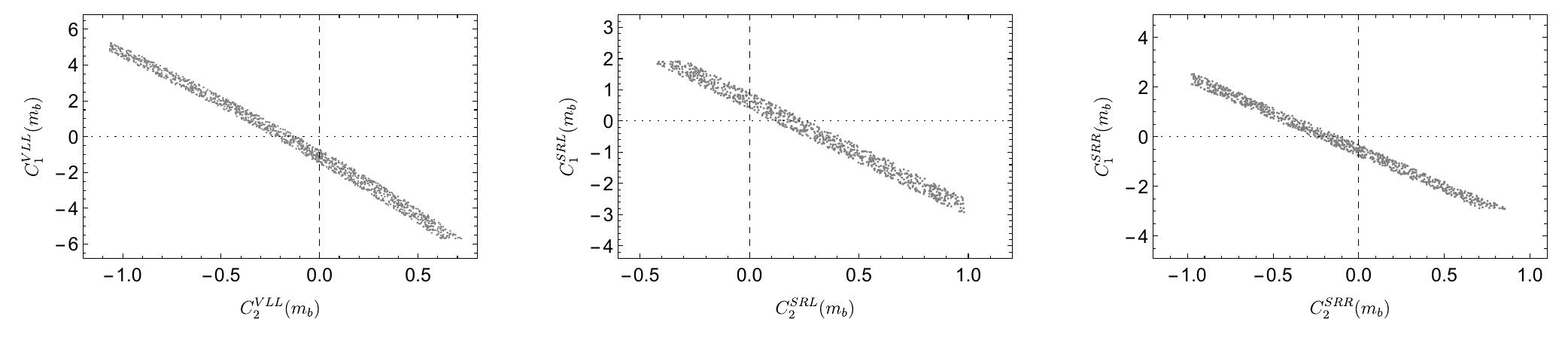}
	\caption{\label{MIVLLVLRSLLSLR} Allowed regions in the $(C_{2}^{VLL}(m_b),C_{1}^{VLL}(m_b))$ (left), $(C_{2}^{SRL}(m_b),C_{1}^{SRL}(m_b))$ (middle), and $(C_{2}^{SRR}(m_b),C_{1}^{SRR}(m_b))$ (right) planes, under the combined constraints from the ratios $R_{(s)L}^{(\ast)}$ varied within $2\sigma$ error bars.}
\end{figure}

In the case where two NP Wilson coefficients are present simultaneously, we show in Fig.~\ref{MIVLLVLRSLLSLR} the allowed regions in the $(C_{2}^{VLL}(m_b),C_{1}^{VLL}(m_b))$, $(C_{2}^{SRL}(m_b),C_{1}^{SRL}(m_b))$, and $(C_{2}^{SRR}(m_b),C_{1}^{SRR}(m_b))$ planes, under the combined constraints from the ten ratios $R_{(s)L}^{(\ast)}$ varied within $2\sigma$ error bars. It is readily to see that, due to the partial cancellation between contributions from the two NP Wilson coefficients, the allowed regions for the NP parameter space become potentially larger than in the case where only one NP Wilson coefficient is present. In the presence of two NP four-quark operators with other Dirac structures, on the other hand, there exist no allowed regions for the corresponding NP Wilson coefficients that can provide a simultaneous explanation of the ratios $R_{(s)L}^{(\ast)}$, even at the $2\sigma$ level.

\subsubsection{High-scale scenario}

\begin{figure}[t]\vspace{-1.2cm}
	\centering	
	\centerline{\hspace{0.5cm}
		\includegraphics[width=0.55\textwidth]{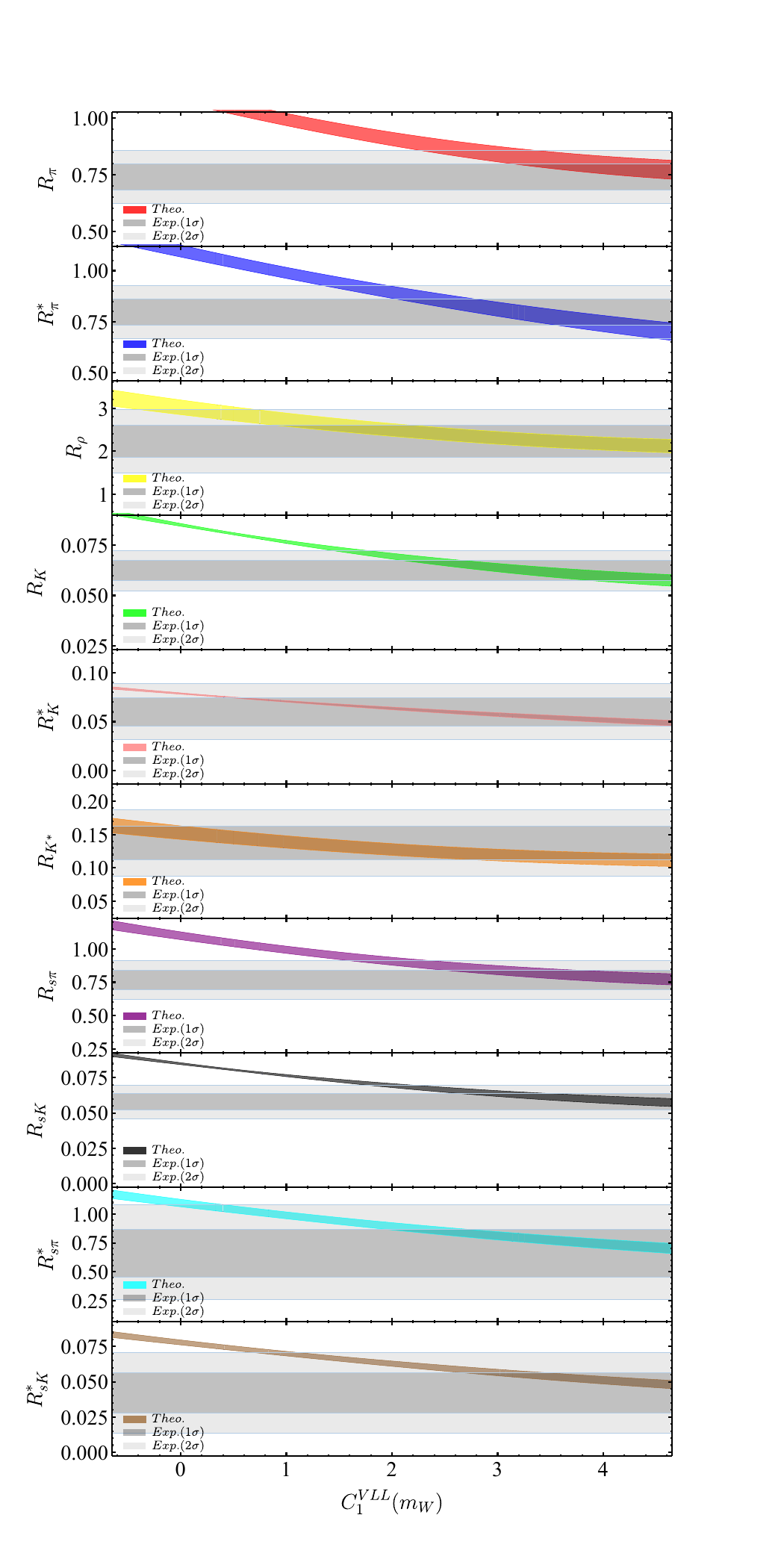}
		\hspace{-1.0cm}
		\includegraphics[width=0.55\textwidth]{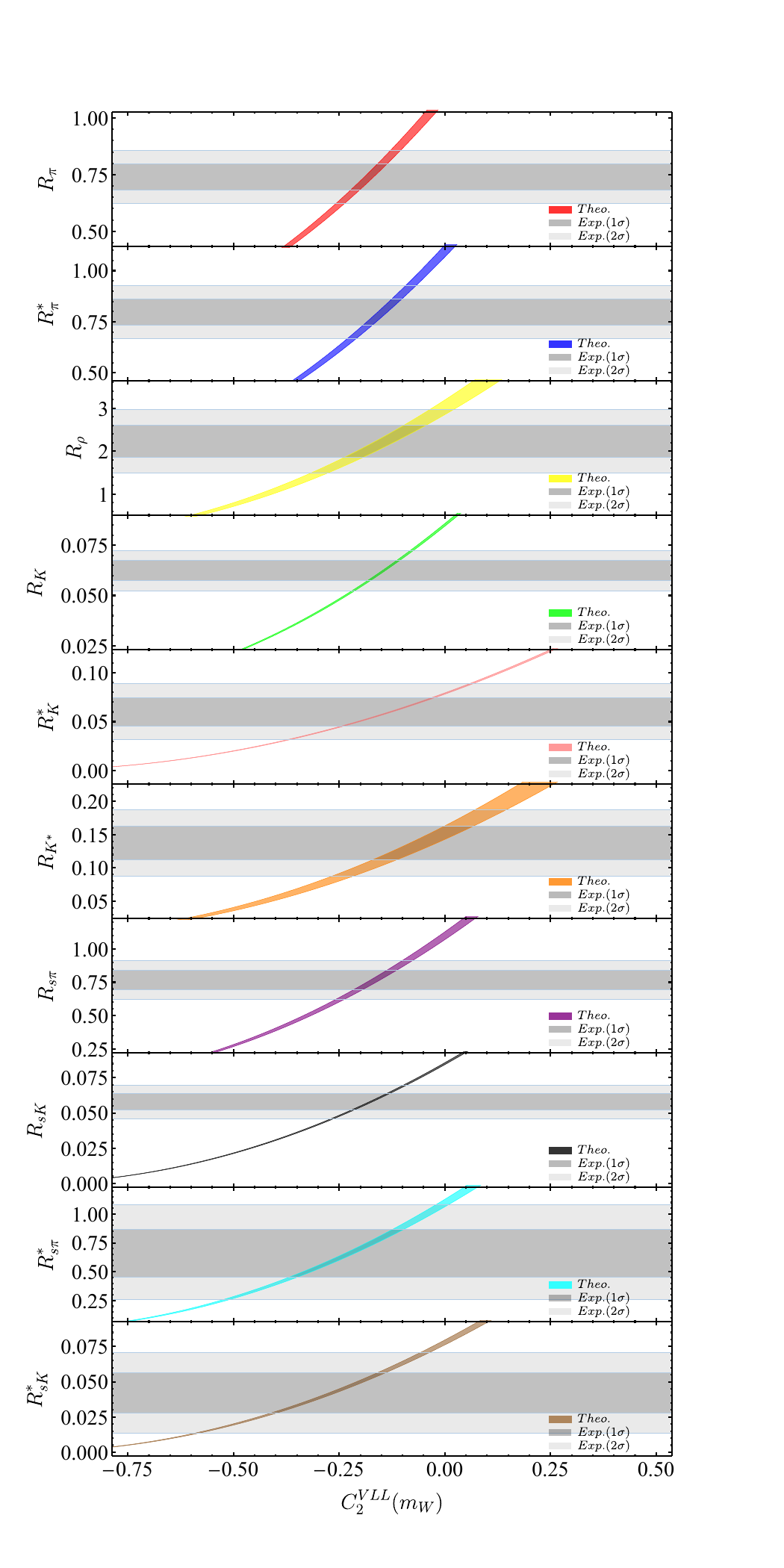}}
	\vspace{-0.5cm}
	\caption{\label{MIVLLatmW} Constraints on the NP Wilson coefficients $C_{1}^{VLL}(m_W)$ (left) and $C_{2}^{VLL}(m_W)$ (right). The other captions are the same as in Fig.~\ref{MIVLL}.}
\end{figure}

\begin{figure}[t]\vspace{-1.2cm}
	\centering	
	\centerline{\hspace{0.5cm}
		\includegraphics[width=0.55\textwidth]{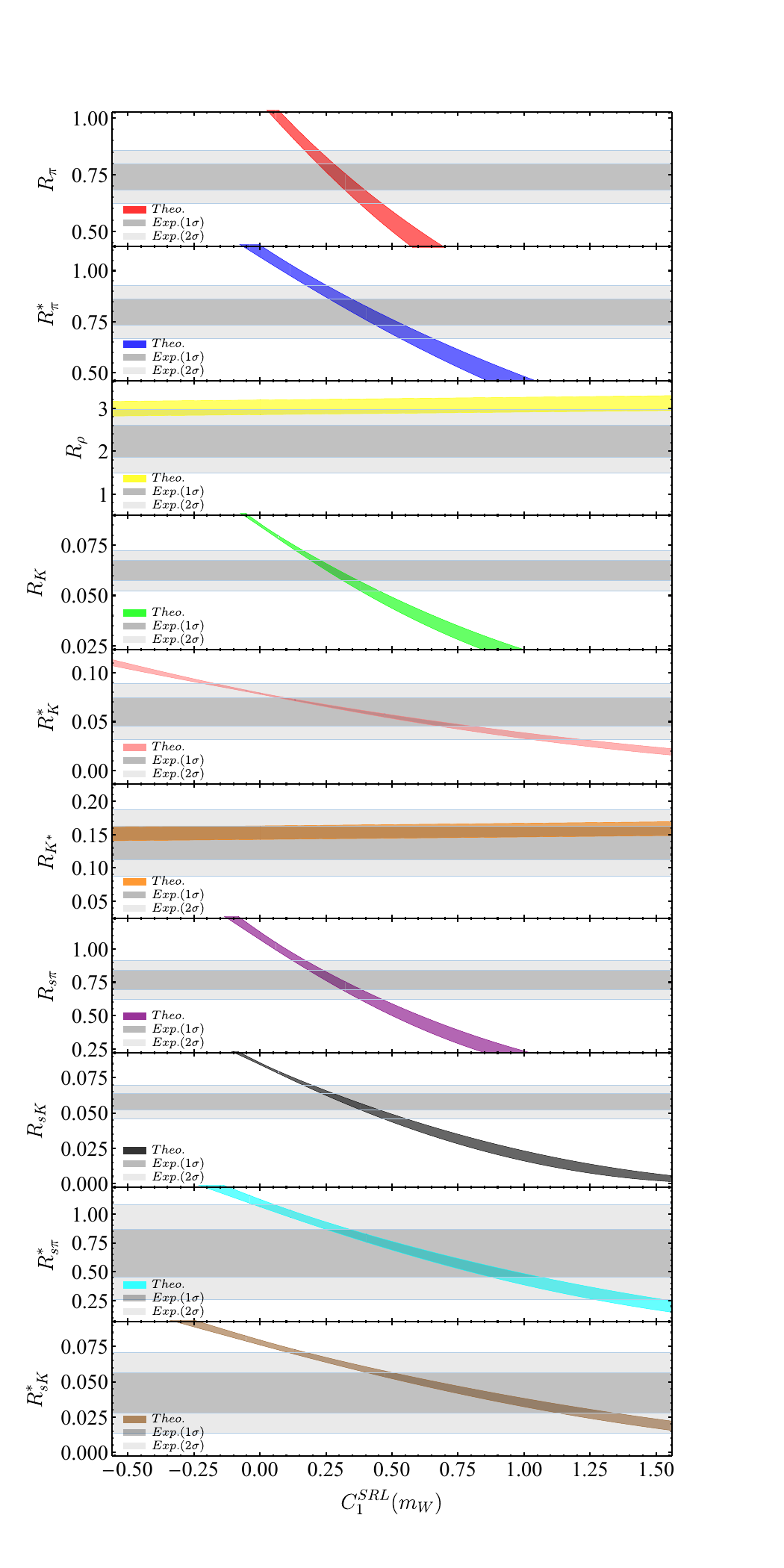}
		\hspace{-1.0cm}
		\includegraphics[width=0.55\textwidth]{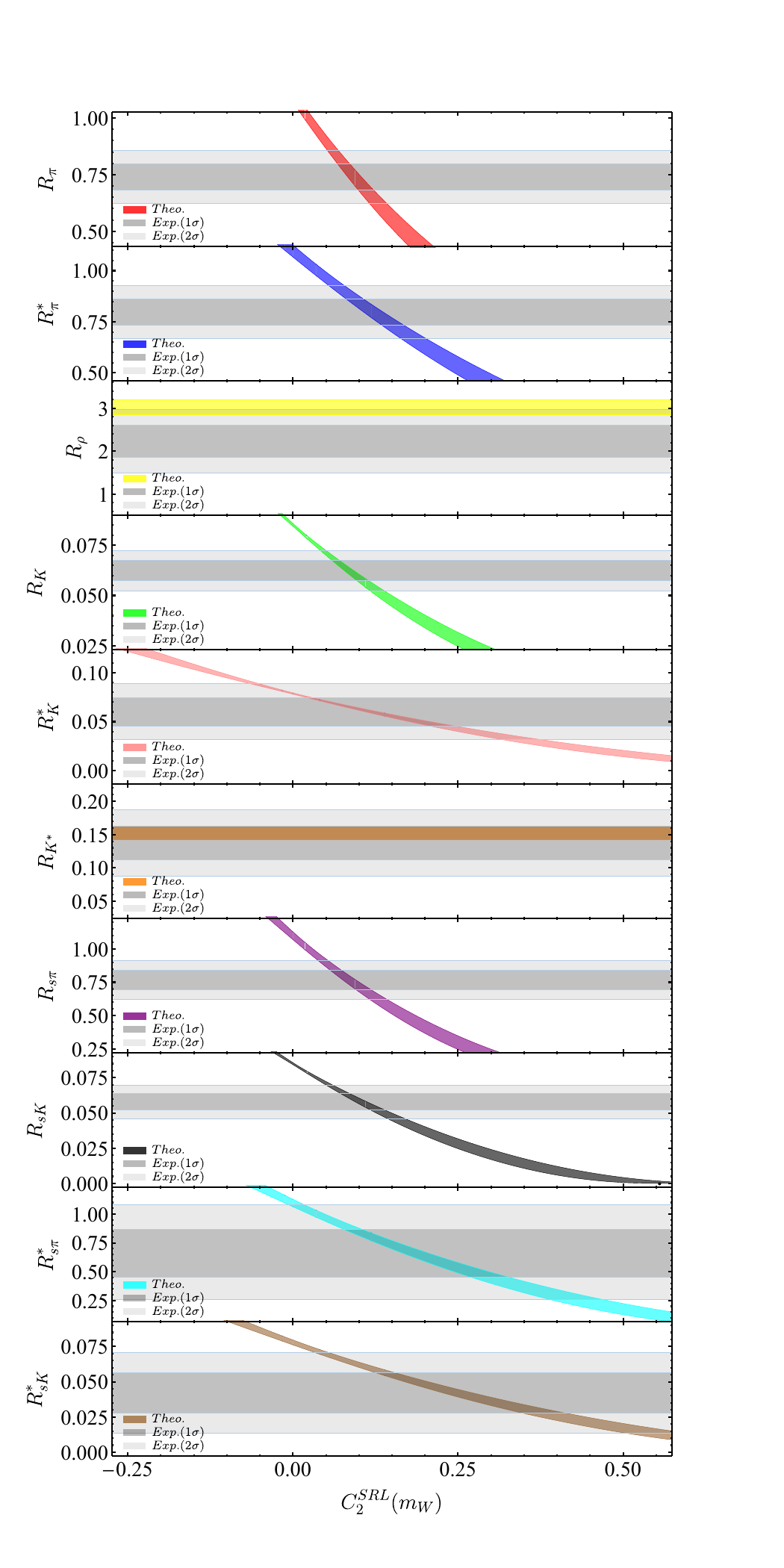}}
	\vspace{-0.5cm}
	\caption{\label{MISRLatmW} Constraints on the NP Wilson coefficients $C_{1}^{SRL}(m_W)$ (left) and $C_{2}^{SRL}(m_W)$ (right). The other captions are the same as in Fig.~\ref{MIVLL}.}
\end{figure}

\begin{figure}[t]\vspace{-1.2cm}
	\centering	
	\centerline{\hspace{0.5cm}
		\includegraphics[width=0.55\textwidth]{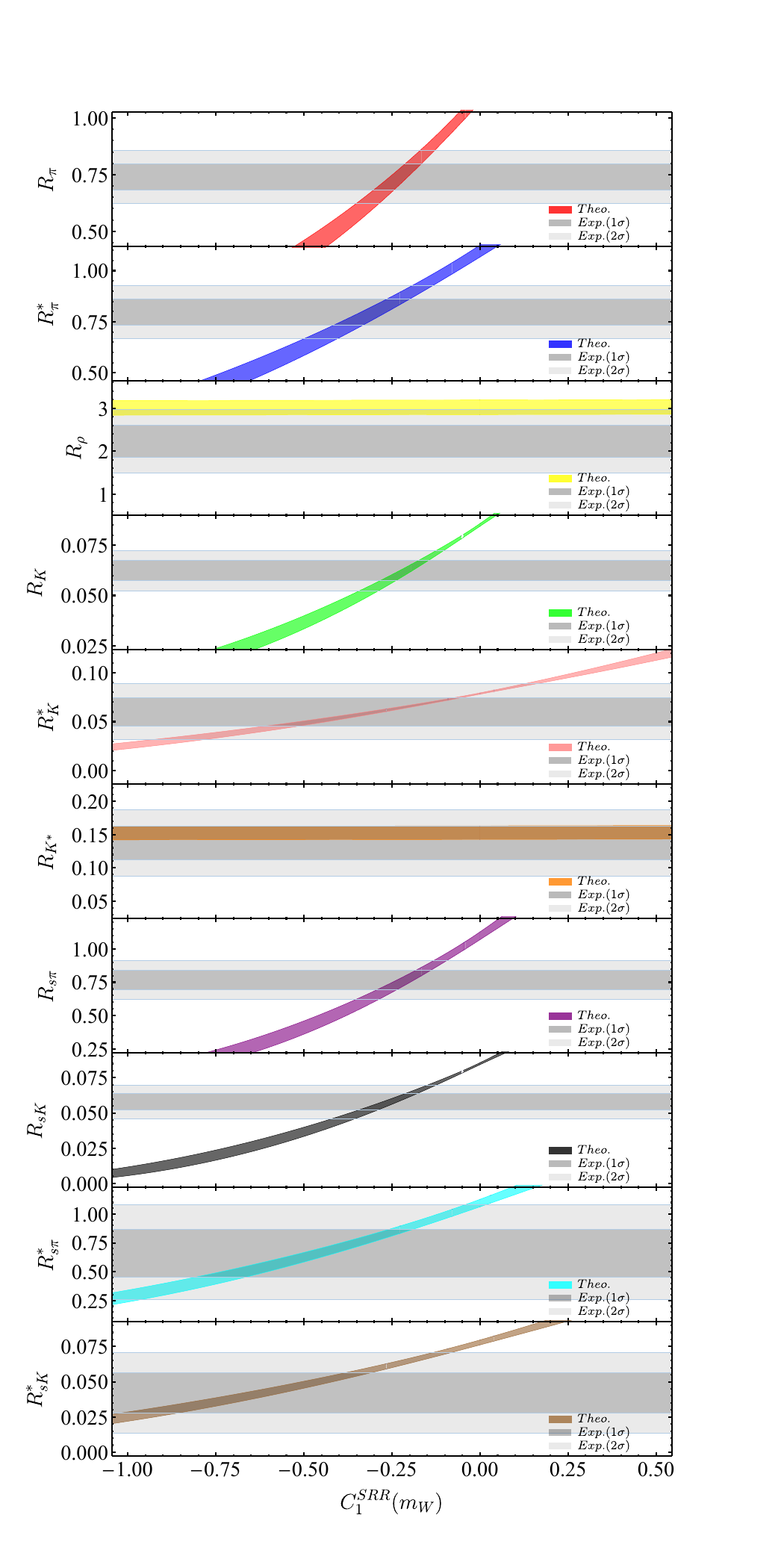}
		\hspace{-1cm}
		\includegraphics[width=0.55\textwidth]{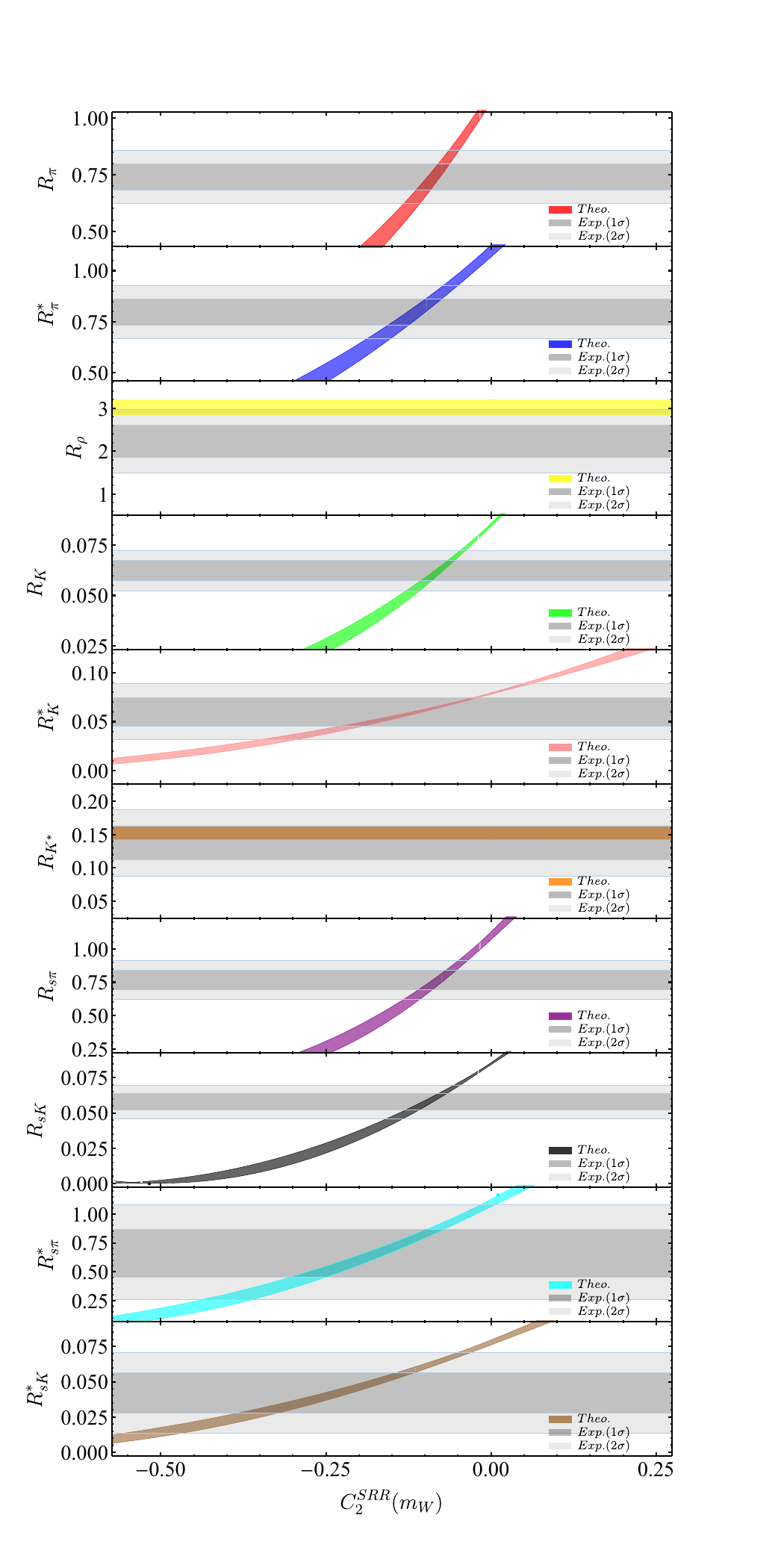}}
	\vspace{-0.5cm}
	\caption{\label{MISRRatmW} Constraints on the NP Wilson coefficients $C_{1}^{SRR}(m_W)$ (left) and $C_{2}^{SRR}(m_W)$ (right). The other captions are the same as in Fig.~\ref{MIVLL}. }
\end{figure}

From the point of view of constructing specific NP models and correlating the low-energy constraints with the direct searches performed at high-energy frontiers, it is also interesting to provide constraints on the NP Wilson coefficients $C_i(\mu_W)$ that are given at the electroweak scale $\mu_W=m_{W}$. To this end, we must take into account the RG evolution of these short-distance Wilson coefficients from $\mu_W$ down to the low-energy scale $\mu_b=m_{b}$, at which the hadronic matrix elements of the NP four-quark operators are evaluated. The most generic formulae for the RG equations satisfied by the NP Wilson coefficients $C_i(\mu)$ can be written as
\begin{align}\label{eq:RGE}
	\mu\frac{dC_{j}(\mu)}{d\mu}=\gamma_{ij}(\mu)C_{i}(\mu)\,,
\end{align}
where $\gamma_{ij}$ are the QCD ADMs of the NP four-quark operators, with their one- and two-loop results given already in refs.~\cite{Ciuchini:1997bw,Ciuchini:1998ix,Buras:2000if}. By solving eq.~\eqref{eq:RGE}, one can then obtain the evolution matrices $\hat{U}(\mu_b,\mu_W)$, which connect the Wilson coefficients at different scales~\cite{Buras:1998raa,Buchalla:1995vs}:
\begin{align}
	\vec{C}(\mu_b)=\hat{U}(\mu_b,\mu_W)\,\vec{C}(\mu_W)\,,
\end{align}
where, once specific to our case with the effective weak Hamiltonian given by eq.~\eqref{eq:Hamiltonian}, $\vec{C}$ is a two-dimensional column vector and $\hat{U}(\mu_b,\mu_W)$ a $2\times2$ matrix for each $VLL$ ($VRR$), $VLR$ ($VRL$), $SLR$ ($SRL$) sector, while $\vec{C}$ is a four-dimensional column vector and $\hat{U}(\mu_b,\mu_W)$ a $4\times4$ matrix in the $SLL$ ($SRR$) sector~\cite{Buras:2000if}.

Here, instead of re-performing a detailed analysis of the NP effects at the electroweak scale, we focus only on the case where only a single NP four-quark operator is present in eq.~\eqref{eq:Hamiltonian}, and investigate how the three solutions obtained in the low-scale scenario change when looked at the electroweak scale. Following the same way as in the low-scale scenario, we show in Figs.~\ref{MIVLLatmW}--\ref{MISRRatmW} the allowed ranges for the NP Wilson coefficients $C_i(m_W)$, under the constraints from the ratios $R_{\pi}$, $R_{\pi}^{*}$, $R_{\rho}$, $R_{K}$, $R_{K}^{*}$, $R_{K^{*}}$, $R_{s\pi}$, $R_{sK}$, $R_{s\pi}^{\ast}$, as well as $R_{sK}^{\ast}$. It is found that, due to the RG evolution, the resulting allowed range for  $C_{1}^{VLL}(m_{W})$ is now given by $C_{1}^{VLL}(m_{W})\in[3.12, 4.86]$, and should be therefore discarded as it is much larger than the SM case with $\mC_{1}(m_{W})=0$ and $\mC_{2}(m_{W})=1$ at the LO in $\alpha_s$~\cite{Buras:1998raa}, while the allowed range for $C_{2}^{VLL}(m_{W})$ remains almost the same as in the low-scale scenario (see eq.~\eqref{eq:C1C2VLLmb}), with
\begin{equation}\label{eq:C1C2VLLmW}
	C_{2}^{VLL}(m_{W})\in[-0.169,-0.138]\,,
\end{equation}
under the combined constraints from the ten ratios $R_{(s)L}^{(\ast)}$ at the $1\sigma$ level. On the other hand, the NP four-quark operators with either $(1+\gam_5)\otimes(1-\gam_5)$ or $(1+\gam_5)\otimes(1+\gam_5)$ structure, could still provide a reasonable explanation of the deviations observed in $\bar{B}_{(s)}^0\to D_{(s)}^{(\ast)+}L^-$ decays at the $2\sigma$ level, with the resulting allowed ranges for the NP Wilson coefficients given, respectively, by
\begin{align}\label{eq:C1C2SRLSRRmW}
	& C_{1}^{SRL}(m_W)\in[0.177,0.448]\,, && C_{2}^{SRL}(m_W)\in[0.054,0.138]\,,\nn \\[0.2cm]
	& C_{1}^{SRR}(m_W)\in[-0.343,-0.160]\,, && C_{2}^{SRR}(m_W)\in[-0.128,-0.051]\,,
\end{align}
which, compared with the results obtained in the low-scale scenario (see eq.~\eqref{eq:C1C2SRLSRRmb}), indicate a large RG evolution effect in these (pseudo-)scalar four-quark operators~\cite{Buras:2000if}.

\subsection{Model-dependent analysis}
\label{subsec:model-dependent}

As found in the last subsection, the deviations observed in $\bar{B}_{(s)}^0\to D_{(s)}^{(\ast)+}L^-$ decays could be well explained by the NP four-quark operators with $\gam^{\mu}(1-\gam_5)\otimes\gam_{\mu} (1-\gam_5)$ structure at the $1\sigma$ level, and also by the operators with $(1+\gam_5)\otimes(1-\gam_5)$ and $(1+\gam_5)\otimes(1+\gam_5)$ structures at the $2\sigma$ level, in a most general model-independent way. In this subsection, as two specific examples of model-dependent considerations, we shall investigate the case where the NP four-quark operators are generated by either a colorless charged gauge boson or a colorless charged scalar, with their masses being in the ballpark of a few TeV. Fitting to the current experimental data on the ratios $R_{(s)L}^{(\ast)}$ collected in Table~\ref{tab:nonlep2semilep}, we can then obtain constraints on the effective coefficients describing the couplings of these mediators to the relevant quarks (see Fig.~\ref{Feynrule}).

\subsubsection{Colorless charged gauge boson}

\begin{figure}[t]\vspace{-1.2cm}
	\centering	
	\includegraphics[width=0.99\textwidth]{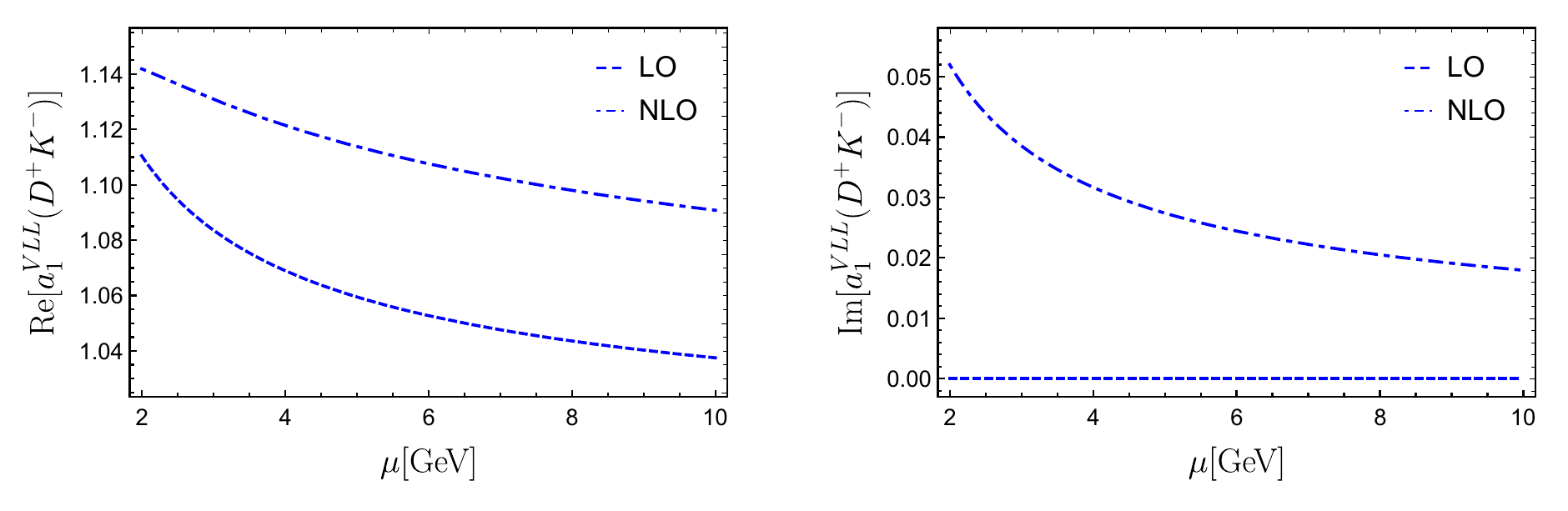}
	\hspace{0.2cm}
	\includegraphics[width=0.99\textwidth]{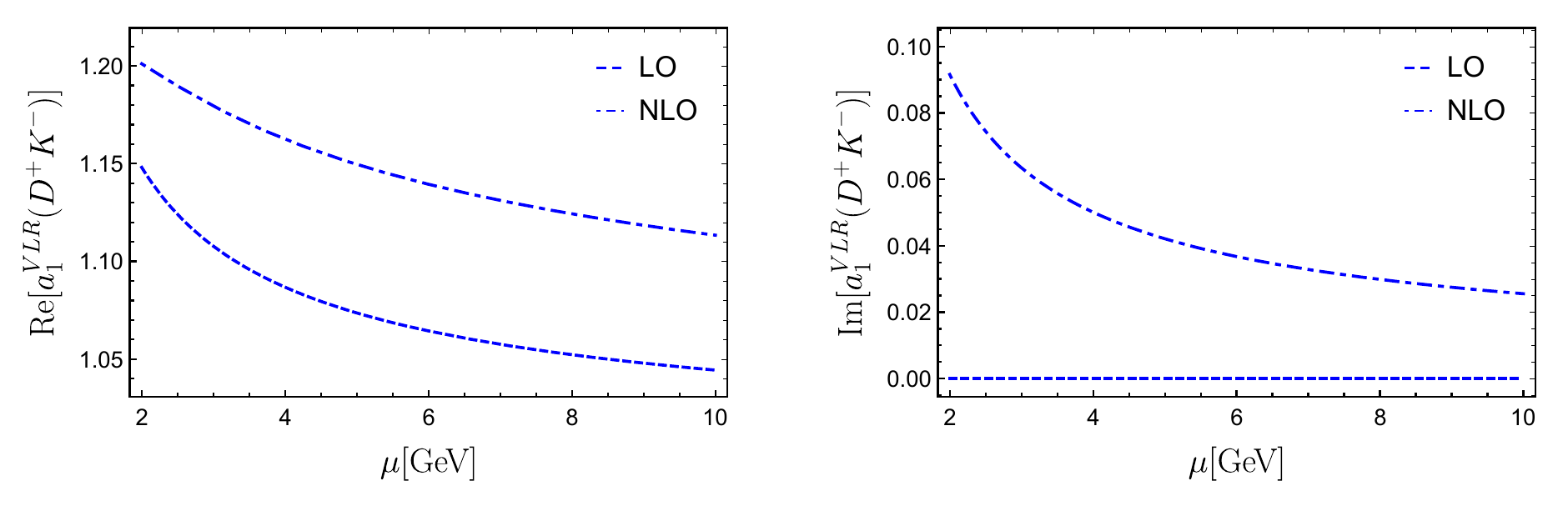}
	\vspace{-0.0cm}
	\caption{\label{scaledependent_VLL&VLR} Dependence of the efficient coefficients $a_{1}^{VLL}(D^{+}K^{-})$ and $a_{1}^{VLR}(D^{+}K^{-})$ on the renormalization scale $\mu$, both at the LO (dashed lines) and at the NLO order (dash-dotted lines).}
\end{figure}

Starting with the Feynman rules given in Fig.~\ref{Feynrule} and after integrating out the heavy colorless charged gauge boson $A^+$, we can obtain the effective weak Hamiltonian describing the quark-level $b\to c\bar{u}d(s)$ transitions mediated by $A^+$~\cite{Buras:2012gm}: 
\begin{align}\label{eq:HamiltonianA}
	\mathcal{H}_\text{eff}^{\rm gauge} &=
	\frac{G_F}{\sqrt{2}}\,V_{cb}V^{\ast}_{uq}\,\bigg\lbrace\lambda_{LL}(A)\,\Big[ C_1^{VLL}(\mu)Q_1^{VLL}(\mu) + C_2^{VLL}(\mu)Q_2^{VLL}(\mu)\Big]\,\nn \\[0.1cm] 
	& \hspace{0.5cm} +\lambda_{LR}(A)\,\Big[C_1^{VLR}(\mu)Q_1^{VLR}(\mu) +C_2^{VLR}(\mu)Q_2^{VLR}(\mu) \Big] +\left(L\leftrightarrow R\right) \bigg\rbrace +\text{h.c.}\,,
\end{align}
with 
\begin{equation}
	\lambda_{LL}(A)=\frac{m_{W}^2}{m_{A}^2}\,\Delta_{cb}^{L}(A)\,\left(\Delta_{uq}^{L}(A)\right)^\ast, \qquad
	\lambda_{LR}(A)=\frac{m_{W}^2}{m_{A}^2}\,\Delta_{cb}^{L}(A)\,\left(\Delta_{uq}^{R}(A)\right)^\ast\,,
\end{equation}
where $m_{A}$ is the mass of the colorless charged gauge boson $A^+$, and $\Delta_{i,j}^{L,R}(A)$ represent the reduced couplings of $A^+$ to an up- and a down-type quark. The short-distance Wilson coefficients $C_i(\mu_b)$ at the low-energy scale $\mu_b=m_b$ can be obtained through a two-step RG evolution~\cite{Ciuchini:1998ix,Buras:2001ra}
\begin{align}\label{eq:twosteprge}
	\vec{C}(\mu_b)=\hat{U}(\mu_b,\mu_W)\,\hat{U}(\mu_W,\mu_0)\,\vec{C}(\mu_0)\,,
\end{align}
where the evolution matrices $\hat{U}(\mu_b,\mu_W)$ and $\hat{U}(\mu_W,\mu_0)$ are evaluated in an effective theory with $f=5$ and $f=6$ quark flavors, respectively. Analytic expressions for these evolution matrices can be found in ref.~\cite{Buras:2001ra}. The matching conditions for the short-distance Wilson coefficients $C_i(\mu_0)$, including the $\mathcal{O}(\alpha_s)$ corrections, at the initial scale $\mu_0=m_{A}$ have been calculated in ref.~\cite{Buras:2012gm}. Together with the one-loop vertex corrections to the hard kernels $T_{ij}(u)$ calculated in subsection~\ref{subsec:HSK}, this enables us to perform a full NLO analysis of the NP effects in the class-I non-leptonic $\bar{B}_{(s)}^0\to D_{(s)}^{(*)+} L^-$ decays. Especially, such a full NLO analysis is helpful for reducing the dependence of the effective coefficients $a_{1}^{VLL(R)}(D_{(s)}^{(*)+} L^-)$ and $a_{1}^{VRR(L)}(D_{(s)}^{(*)+} L^-)$ on the renormalization scale $\mu$. This is illustrated in Fig.~\ref{scaledependent_VLL&VLR} for the effective coefficients $a_{1}^{VLL}(D^{+}K^{-})$ (normalized by $\lambda_{LL}(A)$) and $a_{1}^{VLR}(D^{+}K^{-})$ (normalized by $\lambda_{LR}(A)$), with 
\begin{align}\label{eq:a1_VLL_VLR}
a_{1}^{VLL(R)}(D^{+}K^{-})=C_2^{VLL(R)}(\mu)+\frac{C_1^{VLL(R)}(\mu)}{N_c}+ C_1^{VLL(R)}(\mu)\int^1_0 du\,T^{VLL(R)}(u,z)\Phi_{K}(u)\,,
\end{align}
where the one-loop hard kernels $T^{VLL}(u,z)$ and $T^{VLR}(u,z)$ are given, respectively, by eqs.~\eqref{t8VLL} and \eqref{t8VLR}. It can be seen that the scale dependence is reduced for the real part, but not for the imaginary part, since the latter vanishes at the LO in $\alpha_s$.

\begin{figure}[t]\vspace{-1.2cm}
	\centering	
	\centerline{\hspace{0.5cm}
		\includegraphics[width=0.55\textwidth]{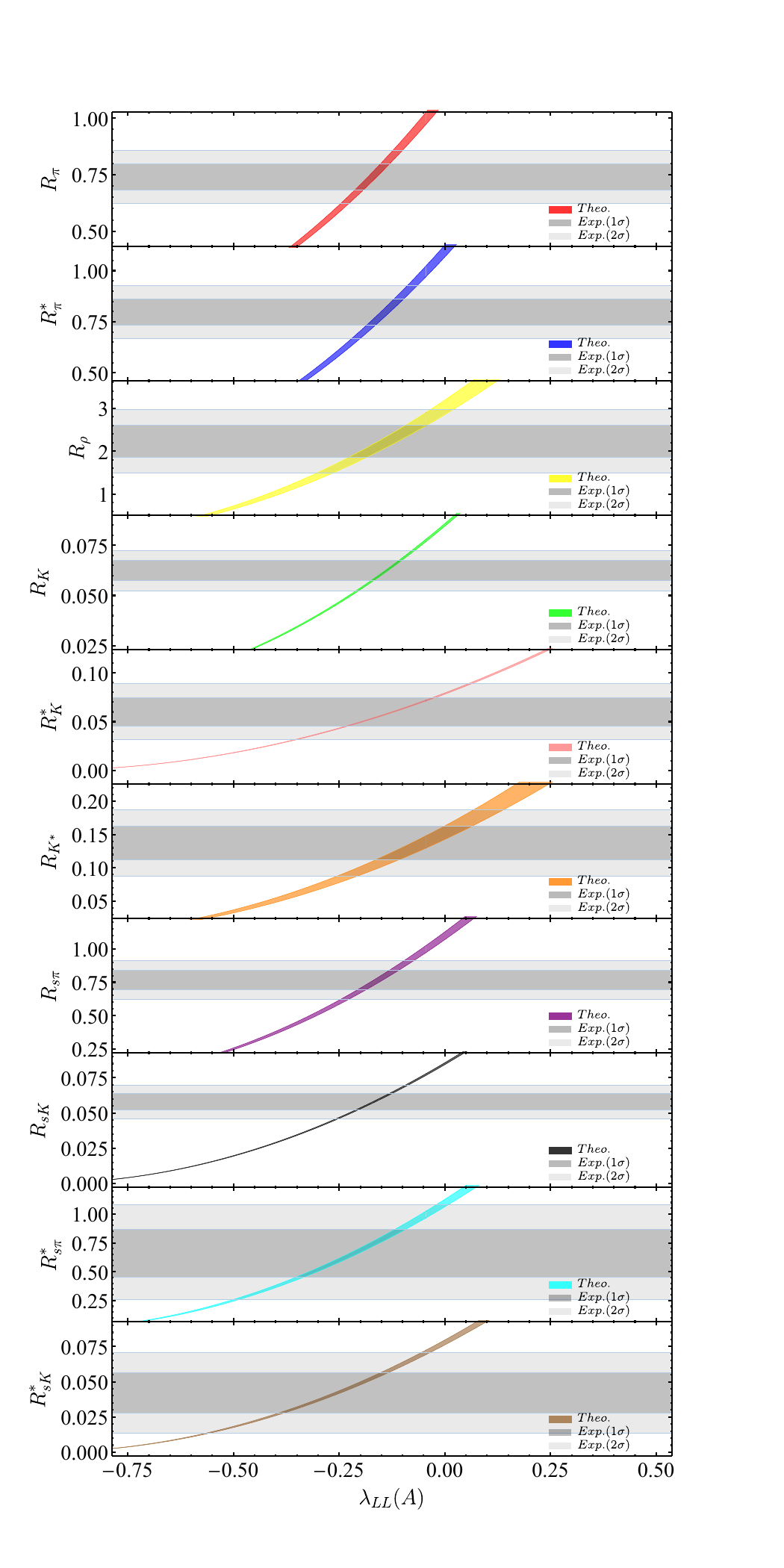}
		\hspace{-1.0cm}
		\includegraphics[width=0.55\textwidth]{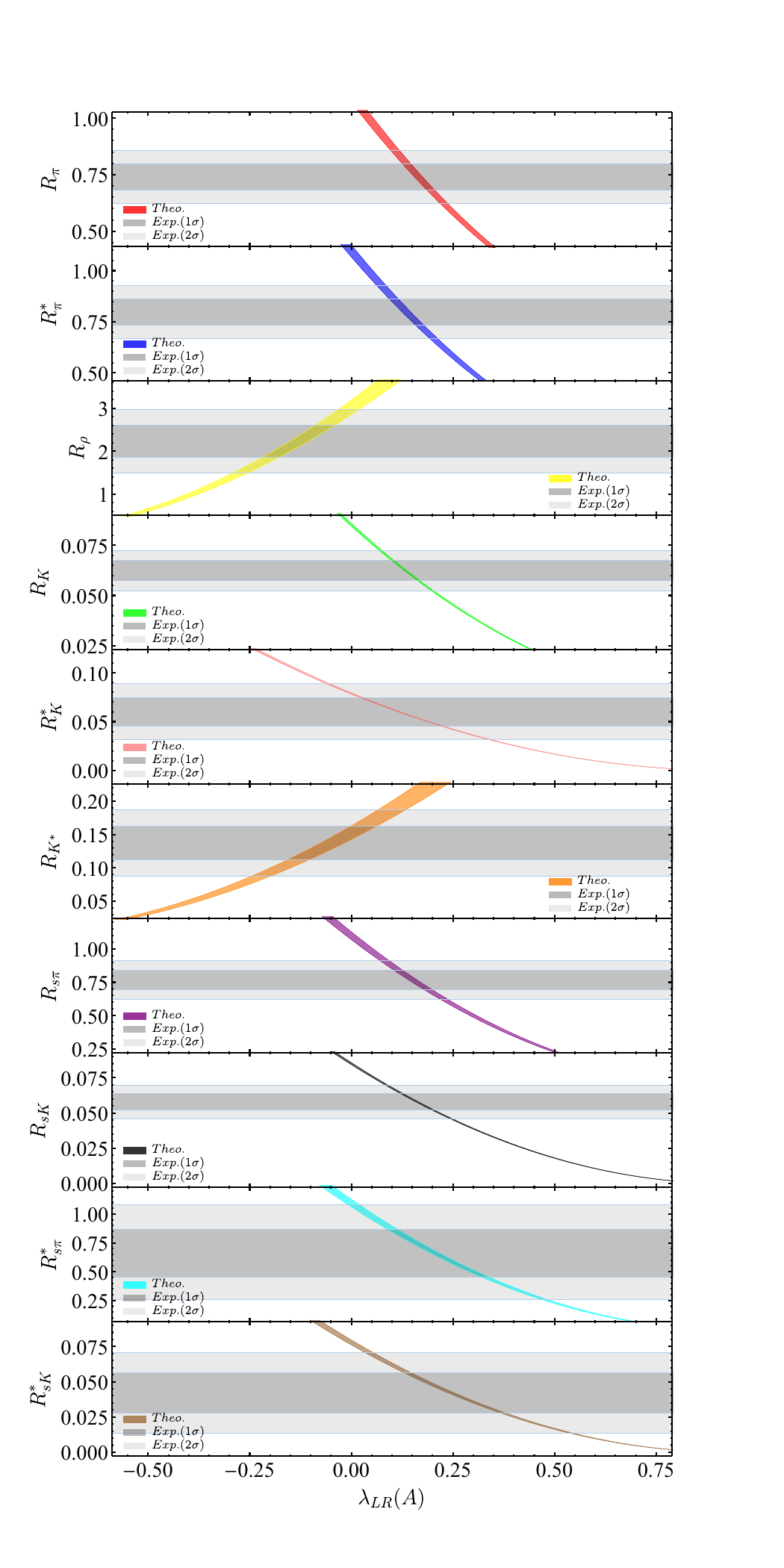}}
	\vspace{-0.5cm}
	\caption{\label{VLL&VLR} Constraints on the effective coefficients $\lambda_{LL}(A)$ (left) and $\lambda_{LR}(A)$ (right) from the ratios $R_{(s)L}^{(\ast)}$ collected in Table~\ref{tab:nonlep2semilep}. The other captions are the same as in Fig.~\ref{MIVLL}.}
\end{figure}

\begin{figure}[t]\vspace{-1.2cm}
	\centering	
	\centerline{\hspace{0.5cm}
		\includegraphics[width=0.55\textwidth]{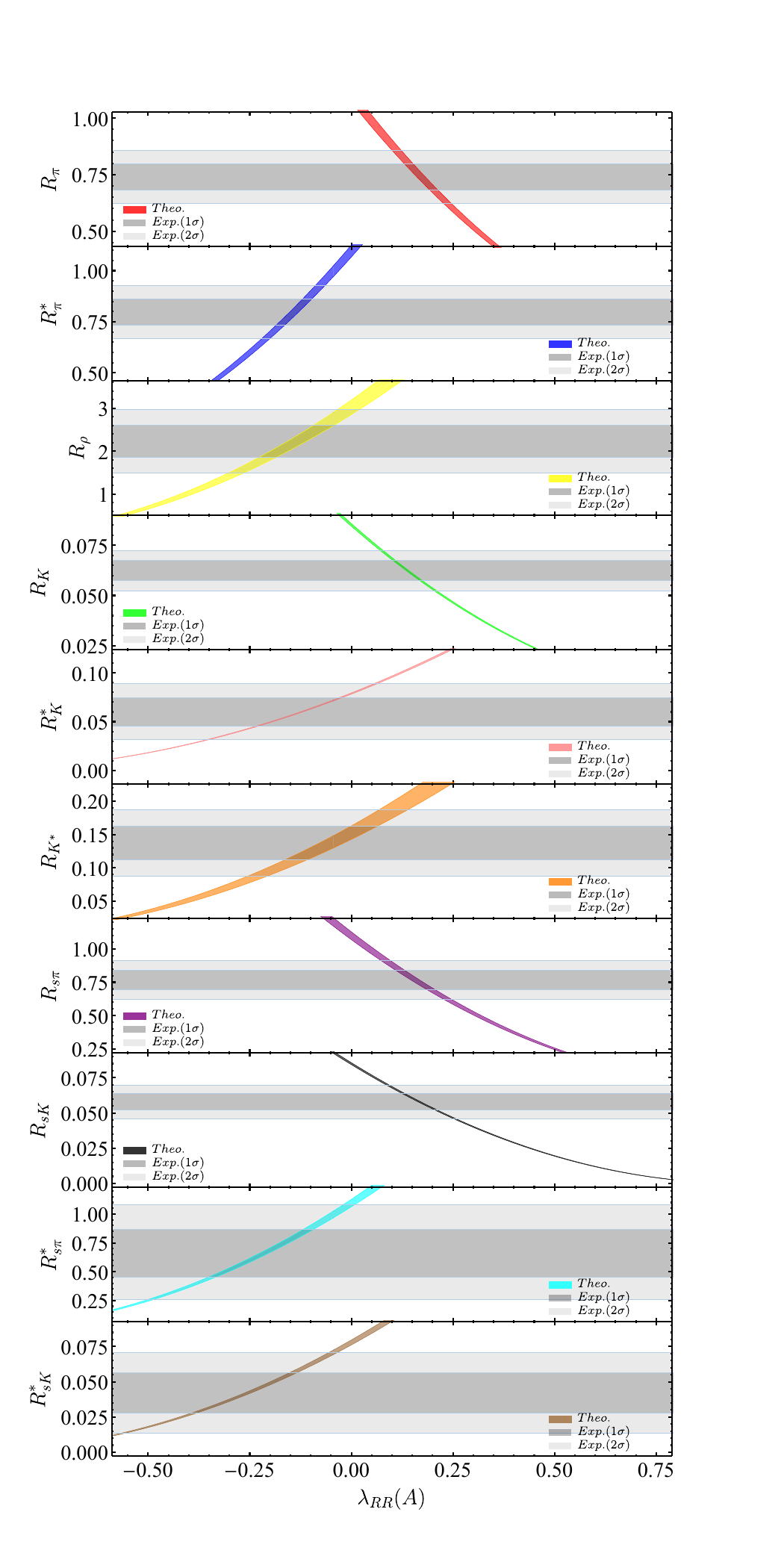}
		\hspace{-1.0cm}
		\includegraphics[width=0.55\textwidth]{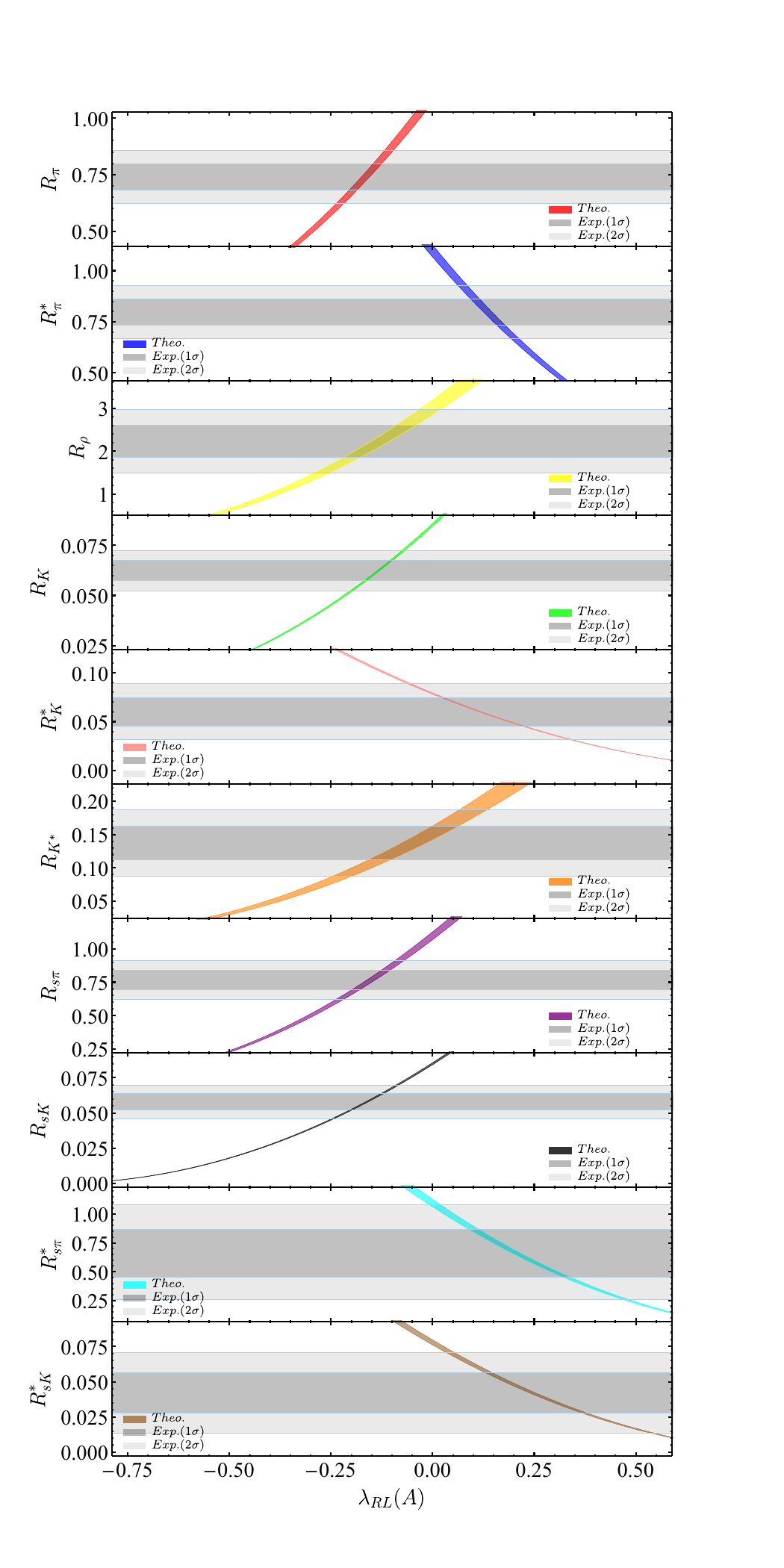}}
	\vspace{-0.5cm}
	\caption{\label{VRR&VRL} Constraints on the effective coefficients $\lambda_{RR}(A)$ (left) and $\lambda_{RL}(A)$ (right) from the ratios $R_{(s)L}^{(\ast)}$ collected in Table~\ref{tab:nonlep2semilep}. The other captions are the same as in Fig.~\ref{MIVLL}.}
\end{figure}

\begin{figure}[t]\vspace{-1.2cm}
	\centering	
	\centerline{\hspace{0.5cm}
		\includegraphics[width=0.55\textwidth]{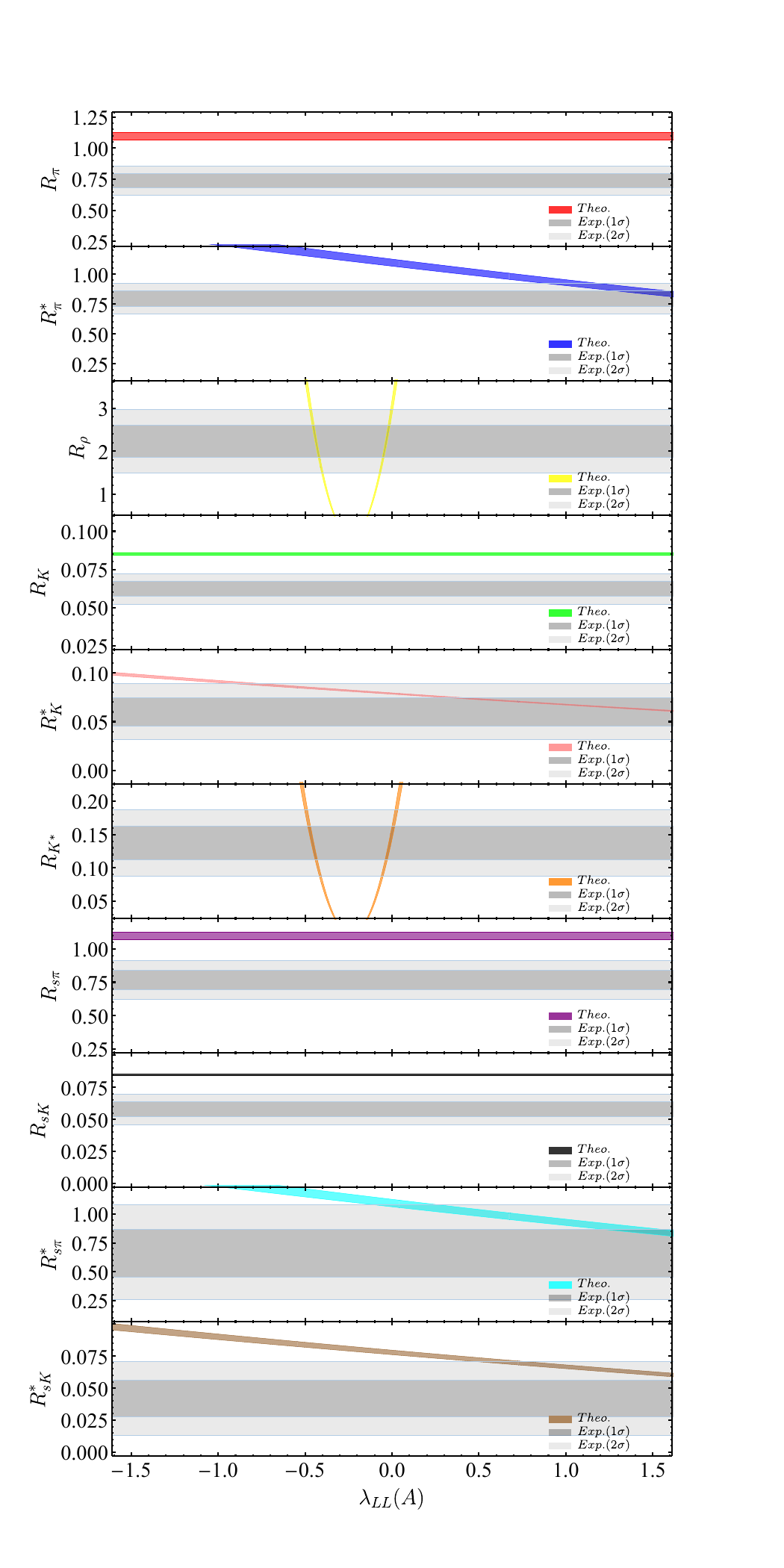}
		\hspace{-1.0cm}
		\includegraphics[width=0.55\textwidth]{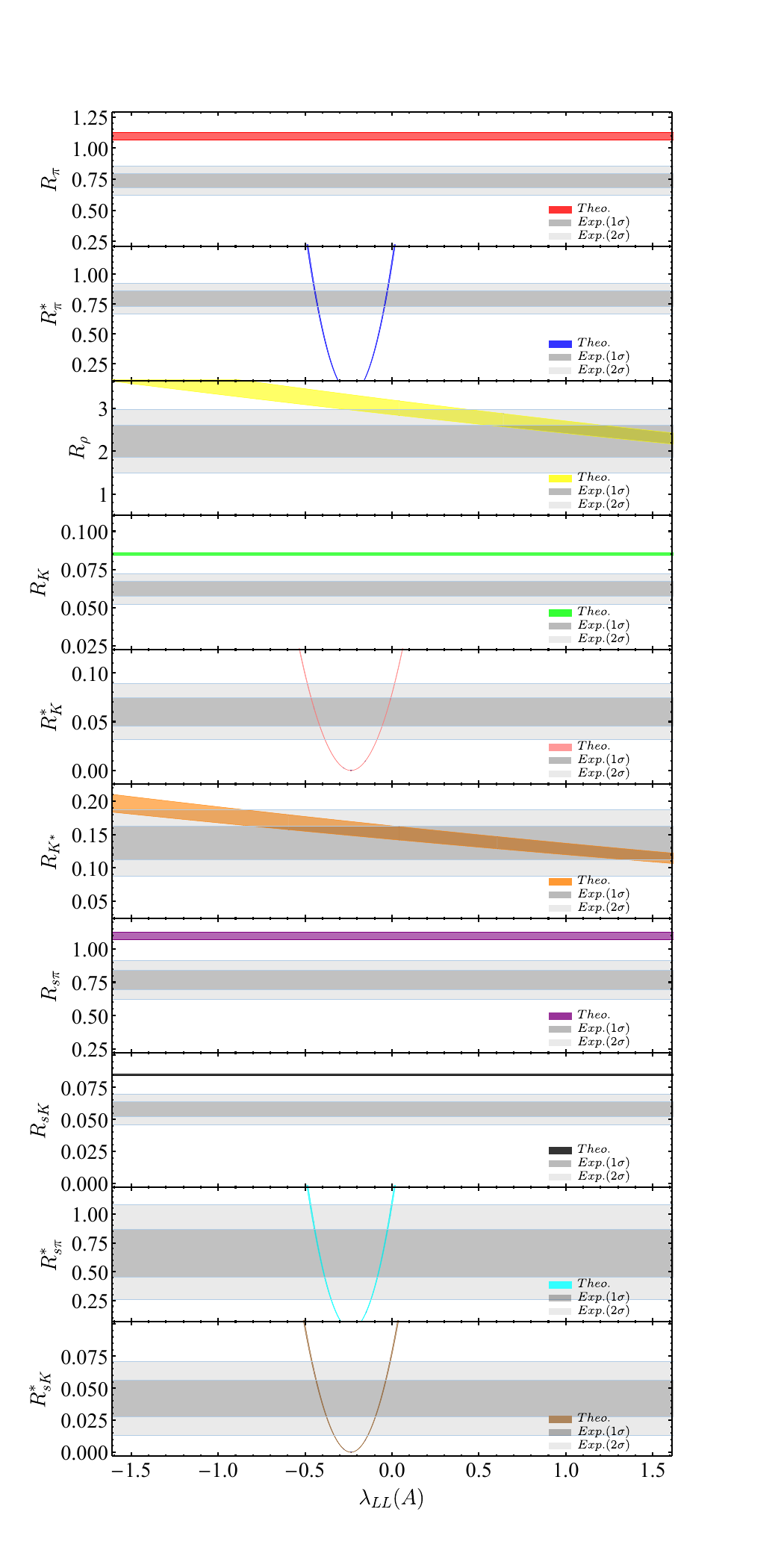}}
	\vspace{-0.5cm}
	\caption{\label{GaugeSym.&Asym.} Constraints on the effective coefficient $\lambda_{LL}(A)$ from the ratios $R_{(s)L}^{(\ast)}$ collected in Table~\ref{tab:nonlep2semilep}, in the scenarios where the left- and right-handed reduced couplings are symmetric (scenario~II, left) and asymmetric (scenario~III, right), defined respectively by eqs.~\eqref{eq:gaugesymmetric} and \eqref{eq:gaugeasymmetric}. The other captions are the same as in Fig.~\ref{MIVLL}.}
\end{figure}

Specific to the case where the NP four-quark operators are mediated by a heavy colorless charged gauge boson $A^+$, with its mass $m_{A}$ fixed at $1~\mathrm{TeV}$, we have generally four nonzero effective couplings, $\lambda_{LL}(A)$, $\lambda_{LR}(A)$, $\lambda_{RR}(A)$, and $\lambda_{RL}(A)$, which might be independent of each other. In order to simplify our analysis and reduce the number of free NP parameters, we shall consider the following three different scenarios:
\begin{enumerate}
	\item[$\bullet$] In scenario~I, we consider the case where only one effective coefficient is nonzero in eq.~\eqref{eq:HamiltonianA}. Under the individual and combined constraints from the ratios $R_{\pi}$, $R_{\pi}^{*}$, $R_{\rho}$, $R_{K}$, $R_{K}^{*}$, $R_{K^{*}}$, $R_{s\pi}$, $R_{sK}$, $R_{s\pi}^{\ast}$, as well as $R_{sK}^{\ast}$ collected in Table~\ref{tab:nonlep2semilep}, we can obtain the allowed ranges for this nonzero effective coefficient, which are shown in Figs.~\ref{VLL&VLR} and \ref{VRR&VRL}. It can be seen that in this scenario only the case with a nonzero $\lambda_{LL}(A)$ could provide a simultaneous account for the deviations observed in $\bar{B}_{(s)}^0\to D_{(s)}^{(\ast)+}L^-$ decays, with the resulting allowed range given by 
	\begin{equation}
		\lambda_{LL}(A)\in[-0.162,-0.132]
	\end{equation}
	at the $1\sigma$ level. Such a conclusion is also consistent with the recent observation made in ref.~\cite{Iguro:2020ndk}, which claims that part of the deviations can be reduced by a left-handed $W^{\prime}$ model through a $-10\%$ shift in the $b\to c\bar{u}d(s)$ decay amplitudes. All the other three cases are, however, ruled out already by the combined constraints from the ratios $R_{(s)L}^{(\ast)}$, even at the $2\sigma$ level. 
	
	\item[$\bullet$] In scenario~II, we consider the case where all the four effective coefficients are nonzero, but with the additional left-right symmetric assumption on the reduced couplings~\cite{Buras:2014sba}:
	\begin{equation}\label{eq:gaugesymmetric}
		\Delta_{cb}^{L}(A)=\Delta_{cb}^{R}(A)\,,\qquad \Delta_{uq}^{L}(A)=\Delta_{uq}^{R}(A)\,,
	\end{equation}
	which implies that the four effective couplings are all equal to each other,  $\lambda_{LL}(A)=\lambda_{LR}(A)=\lambda_{RR}(A)=\lambda_{RL}(A)$. The resulting constraints on the effective coefficient $\lambda_{LL}(A)$ in this case are shown in the left panel of Fig.~\ref{GaugeSym.&Asym.}. One can see clearly that such a scenario fails to provide a simultaneous explanation of the deviations observed in $\bar{B}_{(s)}^0\to D_{(s)}^{(\ast)+}L^-$ decays, even at the $2\sigma$ level. It is also observed that, due to
	$\lambda_{LL}(A)=\lambda_{LR}(A)$ and $\lambda_{RR}(A)=\lambda_{RL}(A)$, the hadronic matrix elements of $Q_{1,2}^{VLL}$ and $Q_{1,2}^{VRR}$ are exactly canceled, respectively, by that of $Q_{1,2}^{VLR}$ and $Q_{1,2}^{VRL}$ for the $\bar{B}_{(s)}^0\to D_{(s)}^{+}\pi^-$ and $\bar{B}_{(s)}^0\to D_{(s)}^{+}K^{-}$ decays. This explains why the ratios $R_{(s)\pi}$ and $R_{(s)K}$ are insensitive to the NP contributions, as shown in the left panel of Fig.~\ref{GaugeSym.&Asym.}.
	
	\item[$\bullet$] In scenario~III, we consider instead the case where the left- and right-handed reduced couplings are asymmetric~\cite{Buras:2014sba}:
	\begin{equation}\label{eq:gaugeasymmetric}
		\Delta_{cb}^{L}(A)=-\Delta_{cb}^{R}(A)\,,\qquad \Delta_{uq}^{L}(A)=-\Delta_{uq}^{R}(A)\,,
	\end{equation}
	which implies that, while all the four effective couplings are still nonzero, they satisfy now the relation $\lambda_{LL}(A)=\lambda_{RR}(A)=-\lambda_{LR}(A)=-\lambda_{RL}(A)$. As shown in the right panel of Fig.~\ref{GaugeSym.&Asym.}, such a scenario also fails to provide a simultaneous account for the ten ratios $R_{(s)L}^{(\ast)}$ collected in Table~\ref{tab:nonlep2semilep}, even at the $2\sigma$ level. Note that in this case the ratios $R_{(s)\pi}$ and $R_{(s)K}$ receive no contributions from the NP four-quark operators, which is now due to $\lambda_{LL}(A)=\lambda_{RR}(A)$ and $\lambda_{LR}(A)=\lambda_{RL}(A)$, resulting in therefore an exact cancellation between the hadronic matrix elements of $Q_{1,2}^{VLL(R)}$ and $Q_{1,2}^{VRR(L)}$ for the decay modes involved.
\end{enumerate} 

\subsubsection{Colorless charged scalar}

\begin{figure}[t]\vspace{-1.2cm}
	\centering	
		\includegraphics[width=0.99\textwidth]{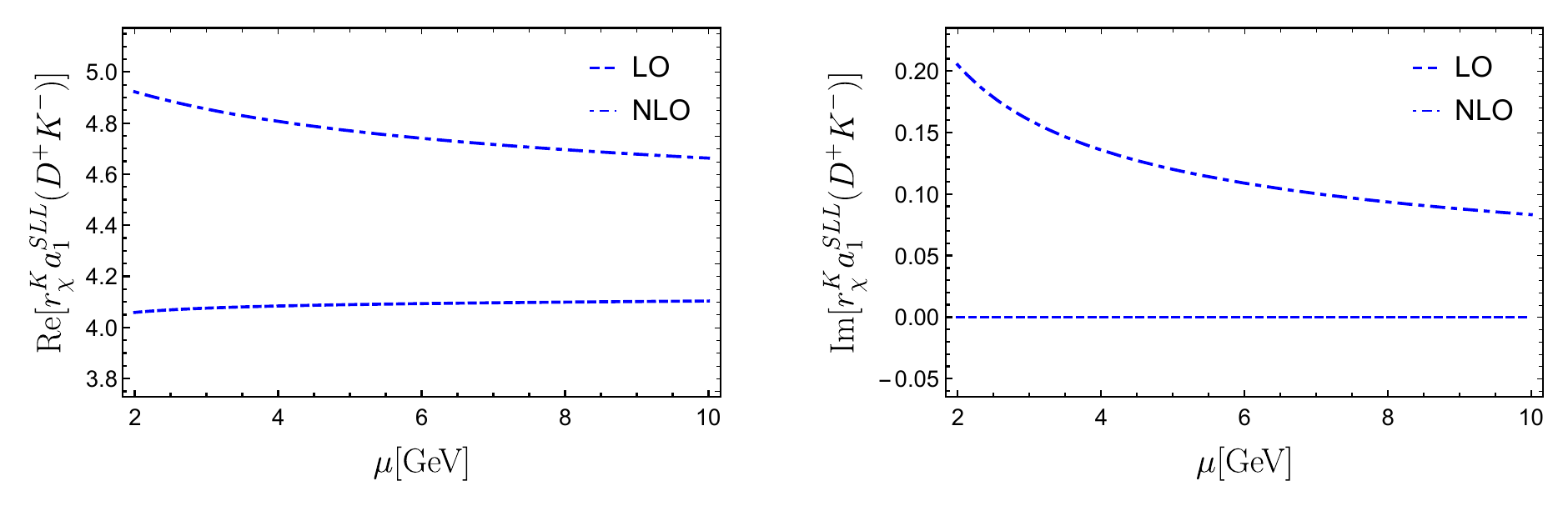}
	\vspace{-0.5cm}
	    \includegraphics[width=0.99\textwidth]{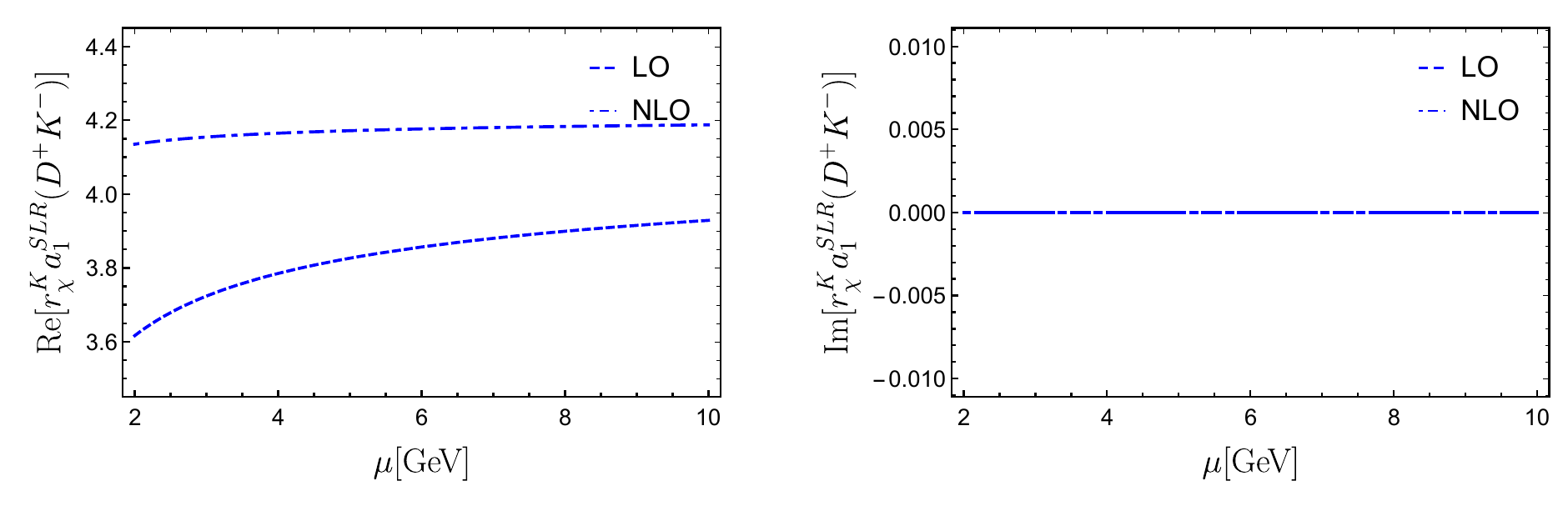}
	\vspace{0.5cm}
	\caption{\label{scaledependent_SLL&SLR} Dependence of the effective coefficients $r_{\chi}^{K}a_{1}^{SLL}(D^{+}K^{-})$ and $r_{\chi}^{K}a_{1}^{SLR}(D^{+}K^{-})$ on the renormalization scale $\mu$. The other captions are the same as in Fig.~\ref{scaledependent_VLL&VLR}.}
\end{figure}

Let us now proceed to discuss the case where the NP four-quark operators are generated by a heavy colorless charged scalar $H^+$, with its mass $m_{H}$ fixed also at $1~\mathrm{TeV}$. The resulting effective weak Hamiltonian for the quark-level $b\to c\bar{u}d(s)$ transitions mediated by such a charged scalar is now given by~\cite{Buras:2012gm}
\begin{align}\label{eq:HamiltonianH}
	\mathcal{H}_\text{eff}^{\rm scalar} &= -\frac{G_F}{\sqrt{2}}\,V_{cb}V^{\ast}_{uq}\,\bigg\lbrace\lambda_{LL}(H)\,\Big[ C_1^{SLL}(\mu)Q_1^{SLL}(\mu) + C_2^{SLL}(\mu)Q_2^{SLL}(\mu) \,\nn \\[0.1cm] 
	& \hspace{4.0cm} +  C_3^{SLL}(\mu)Q_3^{SLL}(\mu) + C_4^{SLL}(\mu)Q_4^{SLL}(\mu)\Big]\,\nn \\[0.1cm] 
	& \hspace{0.5cm} +\lambda_{LR}(H)\,\Big[C_1^{SLR}(\mu)Q_1^{SLR}(\mu) +C_2^{SLR}(\mu)Q_2^{SLR}(\mu) \Big] + \left(L\leftrightarrow R\right) \bigg\rbrace +\text{h.c.}\,,
\end{align}
where 
\begin{equation}
	\lambda_{LL}(H)=\frac{m_{W}^2}{m_{H}^2}\,\Delta_{cb}^{L}(H)\,\left(\Delta_{uq}^{L}(H)\right)^\ast\,, \qquad \lambda_{LR}(H)=\frac{m_{W}^2}{m_{H}^2}\,\Delta_{cb}^{L}(H)\,\left(\Delta_{uq}^{R}(H)\right)^\ast\,,
\end{equation}
and $\Delta_{i,j}^{L,R}(H)$ represent the reduced couplings of $H^+$ to an up- and a down-type quark, as defined in Fig.~\ref{Feynrule}. It should be noted that, at the matching scale $\mu_0=m_{H}$, only the Wilson coefficients $C_2^{SLL}(\mu_0)$, $C_2^{SLR}(\mu_0)$, $C_2^{SRR}(\mu_0)$, and $C_2^{SRL}(\mu_0)$ are nonzero at the LO, while all the remaining ones appear firstly at the NLO in $\alpha_s$, with their explicit expressions given already in ref.~\cite{Buras:2012gm}. To get their values at the low-energy scale $\mu_b=m_b$, we should also perform a two-step RG evolution as in eq.~\eqref{eq:twosteprge}, where the analytic formulae for the evolution matrices $\hat{U}(\mu_b,\mu_W)$ and $\hat{U}(\mu_W,\mu_0)$ can be found in ref.~\cite{Buras:2001ra}. This, together with the $\mathcal{O}(\alpha_s)$ vertex corrections to the hard kernels $T_{ij}(u)$ presented in subsection~\ref{subsec:HSK}, makes it possible to investigate the NP effects on the class-I non-leptonic $\bar{B}_{(s)}^0\to D_{(s)}^{(*)+} L^-$ decays, in a RG-improved way completely at the NLO in $\alpha_s$. The reduction of the scale-dependence of the effective coefficients $r_{\chi}^{K}a_{1}^{SLL}(D^{+}K^{-})$ (normalized by $-\lambda_{LL}(H)$) and $r_{\chi}^{K}a_{1}^{SLR}(D^{+}K^{-})$ (normalized by $-\lambda_{LR}(H)$), as an example, is shown in Fig.~\ref{scaledependent_SLL&SLR}, with
\begin{align}\label{eq:a1_SLL_SLR}
r_{\chi}^{K}a_{1}^{SLL}(D^{+}K^{-})=&\frac{2m_{K^-}^2}{\overline{m}_b(\mu)[\overline{m}_{u}(\mu)+\overline{m}_{s}(\mu)]}\bigg[ C_2^{SLL}(\mu)+\frac{C_1^{SLL}(\mu)}{N_c} +C_4^{SLL}(\mu)+\frac{C_3^{SLL}(\mu)}{N_c} \nn\\[0.1cm]
&\hspace{-0.8cm} + C_1^{SLL}(\mu)\int^1_0 du\,T^{SLL}(u,z)\Phi_p(u) + C_3^{SLL}(\mu)\int^1_0 du\,T^{TLL}(u,z)\Phi_p(u)\bigg]\,,\nn\\[0.2cm]
r_{\chi}^{K}a_{1}^{SLR}(D^{+}K^{-})=&\frac{2m_{K^-}^2}{\overline{m}_b(\mu)[\overline{m}_{u}(\mu)+\overline{m}_{s}(\mu)]}\bigg[ C_2^{SLR}(\mu)+\frac{C_1^{SLR}(\mu)}{N_c}\nn\\[0.1cm]
&\hspace{3.6cm} + C_1^{SLR}(\mu)\int^1_0 du\,T^{SLR}(u,z)\Phi_p(u)\bigg]\,. 
\end{align}
A similar behavior is also observed in the chirality-flipped sectors ($SRR$ and $SRL$). 

\begin{figure}[t]\vspace{-1.2cm}
	\centering	
	\centerline{\hspace{0.5cm}
		\includegraphics[width=0.55\textwidth]{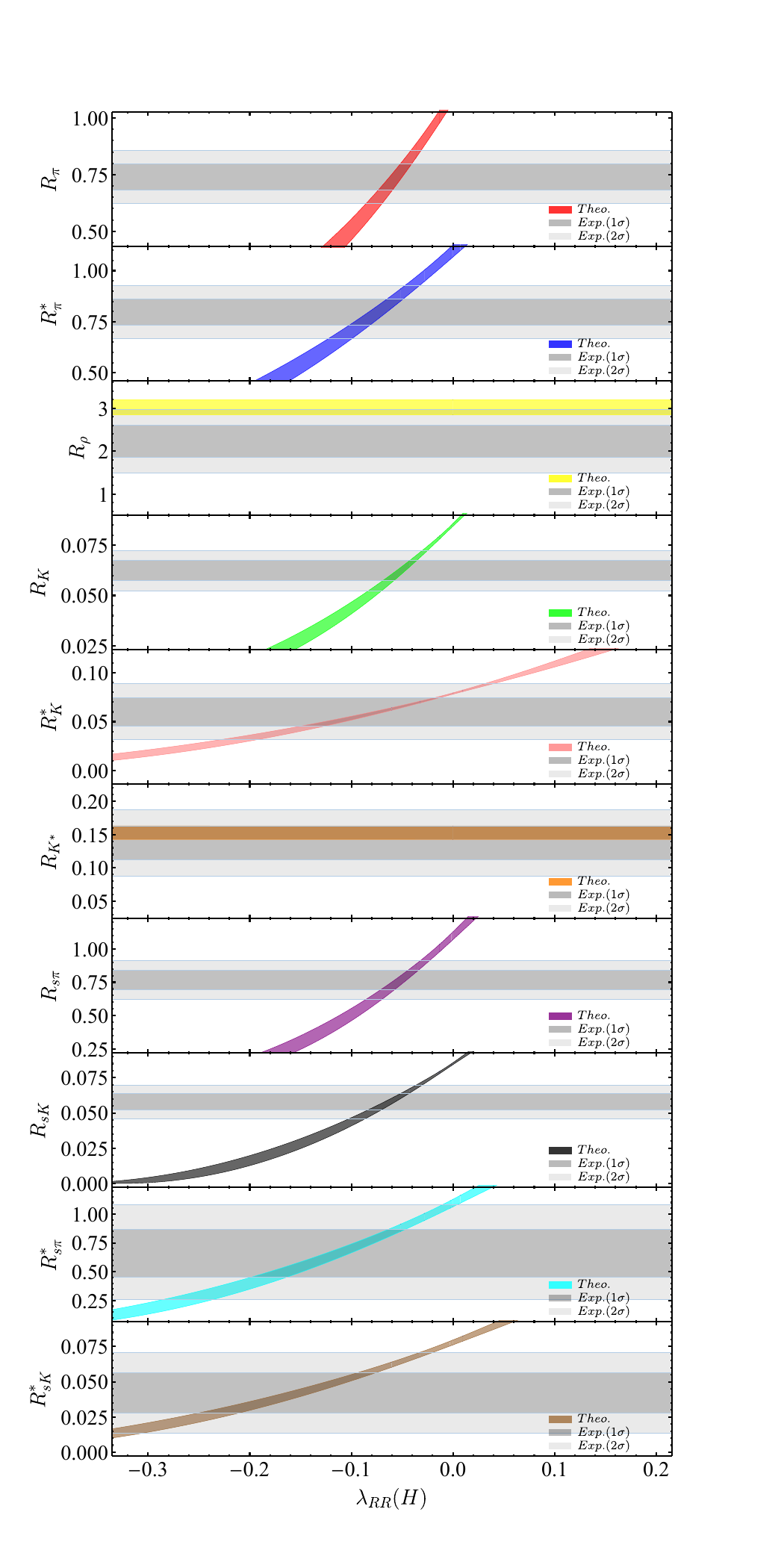}
		\hspace{-1.0cm}
		\includegraphics[width=0.55\textwidth]{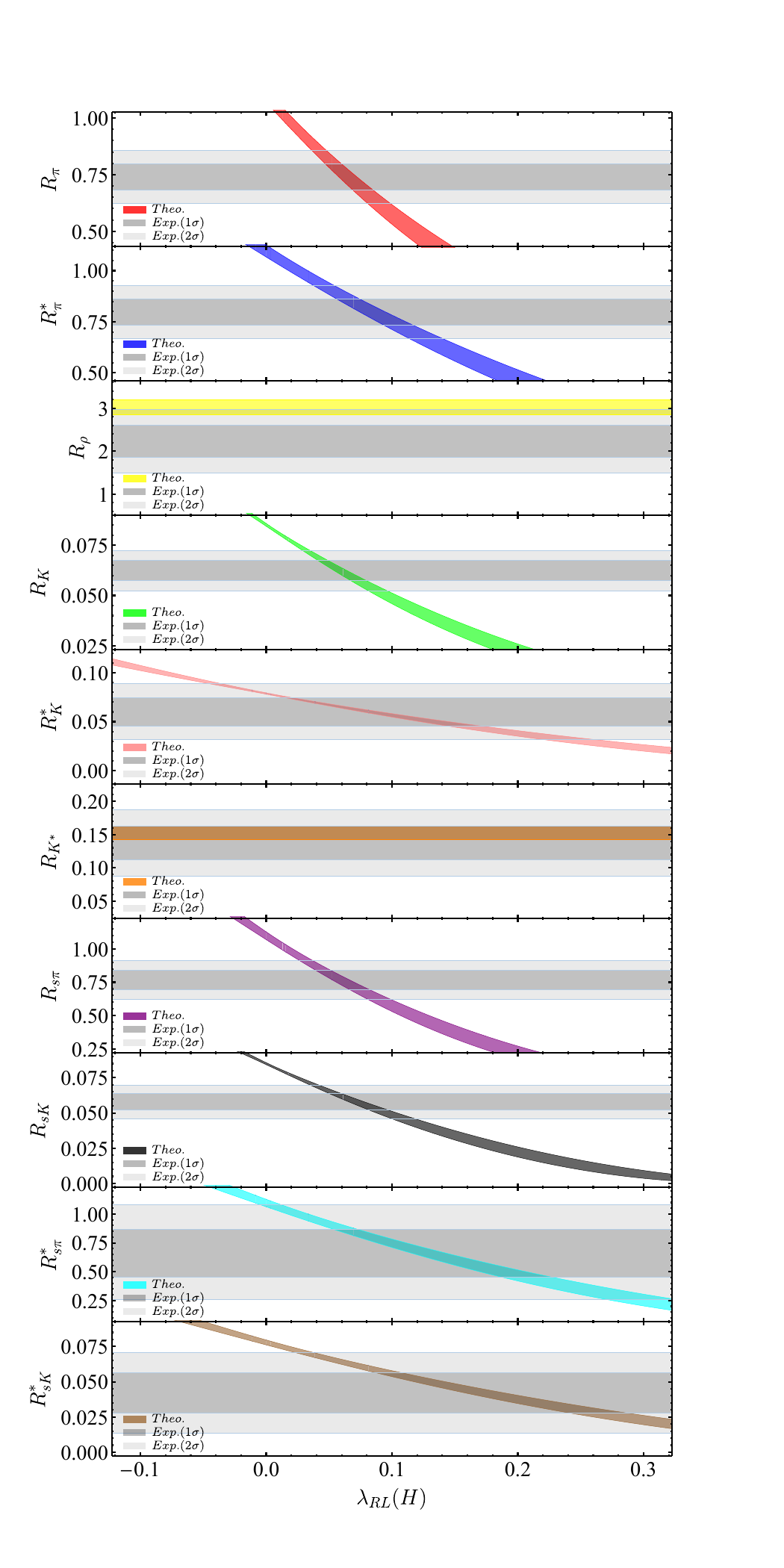}}
	\vspace{-0.5cm}
	\caption{\label{SRR&SRL} Constraints on the effective coefficients $\lambda_{RR}(H)$ (left) and $\lambda_{RL}(H)$ (right) from the ratios $R_{(s)L}^{(\ast)}$ collected in Table~\ref{tab:nonlep2semilep}. The other captions are the same as in Fig.~\ref{MIVLL}.}
\end{figure}

\begin{figure}[t]\vspace{-1.2cm}
	\centering	
	\centerline{\hspace{0.5cm}
		\includegraphics[width=0.55\textwidth]{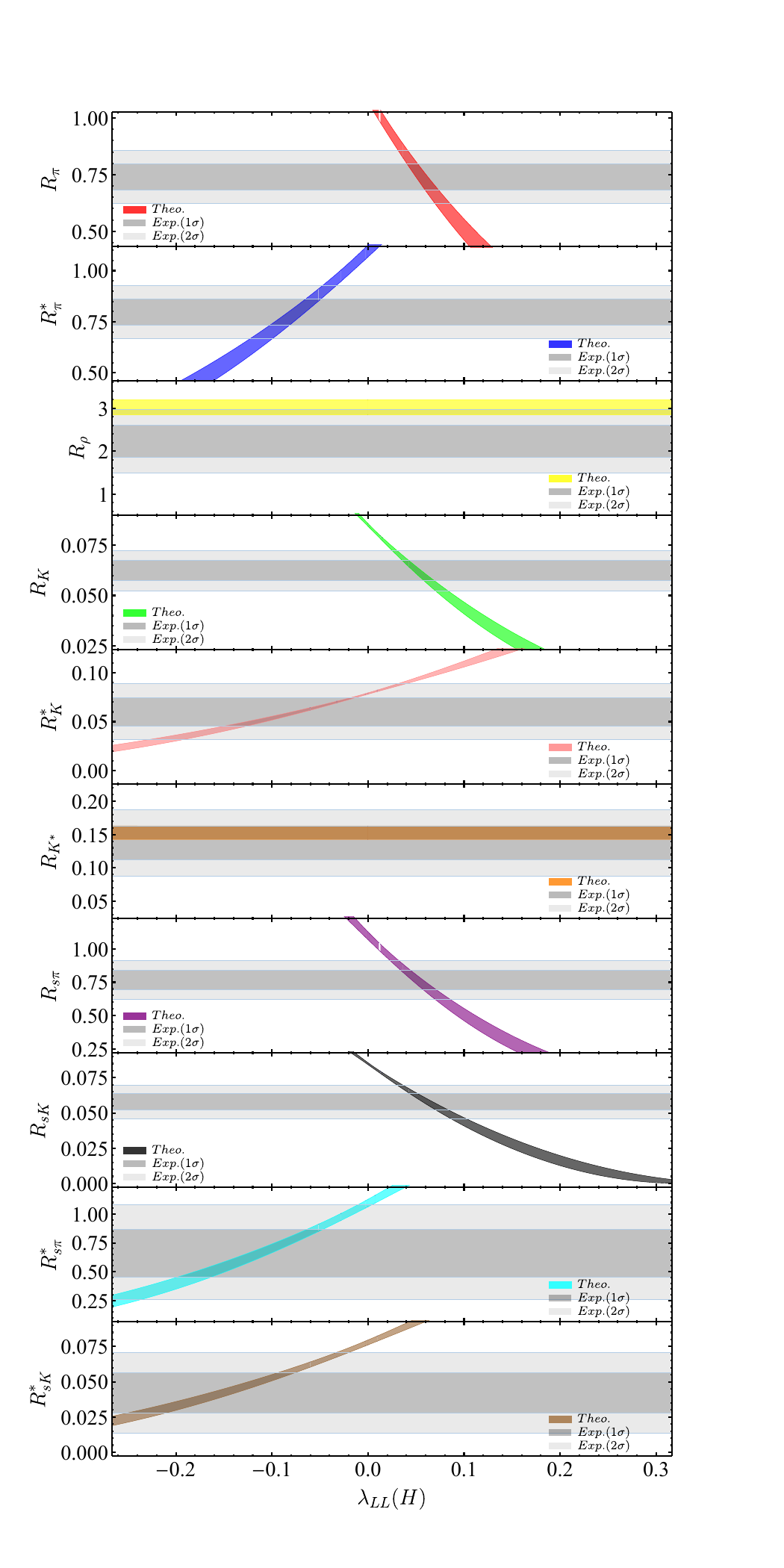}
		\hspace{-1.0cm}
		\includegraphics[width=0.55\textwidth]{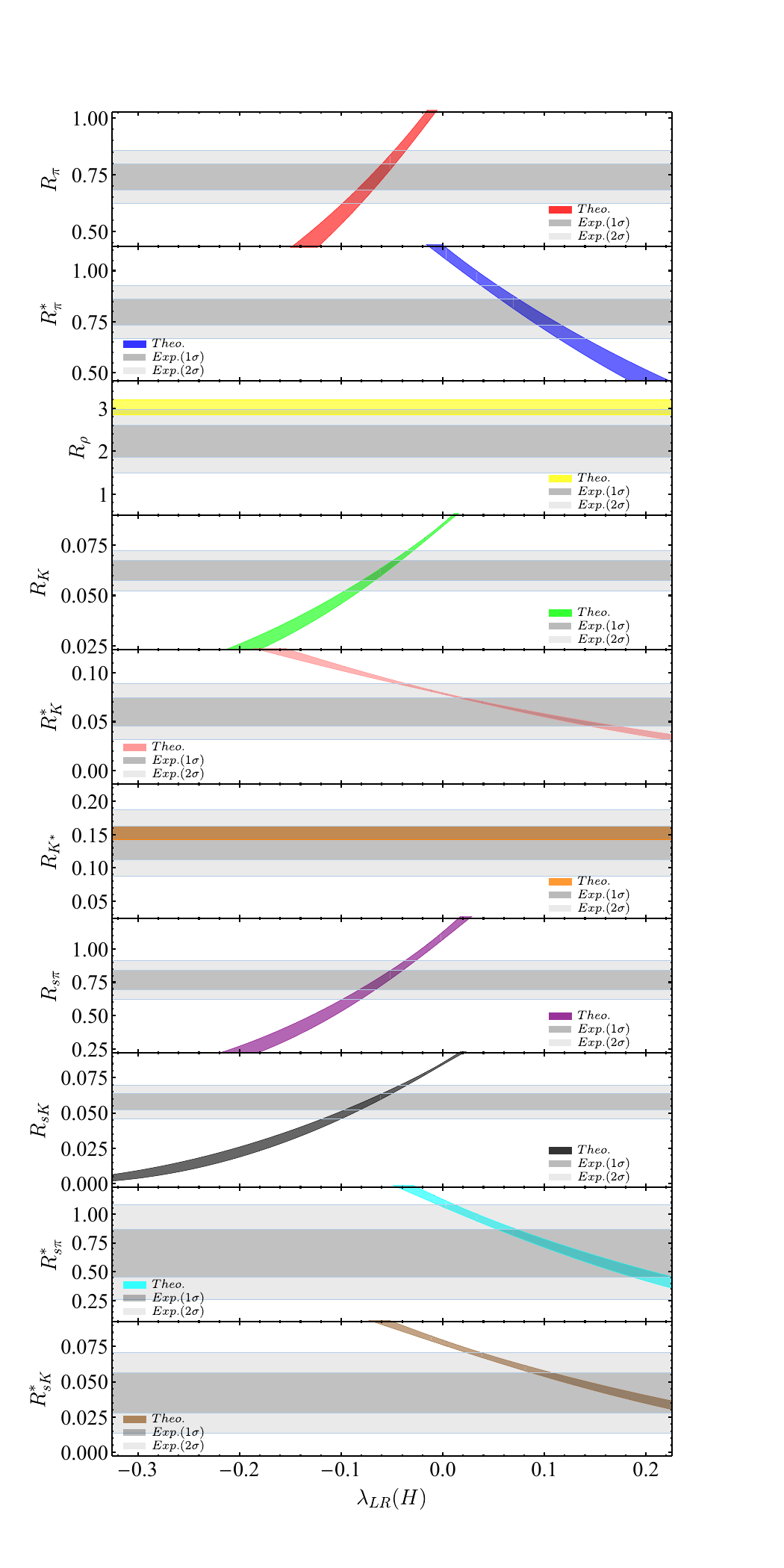}}
	\vspace{-0.5cm}
	\caption{\label{SLL&SLR} Constraints on the effective coefficients $\lambda_{LL}(H)$ (left) and $\lambda_{LR}(H)$ (right) from the ratios $R_{(s)L}^{(\ast)}$ collected in Table~\ref{tab:nonlep2semilep}. The other captions are the same as in Fig.~\ref{MIVLL}. }
\end{figure}

\begin{figure}[t]\vspace{-1.2cm}
	\centering	
	\centerline{\hspace{0.5cm}
		\includegraphics[width=0.55\textwidth]{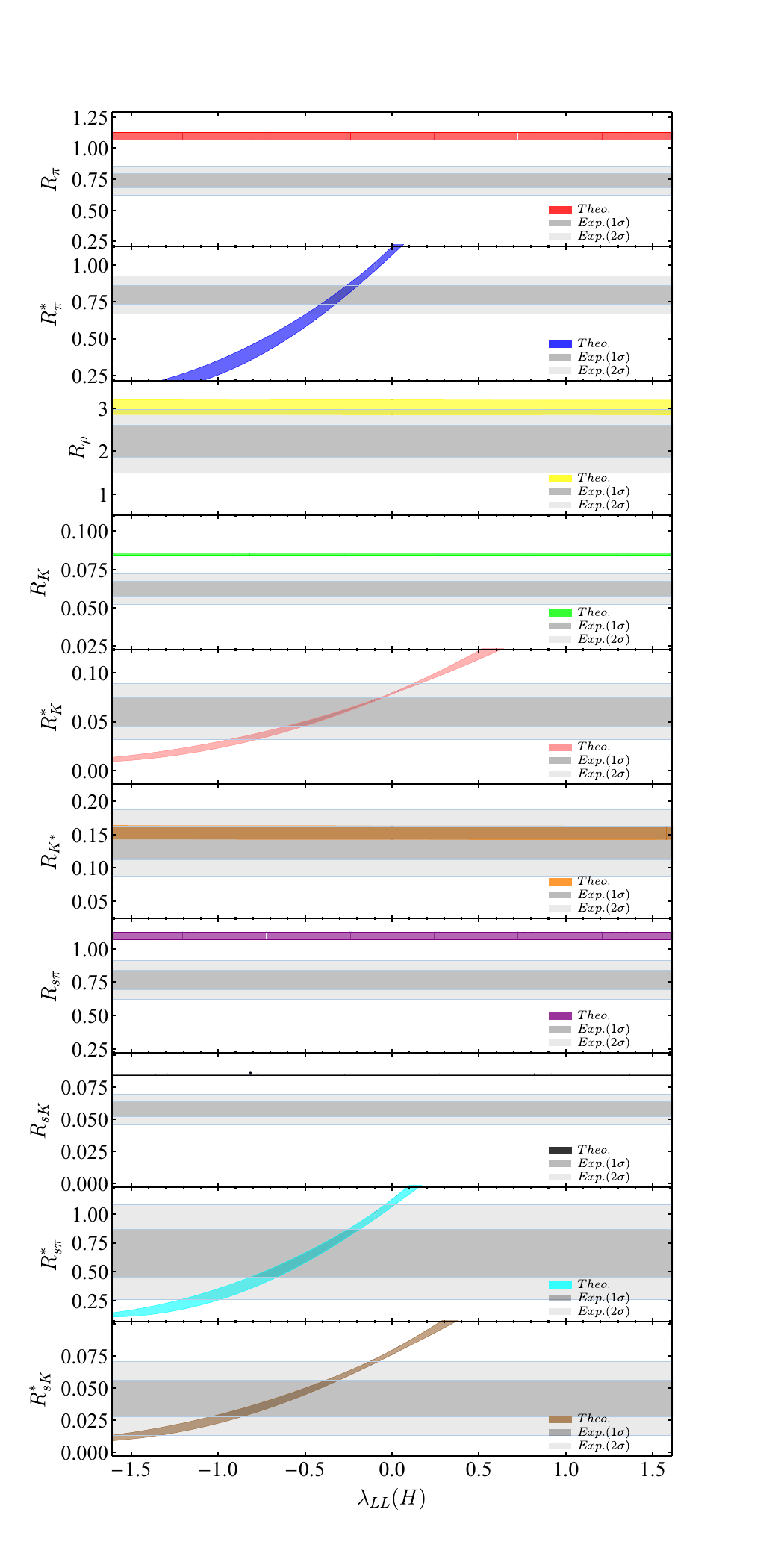}
		\hspace{-1.0cm}
		\includegraphics[width=0.55\textwidth]{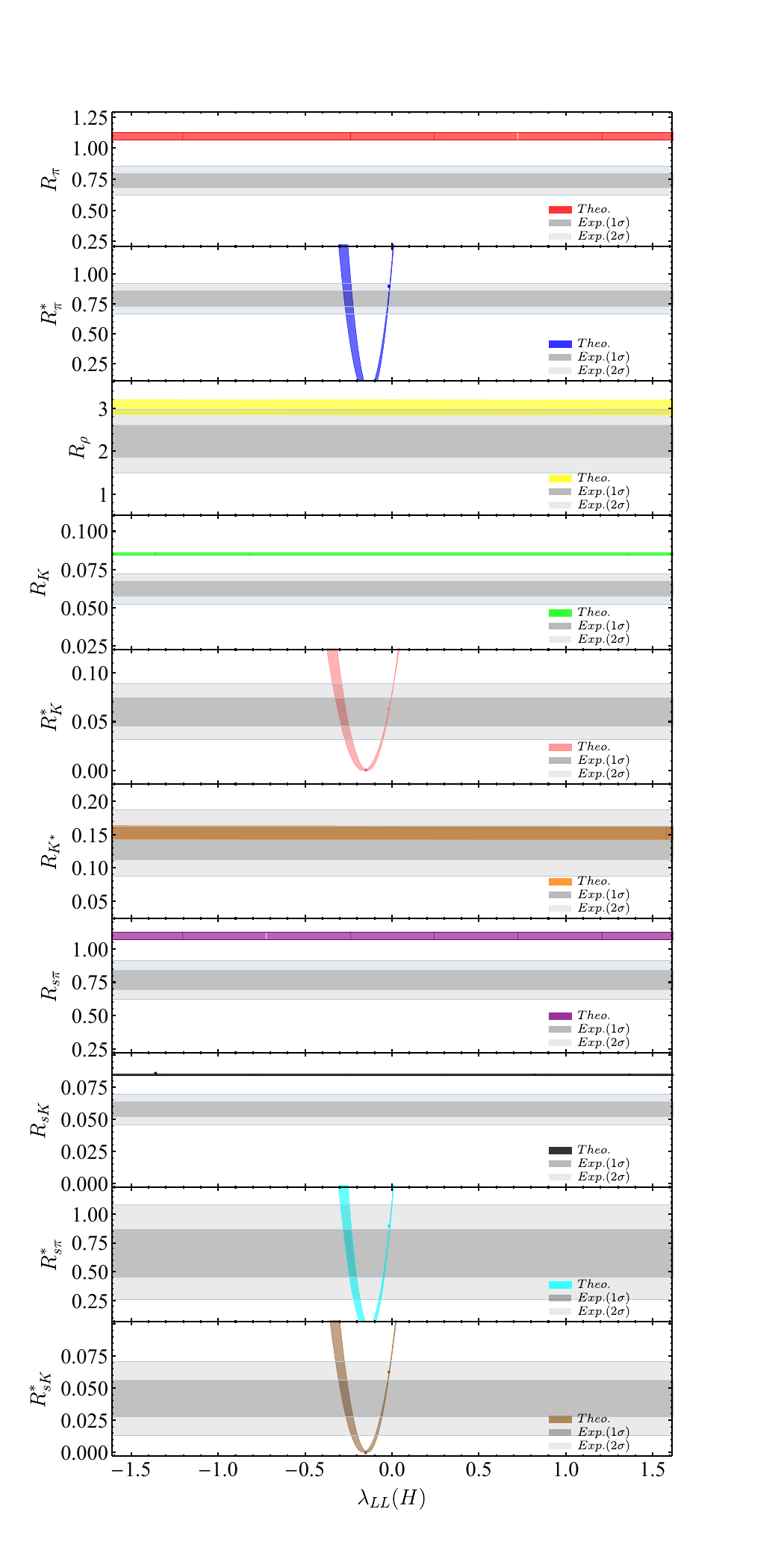}}
	\vspace{-0.5cm}
	\caption{\label{ScalarSym.&Asym.} Same as in Fig.~\ref{GaugeSym.&Asym.} but for the effective coefficient $\lambda_{LL}(H)$ in the colorless scalar case.}
\end{figure}

As in the case for the charged gauge boson, we shall also split the discussions into three different scenarios. Firstly, in scenario-I where only one nonzero effective coefficient is present in eq.~\eqref{eq:HamiltonianH}, it is found that all the deviations observed in $\bar{B}_{(s)}^0\to D_{(s)}^{(\ast)+}L^-$ decays could be explained simultaneously only in the presence of a nonzero $\lambda_{RR}(H)$ or $\lambda_{RL}(H)$, as shown in Fig.~\ref{SRR&SRL}. The resulting allowed ranges for $\lambda_{RR}(H)$ and $\lambda_{RL}(H)$ are given, respectively, as
\begin{equation}
	\lambda_{RR}(H)\in[-0.083,-0.033]\,,\qquad \lambda_{RL}(H)\in[0.038,0.096]\,.
\end{equation}
All the other cases in the presence of only a single effective coefficient are, however, ruled out already by the combined constraints from the ratios $R_{(s)L}^{(\ast)}$ collected in Table~\ref{tab:nonlep2semilep} at the $2\sigma$ level. As an explicit example, we show in Fig.~\ref{SLL&SLR} the individual constraint on the two effective coefficients $\lambda_{LL}(H)$ and $\lambda_{LR}(H)$ from the ratios $R_{\pi}$, $R_{\pi}^{*}$, $R_{\rho}$, $R_{K}$, $R_{K}^{*}$, $R_{K^{*}}$, $R_{s\pi}$, $R_{sK}$, $R_{s\pi}^{\ast}$, as well as $R_{sK}^{\ast}$, respectively. Secondly, we show in the left and the right panel of Fig.~\ref{ScalarSym.&Asym.} the individual constraint on the effective coefficient $\lambda_{LL}(H)$ from the ten ratios $R_{(s)L}^{(\ast)}$, in the scenarios where the left- and right-handed reduced couplings are symmetric (scenario-II with $\Delta_{cb}^{L}(H)=\Delta_{cb}^{R}(H)$ and $\Delta_{uq}^{L}(H)=\Delta_{uq}^{R}(H)$) and asymmetric (scenario-III with $\Delta_{cb}^{L}(H)=-\Delta_{cb}^{R}(H)$ and $\Delta_{uq}^{L}(H)=-\Delta_{uq}^{R}(H)$), respectively. One can see clearly that both of these two scenarios fail to provide a simultaneous account for the deviations observed in $\bar{B}_{(s)}^0\to D_{(s)}^{(\ast)+}L^-$ decays.

\section{Conclusions}
\label{sec:conclusions}

In this paper, motivated by the deviations observed between the updated SM predictions and the current experimental measurements of the branching ratios of $\bar{B}_{(s)}^0\to D_{(s)}^{(*)+} L^-$ decays with $L\in\{\pi,\rho,K^{(\ast)}\}$, we have investigated possible NP effects in these class-I non-leptonic $B$-meson decays. In order to facilitate a full NLO analysis, we have also calculated the one-loop vertex corrections to the hadronic matrix elements of the NP four-quark operators involved in these decays, within the QCDF framework.

Firstly, we have performed a model-independent analysis of the effects from twenty linearly independent four-quark operators that can contribute, either directly or through operator mixing, to the quark-level $b\to c\bar{u} d(s)$ transitions. Under the combined constraints from the ten ratios $R_{(s)L}^{(\ast)}$ collected in Table~\ref{tab:nonlep2semilep}, we found that the deviations observed could be well explained at the $1\sigma$ level by the NP four-quark operators with $\gam^{\mu}(1-\gam_5)\otimes\gam_{\mu} (1-\gam_5)$ structure, and also at the $2\sigma$ level by the operators with $(1+\gam_5)\otimes(1-\gam_5)$ and $(1+\gam_5)\otimes(1+\gam_5)$ structures. However, the NP operators with other Dirac structures fail to provide a consistent interpretation, even at the $2\sigma$ level. In the case where only a single nonzero NP Wilson coefficient is present in the effective weak Hamiltonian given by eq.~\eqref{eq:Hamiltonian}, the resulting allowed ranges for the corresponding NP Wilson coefficients are obtained both at the low-energy scale $\mu_b=m_b$ and at the electroweak scale $\mu_W=m_W$. In the case where two NP four-quark operators with the same Dirac but different color structures are present in eq.~\eqref{eq:Hamiltonian}, with the corresponding two NP Wilson coefficients varied simultaneously, it was found that only in the $(C_{2}^{VLL}(m_b),C_{1}^{VLL}(m_b))$, $(C_{2}^{SRL}(m_b),C_{1}^{SRL}(m_b))$, and $(C_{2}^{SRR}(m_b),C_{1}^{SRR}(m_b))$ planes are there allowed regions that can provide a simultaneous explanation of the ten ratios $R_{(s)L}^{(\ast)}$ collected in Table~\ref{tab:nonlep2semilep} at the $2\sigma$ level. 

As two specific examples of model-dependent considerations, we have also performed a full NLO analysis in the case where the NP four-quark operators are mediated by either a colorless charged gauge boson or a colorless charged scalar, with their masses fixed both at $1$~TeV. In each of these two cases, three different scenarios were considered. In scenario~I where only one effective coefficient is nonzero, we found that all the deviations observed in $\bar{B}_{(s)}^0\to D_{(s)}^{(\ast)+}L^-$ decays could be explained simultaneously only in the presence of a nonzero $\lambda_{LL}(A)$ in the case for a colorless charged gauge boson, as well as a nonzero $\lambda_{RR}(H)$ or $\lambda_{RL}(H)$ in the case for a colorless charged scalar, while all the other cases are ruled out already by the combined constraints from the ten ratios $R_{(s)L}^{(\ast)}$, even at the $2\sigma$ level. On the other hand, both of the other two scenarios where the left- and right-handed reduced couplings are symmetric (scenario-II) and asymmetric (scenario-III) fail to provide a simultaneous account for the deviations observed in $\bar{B}_{(s)}^0\to D_{(s)}^{(\ast)+}L^-$ decays.

As a final comment, it should be mentioned that our conclusions about the NP Wilson coefficients in the model-independent framework as well as the effective coefficients in the two model-dependent scenarios are very flavor-specific. If additional flavor-university assumptions were made between the different generations, \textit{e.g.}, between the up and charm quarks, other processes mediated by the tree-level $b\to c\bar{c}d(s)$ and loop-level $b\to d(s)q\bar{q}$ transitions will be involved, which are expected to provide further constraints on the NP parameter space~\cite{Bobeth:2014rda,Brod:2014bfa,Bobeth:2014rra,Lenz:2019lvd}. At the same time, in order to further discriminate the different solutions found for the deviations observed in these class-I non-leptonic $B$-meson decays, more precise measurements, especially of the decay modes involving $\rho$ and $K^\ast$ mesons, are urgently expected from the LHCb~\cite{Bediaga:2018lhg} and Belle II~\cite{Kou:2018nap} experiments. 

\section*{Acknowledgements}
We are very grateful to Nico Gubernari and Martin Jung for useful communications concerning the $B_{(s)}\to D_{(s)}^{(\ast)}$ transition form factors given in refs.~\cite{Bordone:2019vic,Bordone:2019guc}. This work is supported by the National Natural Science Foundation of China under Grant Nos.~12075097, 11675061 and 11775092, as well as by the Fundamental Research Funds for the Central Universities under Grant Nos.~CCNU20TS007 and 2019YBZZ078.

\appendix

\section{\boldmath Allowed ranges for $C_i(m_b)$ under the constraints from $R_{(s)L}^{(\ast)}$}
\label{app:A}

In this appendix, we give in Table~\ref{tab:constraint} the allowed ranges for the NP Wilson coefficients $C_i(m_b)$ under the individual and combined (last column) constraints from the ten ratios $R_{(s)L}^{(\ast)}$ varied within $1\sigma$ and $2\sigma$ error bars, respectively. For further details, the readers are referred to the main text presented in subsection~\ref{subsec:model-independent}.

\begin{landscape}
	\centering
	\tabcolsep0.03cm
	\let\oldarraystretch=\arraystretch
	\renewcommand*{\arraystretch}{1.2}
	\fontsize{7.8}{14.0}\selectfont
	\LTcapwidth=\linewidth
	\begin{longtable}{|c|c|c|c|c|c|c|c|c|c|c|c|c|c|c|}
		\hline
		\diagbox[width=7.8em]{NP Coeff.}{C.L.$\backslash$Obs.}& C.L. &$R_{\pi}$&$R_{\pi}^{*}$ &$R_{\rho}$ &$R_{K}$ &$R_{K}^{*}$ &$R_{K^{*}}$ &$R_{s\pi}$&$R_{sK}$&$R_{s\pi}^{\ast}$&$R_{sK}^{\ast}$&Combined\cr
		\hline
		\multirow{2}{*}{$C_{1}^{VLL}$}
		&$1\sigma$ &[-1.41,-0.836]&[-1.19,-0.615]&[-1.50,-0.267]&[-1.17,-0.654]
		&[-1.53,-0.136]&[-1.04,0.403]&[-1.37,-0.694] &[-1.42,-0.793]&[-2.31,-0.585] &[-2.65,-0.849] &[-1.04,-0.849]\cr
		\cline{2-13}
		&$2\sigma$ &[-1.63,-0.645]&[-1.42,-0.414]&[-2.06,0.135]&[-1.41,-0.453]
		&[-2.40,0.412]&[-1.69,0.868]&[-1.65,-0.459] &[-1.74,-0.548]&[-3.42,0.064] &[-4.06,-0.209] &[-1.41,-0.645]\cr
		\hline
		\multirow{2}{*}{$C_{2}^{VLL}$}
		&$1\sigma$ &[-0.238,-0.146]&[-0.207,-0.109]&[-0.254,-0.047]&[-0.196,-0.114]
		&[-0.259,-0.024]&[-0.181,0.073]&[-0.232,-0.121] &[-0.238,-0.138]&[-0.392,-0.103] &[-0.442,-0.148] &[-0.181,-0.148]\cr
		\cline{2-13}
		&$2\sigma$ &[-0.275,-0.113]&[-0.246,-0.073]&[-0.340,0.024]&[-0.235,-0.079]
		&[-0.400,0.073]&[-0.286,0.157]&[-0.278,-0.081] &[-0.287,-0.096]&[-0.558,0.011] &[-0.637,-0.037] &[-0.235,-0.113]\cr
		\hline
		\multirow{2}{*}{$C_{1}^{VLR}$}
		&$1\sigma$ &[0.370,0.605]&[0.275,0.525]&[-0.644,-0.119]&[0.292,0.499]
		&[0.061,0.661]&[-0.456,0.182]&[0.307,0.588] &[0.353,0.606]&[0.262,0.994] &[0.378,1.13] &$\varnothing$\cr
		\cline{2-13}
		&$2\sigma$ &[0.285,0.697]&[0.186,0.624]&[-0.865,0.060]&[0.203,0.599]
		&[-0.186,1.02]&[-0.722,0.396]&[0.204,0.705] &[0.245,0.733]&[-0.029,1.42] &[0.094,1.64] &$\varnothing$\cr
		\hline
		\multirow{2}{*}{$C_{2}^{VLR}$}
		&$1\sigma$ &[0.146,0.238]&[0.109,0.207]&[-0.254,-0.047]&[0.114,0.196]
		&[0.024,0.259]&[-0.181,0.073]&[0.121,0.232] &[0.138,0.238]&[0.103,0.392] &[0.148,0.442] &$\varnothing$\cr
		\cline{2-13}
		&$2\sigma$ &[0.113,0.275]&[0.073,0.246]&[-0.340,0.024]&[0.079,0.235]
		&[-0.073,0.400]&[-0.286,0.157]&[0.081,0.278] &[0.096,0.287]&[-0.011,0.558] &[0.037,0.637] &$\varnothing$\cr
		\hline
		\multirow{2}{*}{$C_{1}^{VRR}$}
		&$1\sigma$ &[0.836,1.41]&[-1.19,-0.615]&[-1.50,-0.267]&[0.654,1.17]
		&[-1.53,-0.136]&[-1.04,0.403]&[0.694,1.37] &[0.793,1.42]&[-2.31,-0.585] &[-2.65,-0.849] &$\varnothing$\cr
		\cline{2-13}
		&$2\sigma$ &[0.645,1.63]&[-1.42,-0.414]&[-2.06,0.135]&[0.453,1.41]
		&[-2.40,0.412]&[-1.69,0.868]&[0.459,1.65] &[0.548,1.74]&[-3.42,0.064] &[-4.06,-0.209] &$\varnothing$\cr
		\hline
		\multirow{2}{*}{$C_{2}^{VRR}$}
		&$1\sigma$ &[0.146,0.238]&[-0.207,-0.109]&[-0.254,-0.047]&[0.114,0.196]
		&[-0.259,-0.024]&[-0.181,0.073]&[0.121,0.232] &[0.138,0.238]&[-0.392,-0.103] &[-0.442,-0.148] &$\varnothing$\cr
		\cline{2-13}
		&$2\sigma$ &[0.113,0.275]&[-0.246,-0.073]&[-0.340,0.024]&[0.079,0.235]
		&[-0.400,0.073]&[-0.286,0.157]&[0.081,0.278] &[0.096,0.287]&[-0.558,0.011] &[-0.637,-0.037] &$\varnothing$\cr
		\hline
		\multirow{2}{*}{$C_{1}^{VRL}$}
		&$1\sigma$ &[-0.605,-0.370]&[0.275,0.525]&[-0.644,-0.119]&[-0.499,-0.292]
		&[0.061,0.661]&[-0.456,0.182]&[-0.588,-0.308] &[-0.606,-0.353]&[0.262,0.994] &[0.378,1.13] &$\varnothing$\cr
		\cline{2-13}
		&$2\sigma$ &[-0.697,-0.286]&[0.186,0.624]&[-0.865,0.060]&[-0.599,-0.203]
		&[-0.186,1.02]&[-0.722,0.396]&[-0.705,-0.204] &[-0.733,-0.245]&[-0.029,1.42] &[0.094,1.64] &$\varnothing$\cr
		\hline
		\multirow{2}{*}{$C_{2}^{VRL}$}
		&$1\sigma$ &[-0.238,-0.146]&[0.109,0.207]&[-0.254,-0.047]&[-0.196,-0.114]
		&[0.024,0.259]&[-0.181,0.073]&[-0.232,-0.121] &[-0.238,-0.138]&[0.103,0.392] &[0.148,0.442] &$\varnothing$\cr
		\cline{2-13}
		&$2\sigma$ &[-0.275,-0.113]&[0.073,0.246]&[-0.340,0.024]&[-0.235,-0.079]
		&[-0.072,0.400]&[-0.286,0.157]&[-0.278,0.081] &[-0.287,-0.096]&[-0.011,0.558] &[0.037,0.637] &$\varnothing$\cr
		\hline
		\multirow{2}{*}{$C_{1}^{SLL}$}
		&$1\sigma$ &[0.414,0.750]&[-1.03,-0.494]&$\varnothing$&[0.363,0.705]&[-1.49,-0.125]&R &[0.345,0.728] &[0.438,0.857]&[-1.99,0.471] &[-2.54,-0.759] &$\varnothing$\cr
		\cline{2-13}
		&$2\sigma$ &[0.322,0.869]&[-1.23,-0.335]&R &[0.254,0.848]&[-2.31,0.406]&R &[0.231,0.879] &[0.306,1.04]&[-2.87,0.053] &[-3.69,-0.191] &$\varnothing$\cr
		\hline
		\multirow{2}{*}{$C_{2}^{SLL}$}
		&$1\sigma$ &[0.138,0.250]&[-0.344,-0.165]&$\varnothing$&[0.121,0.235]&[-0.497,-0.042]&R &[0.115,0.243] &[0.146,0.286]&[-0.664,-0.157] &[-0.846,-0.253] &$\varnothing$\cr
		\cline{2-13}
		&$2\sigma$ &[0.107,0.290]&[-0.412,-0.112]&R &[0.085,0.283]&[-0.771,0.135]&R &[0.077,0.293] &[0.102,0.346]&[-0.955,0.018] &[-1.23,-0.064] &$\varnothing$\cr
		\hline
		\multirow{2}{*}{$C_{1}^{SLR}$}
		&$1\sigma$ &[-0.875,-0.482]&[0.575,1.20]&$\varnothing$&[-0.822,-0.423]&[0.145,1.74]&R &[-0.849,-0.403] &[-1.00,-0.511]&[2.32,0.549] &[0.222,4.30] &$\varnothing$\cr
		\cline{2-13}
		&$2\sigma$ &[-1.01,-0.374]&[0.390,1.44]&R &[-0.989,-0.296]&[-0.472,2.70]&R &[-1.03,-0.269] &[-1.21,-0.356]&[3.34,0.062] &[0.884,2.96] &$\varnothing$\cr
		\hline
		\multirow{2}{*}{$C_{2}^{SLR}$}
		&$1\sigma$ &[-0.250,-0.138]&[0.165,0.344]&$\varnothing$&[-0.235,-0.121]&[0.042,0.497]&R &[-0.242,-0.115] &[-0.286,-0.146]&[0.157,0.664] &[0.253,0.846] &$\varnothing$\cr
		\cline{2-13}
		&$2\sigma$ &[-0.290,-0.107]&[0.112,0.412]&R &[-0.283,-0.085]&[-0.135,0.771]&R &[-0.293,-0.077] &[-0.346,-0.102]&[0.955,-0.018] &[0.064,1.23] &$\varnothing$\cr
		\hline
		\multirow{2}{*}{$C_{1}^{SRR}$}
		&$1\sigma$ &[-0.750,-0.414]&[-1.03,-0.494]&$\varnothing$&[-0.705,-0.363]&[-1.49,-0.125]&R &[-0.728,-0.345] &[-0.857,-0.438]&[-1.99,-0.471] &[-2.54,-0.759] &$\varnothing$\cr
		\cline{2-13}
		&$2\sigma$ &[-0.869,-0.322]&[-1.23,-0.335]&R &[-0.848,-0.254]&[-2.31,0.406]&R &[-0.879,-0.231] &[-1.04,-0.306]&[-2.87,0.053] &[-3.69,-0.191] &[-0.848,-0.335]\cr
		\hline
		\multirow{2}{*}{$C_{2}^{SRR}$}
		&$1\sigma$ &[-0.250,-0.138]&[-0.344,-0.165]&$\varnothing$&[-0.235,-0.121]&[-0.497,-0.042]&R &[-0.243,-0.115] &[-0.286,-0.146]&[-0.664,-0.157] &[-0.846,-0.253] &$\varnothing$\cr
		\cline{2-13}
		&$2\sigma$ &[-0.290,-0.107]&[-0.412,-0.112]&R &[-0.283,-0.085]&[-0.771,0.135]&R &[-0.293,-0.077] &[-0.346,-0.102]&[-0.955,0.018] &[-1.23,-0.064] &[-0.283,-0.112]\cr
		\hline
		\multirow{2}{*}{$C_{1}^{SRL}$}
		&$1\sigma$ &[0.482,0.875]&[0.575,1.20]&$\varnothing$&[0.423,0.822]&[0.145,1.74]&R &[0.403,0.849] &[0.511,1.00]&[0.549,2.32] &[0.884,2.96] &$\varnothing$\cr
		\cline{2-13}
		&$2\sigma$ &[0.374,1.01]&[0.390,1.44]&R &[0.296,0.989]&[-0.472,2.70]&R &[0.269,1.03] &[0.356,1.21]&[-0.062,3.34] &[0.222,4.30] &[0.390,0.989]\cr
		\hline
		\multirow{2}{*}{$C_{2}^{SRL}$}
		&$1\sigma$ &[0.138,0.250]&[0.165,0.344]&$\varnothing$&[0.121,0.235]&[0.042,0.497]&R &[0.115,0.243] &[0.146,0.286]&[0.157,0.664] &[0.253,0.846] &$\varnothing$\cr
		\cline{2-13}
		&$2\sigma$ &[0.107,0.290]&[0.112,0.412]&R &[0.085,0.283]&[-0.135,0.771]&R &[0.077,0.293] &[0.102,0.346]&[-0.018,0.955] &[0.064,1.23] &[0.112,0.283]\cr
		\hline       
		\caption{\label{tab:constraint} \setlength{\LTcapwidth}{4in} Allowed ranges for the NP Wilson coefficients $C_i(m_b)$ under the individual and combined constraints (last column) from the ten ratios $R_{(s)L}^{(\ast)}$ varied within $1\sigma$ and $2\sigma$ error bars, respectively. Here ``$\varnothing$'' represents an empty set and ``R'' the set of all real numbers within the plot ranges for $C_i(m_b)$.} 
	\end{longtable}
\end{landscape}

\bibliographystyle{JHEP}
\bibliography{reference}

\end{document}